\documentclass[12pt, a4paper]{book}

\usepackage{makeidx}
\usepackage{booktabs} 
\usepackage{graphicx}
\usepackage{amsmath}
\usepackage{bm}

\makeindex
\title{\large{\textbf{PhD thesis}}\\
\Huge{\textbf{Photon stimulated desorption of and nuclear resonant scattering}\\
\textbf{by noble gas atoms\\at solid surfaces}}\\
	}
	
\author{\large\textbf{Akihiko Ikeda}\\
	{\textbf{Institute of Industrial Science, The University of Tokyo}}}
\date{\textbf{March 2013}}
\begin{document}
\maketitle
\frontmatter

\paragraph{Note:}This is a PhD thesis approved by the University of Tokyo in March 2013. The contents concerning the Knudsen layer formation and the nuclear resonant scattering were later published as J. Chem. Phys. \textbf{138}, 124705 (2013) and as Phys. Rev. B. \textbf{91}, 155402 (2015), respectively. The thesis was uploaded on arXiv in 2015 [arXiv:1501.04297]. Several figures are revised in the uploaded materials.

\begin{flushright}
A. I.\\
April, 2015
\end{flushright}

\chapter{Acknowledgement}
I am most grateful to Prof. K. Fukutani and Prof. T. Okano for their patience,  passionate discussions and kind instructions. Thoughtfull and best supports by Dr. M. Matsumoto, associate Prof. M. Wilde, Mr. T. Kawauchi, Dr. S. Ogura and Ms. T. Nakamura are greatly acknowledged. I would also like to thank Ms. A. Kashifuku, Dr. H. Kasai, Dr. T. Sugimoto, Dr. K. Yamakawa, Mr. S. Ohno, Mr. K. Takeyasu, Mr. Y. Itakura, Mr. Y. Kazama, Mr. S. Watanabe, Mr. T. Namba, Mr. K. Fukada, Ms. Wen Di, Mr. K. Asakawa and Mr. K. Miyao for valuable discussions and pleasant life as students in IIS. Delightful conversations with Dr. Y.-C. Ong, and Dr. H. Yonemura are recognized. Profound insights and advice from emeritus Prof. A. Kimbara, emeritus Prof. Y. Fukai, emeritus Prof. Y. Murata and Prof. T. Kawamura are greatly appreciated. Substantial supports from Dr. Y. Yoda at SPring-8 and encouraging comments from associate Prof. K. Watanabe at Tokyo Univ. Sci. are also acknowledged. I acknowledge Prof. K. Fukutani, Prof. T .Takahashi, Prof. F. Komori, Prof. A. Oshiyama and associate Prof. Y. Hasegawa who thoroughly reviewed the thesis.

Last but not least, I am glad to express best thankfulness to my family who supported me for my life. I am also happy to acknowledge sincere gratitude to my wife Kanako and her family for their continuous support and encouragement.

\begin{flushright}
A. I.\\
February, 2013
\end{flushright}

\begin{figure}
\begin{center}
\includegraphics[scale=.5, clip]{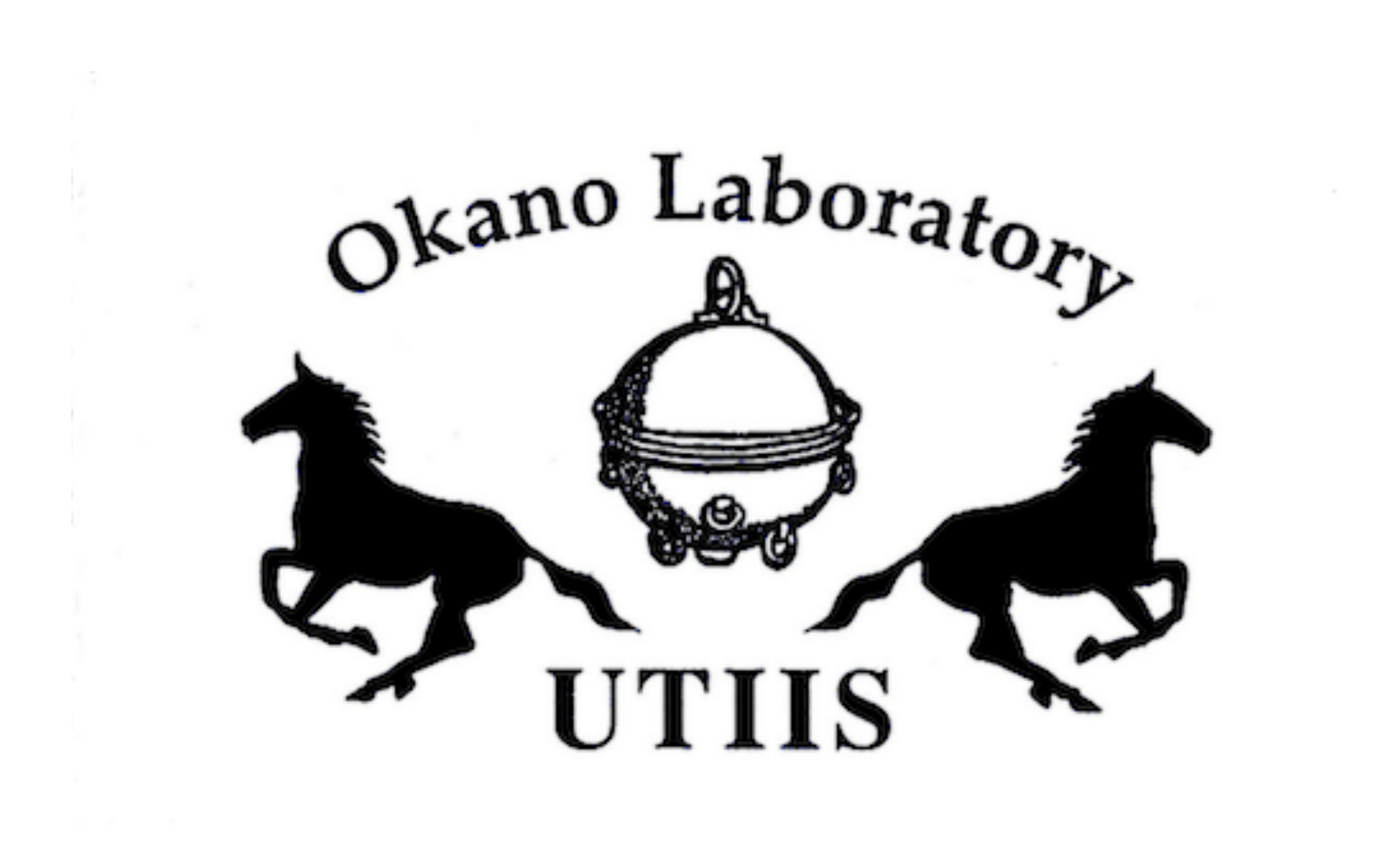} 
\caption{Symbol of Okano laboratory (1982 - 2012).\label{okano}}
\end{center}
\end{figure}

\chapter{Abstract}
When a noble gas atom approaches a solid surface, it is adsorbed via the Van der Waals force, which is called physisorption. In this thesis, several experimental results concerning physisorbed atoms at surfaces are presented. First, photon stimulated desorption of Xe atoms from a Au substrate using nano-second laser is presented. With the time-of-flight measurements, the translational temperature and the desorption yield of desorbing Xe as a function of laser fluence are obtained. It is discovered that there are non-thermal and thermal desorption pathways. It is discussed that the former path involves a transient formation of the negative ion of Xe. The desorption flux dependence of the thermal pathway is also investigated. We found that at a large desorption fluxes the desorption flow is thermalized due to the post-desorption collisions. The resultant velocity and the temperature of the flow is found to be in good agreement with the theoretical predictions based on the Knudsen layer formation. Lastly, nuclear resonant scattering of synchrotron radiation by the multi- and mono-layer of $^{83}$Kr at a surface of the titanium oxide is presented. The use of the noble gas atoms as the probes of the electric field gradient at the solid surfaces is discussed.

\clearpage

This thesis is based on the following writings.

\begin{enumerate}
\item A. Ikeda$^{\mathrm{a}}$, M. Matsumoto$^{\mathrm{a}}$, S. Ogura$^{\mathrm{a}}$, K. Fukutani$^{\mathrm{a}}$ and T. Okano$^{\mathrm{a}}$, ''\textit{Photostimulated desorption of Xe from Au(001) surfaces via transient Xe$^{-}$ formation},'' Phys. Rev. B \textbf{84}, 155412 (2011). 
\item A. Ikeda$^{\mathrm{a}}$, M. Matsumoto$^{\mathrm{a}}$, S. Ogura$^{\mathrm{a}}$, T. Okano$^{\mathrm{a}}$ and K. Fukutani$^{\mathrm{a}}$, ''\textit{Knudsen layer formation in laser induced thermal desorption},'' J. Chem. Phys. in press.
\item A. Ikeda$^{\mathrm{a}}$, T. Kawauchi$^{\mathrm{a}}$, M. Matsumoto$^{\mathrm{a}}$, T. Okano$^{\mathrm{a}}$, K. Fukutani$^{\mathrm{a}}$, X. W. Zhang$^{\mathrm{b}}$ and Y. Yoda$^{\mathrm{c}}$, ''\textit{Nuclear resonant scattering of synchrotron radiation by $^{83}$Kr on TiO$_{2}$(110) surfaces},'' In preparation.
\\
\\
$^{\mathrm{a}}$ Institute of Industrial Science, University of Tokyo, Tokyo, Japan\\
$^{\mathrm{b}}$ High Energy Accelerator Research Organization (KEK), Tsukuba, Japan\\
$^{\mathrm{c}}$ Japan Synchrotron Research Institute (JASRI), Hyogo, Japan

\end{enumerate}

\tableofcontents

\mainmatter

\chapter{Introduction}
In this chapter, the background and the objective of the present thesis are described. The first section focuses on the electronic structures of physisorption systems, photon stimulated process on the solid surfaces, the post-desorption collisions and the introduction of the nuclear resonant scattering (NRS). In the second section, the three objectives of the present study are presented.

\section{Background}
In this section, the several backgrounds of the present studies are described. First, the modifications of electronic structure of rare gas atoms as a result of physisorption are presented based on the previous studies. Besides, the experimental methods for the investigation of electronic structures of physisorption systems is discussed. The following subsection focuses on the photon stimulated desorption (PSD) of rare gas atoms from solid surfaces. The capability of PSD as a probe for the unoccupied electronic states are discussed. The effects of the post-desorption collisions on the laser desorption experiments are also discussed. In the last subsection, NRS is introduced as a unconventional  type of M\"ossbauer spectroscopy making use of the synchrotron radiation. The use of NRS to probe the modification of the electronic structure of physisorbed atoms due to the electric field gradient at the solid surface is discussed.

\begin{figure}
\begin{center}
\includegraphics[scale=.4, clip]{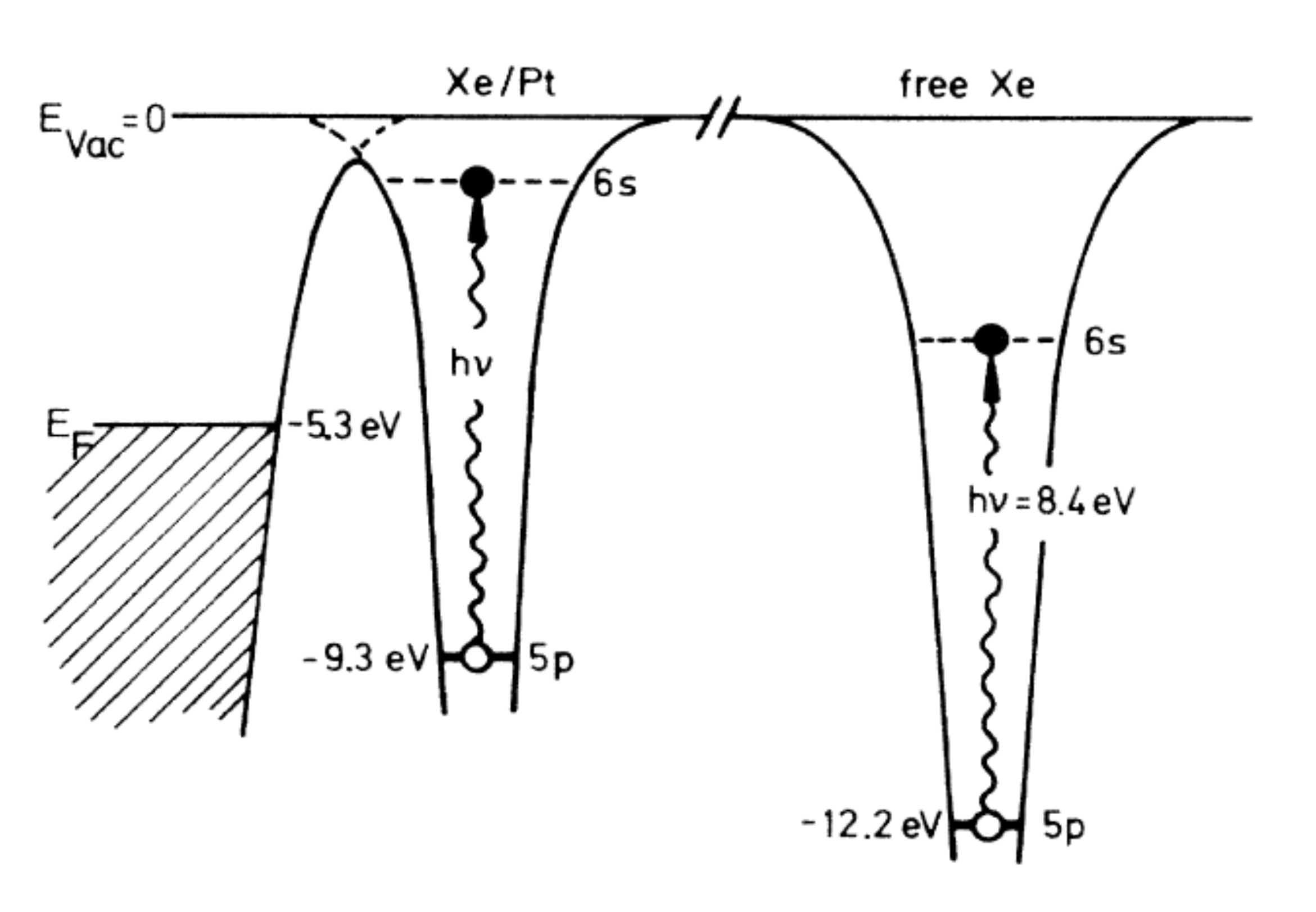}
\caption{A schematic drawing of the effective one-electron potential curves for Xe atoms in a monolayer on the Pt(111) surface and for free Xe atoms. The image is adopted from Ref. \cite{Schonhense}.\label{xept}}
\end{center}
\end{figure}

\subsection{Physisorption}
Although noble gas atoms are not bound to solid surfaces by forming a chemical bond, they are weakly bound to the surfaces through Van der Waals interactions, which is called physisorption. Van der Waals force possesses its roots in the second-order perturbation energy of the Coulomb interactions between neutral atoms \cite{Sakurai}. The attractive potential due to Van der Waals force between the two neutral atoms decays with $1/r^{6}$, where the two atoms are sufficiently separated in distance compared with the Bohr radius. On the other hand, the attractive potential due to the Van der Waals interaction between an atom and a metal surface decays with $1/r^{3}$. This is because in the case of atom-surface system the attraction arise from the interaction between an induced dipole and an image dipole, whereas in the case of two neutral atoms two induced dipoles are in play \cite{Zangwill}. When the atom  gets very close to the surface, it is repelled due to the Pauli repulsion which exponentially increases with decreasing the separation. Therefore, in the physisorption regime, the potential minima appears at some distance from the surface, which is typically at around 3 $\sim$ 4 \AA \ for most of the species.

Physisorption occurs with a weak interaction with the typical values of 30 to 300 meV for the depth of the potential well. Therefore, it has been believed that the electronic structure of the physisorbed species is not significantly modified compared with their electronic structures in the gas phase. The estimation of the interaction energy and the electronic structure with calculations have suffered from the poor accuracy which is comparable to the interaction energies themselves, so far. Actually, it has been speculated that the rare gas atoms adsorbed on a hollow sites on closed packed surfaces because it is most highly coordinated. However, it has been recently found by several experiments \cite{Eigler, Weiss} and ab-initio calculations that Xe atoms adsorb on the atop sites on most metal surfaces \cite{Silva, Kelkkanen, Chen}.

Physisorbed rare gas atoms are simplest amongst the physisorbed species that includes not only various inert gas molecules but also the complex organic molecules. The binding energy of physisrbed atoms is in the order of a few 10 meV to 100 meV. The equilibrium distance of the physisorbed atom from the substrate varies from 3 to 5 \AA\ \cite{Bruch}. Generally speaking, the interactions between physisorbed atoms and the substrate is small compared to the interactions of the ones who form chemical bonds to the substrate, which is called chemisorbed species.

Nevertheless, several studies have shown the trends otherwise. Although Van der Waals interactions are weak, the electronic states of Xe atom were found to appreciably modified as a result of physisorption on metal surfaces by a photoemission spectroscopy \cite{Mandel} and a resonant photoemission study \cite{Schonhense}. Figure \ref{xept} shows a schematic adiabatic potential for the physisorption system of monolayer Xe on Pt(111), which has been clarified by the resonant photoemission study \cite{Schonhense}. The ionization potential and the first excitation energy of a free Xe atom is known to lie 12.2 eV below the vacuum level and 8.4 eV, respectively. On the Pt(111) surface, the ionization potential was shown to relaxed to be 9.3 eV below the vacuum level whereas the excitation energy was almost the same as that of a free Xe. The relaxation was assumed to occur mainly due to the image charge screening. Although the modification of the occupied electronic states has been investigated on physisorption systems, the unoccupied states represented by electron affinities have not been well investigated so far.

Similar to the case of the electronic states of physisorbed atoms, that of the substrate are also reported to be modified. The work function of the substrate were known to greatly reduced as a result of physisorbed layer of rare gas atoms \cite{YCChen}. The reduction of the work function is intuitively explained by either one of the following context. (1) The charge transfer from the substrate to the physisrobed atoms form an inverse dipole with regard to the surface dipole which is theoretically supported by Ref. \cite{Muller}. (2) The push back effect is in play where the physisrobed atom push back the extended substrate electrons back into the bulk, which is theoretically supported by Ref. \cite{Bagus, Silva2008}. At this moment, the latter idea is widely accepted because it was supported by the report that the chemical shift of the physisorbed rare gas atoms was negligibly small with x ray photoemission experiments. The direct observation of the push back effect, however, is not presented so far. Silva and coworkers suggested that in the push back regime not only the substrate but also the physisorbed rare gas atoms are largely polarized after the physisorption \cite{Silva, Silva2, Silva2008}. Therefore, it is possible that the polarization or some kind of modification of the electronic structure of rare gas atoms are observed as a confirmation of the physisorption induced modification of the electronic structure of atoms and the substrates.

There are several ways to detect such electronic properties of physisorbed rare gas atoms. One of the ways to detect the electronic properties of the physisorbed atom is to follow the photo-induced motion of the atom, for the atomic motion is determined by the electronic structure specific to the atom. Another way is observation of the hyperfine splitting of the nucleus of the physisorbed atom due to the strong electric field gradient at the surface as a probe of electronic structure. Both ways are explored in this thesis.

\subsection{Photon stimulated desorption}

\begin{figure}
\begin{center}
\includegraphics[scale=.3, clip]{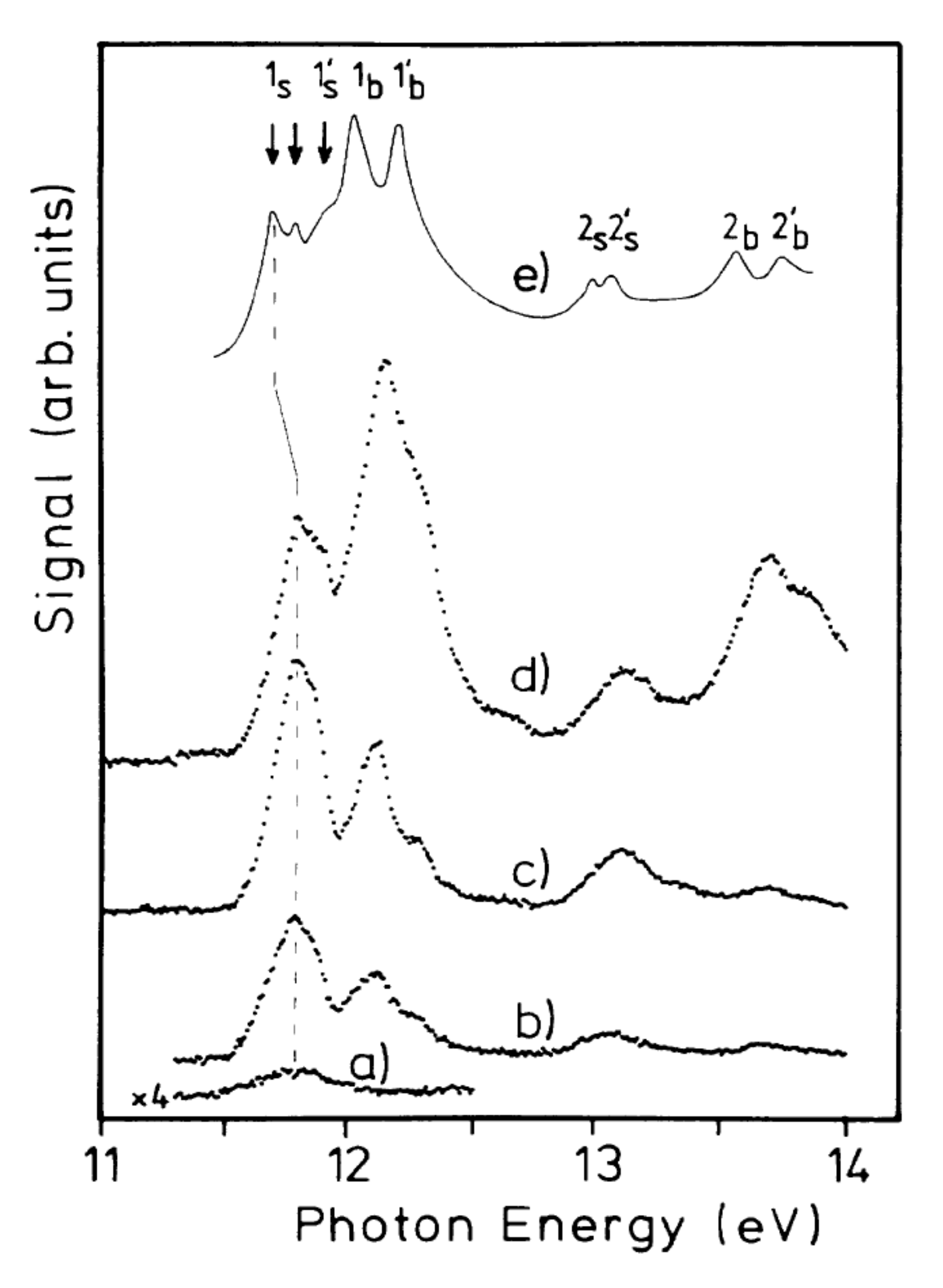} 
\caption{Photon stimulated desorption yield of Ar mono- and multi-layers on Ru(0001) following photon irradiations ranging from 11 to 14 eV generated with synchrotron radiations. The image is adapted from Ref. \cite{Feulner1987}.\label{Feulner}}
\end{center}
\end{figure}

Photo-stimulated processes at solid surfaces have been a topic of extensive studies because they allow us to control adsorbates in either thermal or nonthermal ways \cite{Ho1, Chuang, Fukutani2003}. Laser-induced thermal desorption (LITD) was investigated in detail for the systems of CO/Fe(110) (Ref.~\cite{Wedler}) and Xe/Cu \cite{Hussla}, and has been successfully applied to the studies of surface diffusion combined with low-energy electron microscopy \cite{Yim} or with scanning tunneling microscopy \cite{Schwalb}. Nonthermal photostimulated phenomena, on the other hand, provide us with pathways that are nonaccessible in a thermal process. The nonthermal photostimulated desorption (PSD) of rare gas atoms from metal surfaces has been investigated using photons of two energy regions: At $h\nu>$ 7 eV, the excitonic or ionic excitation of the mono and multi-layers of Ar and Kr induces desorption \cite{Feulner1987} as shown in Fig. \ref{Feulner}, while infrared light at $h\nu<$ 1 eV causes the direct excitation of the vibrational mode in the physisorption well to a continuum state \cite{Pearlstine}.

\begin{figure}
\begin{center}
\includegraphics[scale=.85, trim = 0 10 0 0]{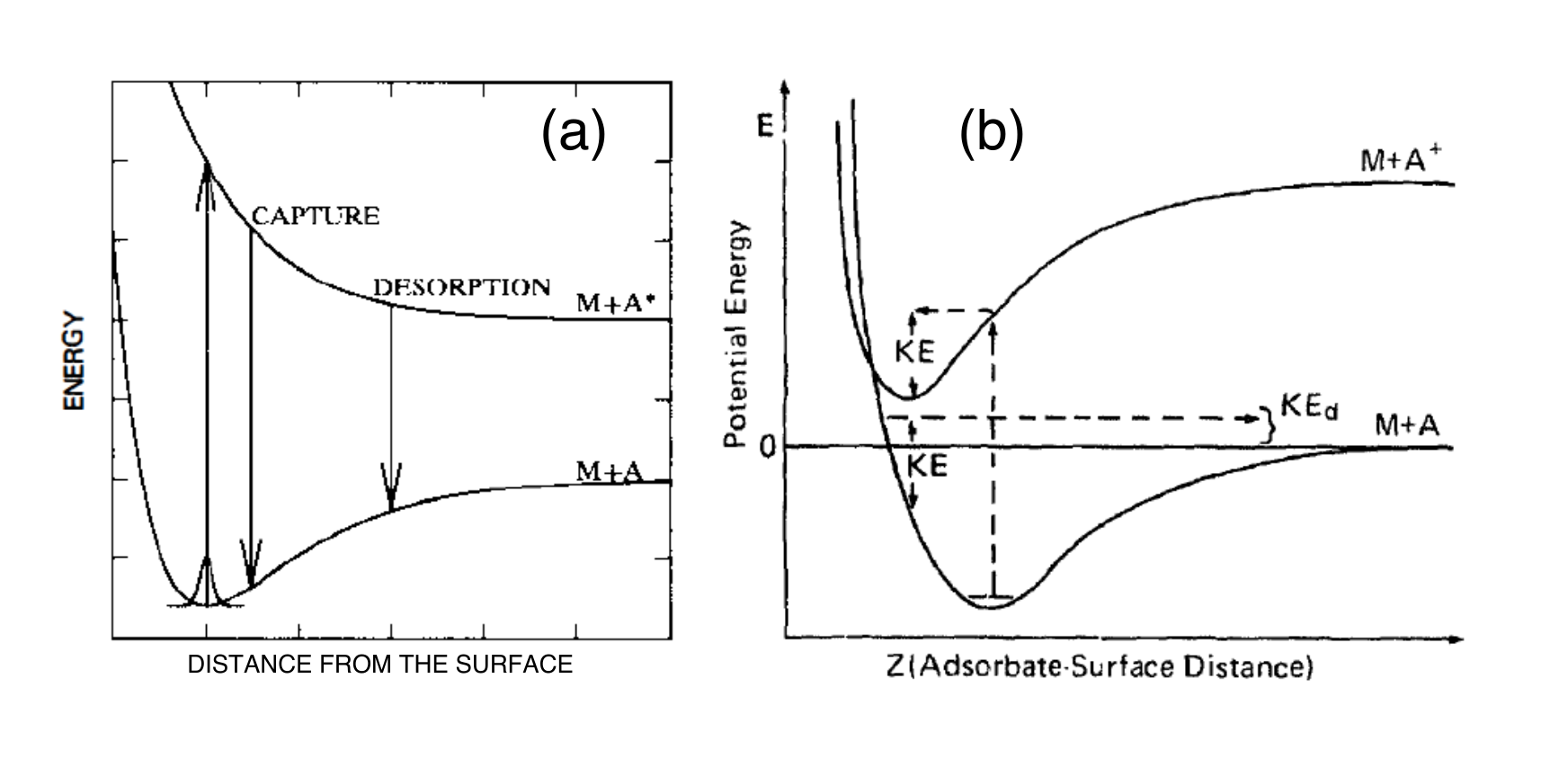} 
\caption{Two models of desorption dynamics of atoms from surfaces following the excitation to the intermediate states (a) proposed by Menzel, Gomer and Redhead \cite{MGR1, MGR2}, and (b) proposed by Antoniewicz \cite{Antoniewicz}. The images are adopted from Ref. \cite{Avouris} and Ref. \cite{Chuang}, respectively. \label{ant00}}
\end{center}
\end{figure}

The non-thermal PSD of Xe/metal using 1$-$7 eV photons has been considered not to occur. Generally, the mechanism of nonthermal PSD using photons of 1$-$7 eV is understood in terms of formation of the transient negative ion (TNI) \cite{Richter} and the Antoniewicz model \cite{Antoniewicz} as shown in Fig. \ref{ant00}, where a substrate conduction electron is photoexcited to the adsorbate affinity level. The ground state Xe in the gas phase does not bind an electron stably \cite{Buckman, Nicolaides, Bae, Hird}, which has been confirmed both theoretically and experimentally with the exception of Ref.~\cite{Haberland}. Xe atoms physisorb on a metal surface. Physisorption is assumed to occur with little influence on the electronic states. Hence, it has been anticipated that the PSD of Xe/metal via TNI is absent. Condensed Xe, however, has been reported to have modified electronic states compared with the isolated ones due to hybridization with the orbitals of neighboring atoms. It is known that it takes 0.5 eV to remove an excess electron from the bulk Xe \cite{Schwentner1975}, and also that the ground state Xe$_{N}$ clusters with $N>6$ stably bind an electron \cite{Haberland}, indicating that the electron affinity level of Xe is shifted downward or broadened depending on the phase of Xe. In this sense, adsorption of Xe onto metal surfaces may well result in a shift and/or broadening or even narrowing of its affinity level by hybridization of the unoccupied orbitals with the substrate electronic states, as is predicted by theoretical studies \cite{Nordlander, Silva}.

Laser induced desorption of atoms and molecules from solid surfaces is a vital phenomenon to investigate fundamentals such as surface electronic structures and dynamics \cite{Ho1, Zheng, Wang2009, Klass, Toker}. Besides, laser desorption is an essential technique for pulsed laser deposition used in thin film growth \cite{Kools1992, Kools1993, Konomi2009, Konomi2010, Singh} and mass spectrometry of protein employing matrix-assisted laser desorption ionization \cite{Puretzky}. When desorption flux is small, the velocity distribution of desorbed atoms is directly governed by the desorption mechanism. When the desorption flux is large enough, on the other hand, the post-desorption collision between desorbed particles may become significant and modify the velocity distribution in the vicinity of the surface after the desorption.

Manifestations of the collision effect in laser induced desorption have been reported both by experiments and simulations as the modifications of the angular and velocity distribution of desorbing atoms and molecules \cite{Cowin1978, NoorbatchaJCP1987, NoorbatchaPRB1987, NoorbatchaSS1988}. Cowin \textit{et al.} investigated the angular dependence of the translational temperature of D$_{2}$ desorbed from tungsten surfaces under a pulsed laser irradiation \cite{Cowin1978}. The translational temperature of D$_{2}$ desorbed in the surface normal direction was higher than those in oblique directions. They attributed the variation of the translational temperature to the collision effect. Noorbatcha \textit{et al.} used the direct Monte Carlo simulation of desorbing atoms to investigate the collision effect. They showed that even in the sub-monolayer regime the collision noticeably modifies the final angular, velocity and rotational-energy distributions \cite{NoorbatchaJCP1987, NoorbatchaPRB1987, NoorbatchaSS1988}. However, there has not been any model that can quantitatively estimate the degree of modification by the post-desorption collision in laser desorption so far.

\begin{figure}
\begin{center}
\includegraphics[scale=.9, clip]{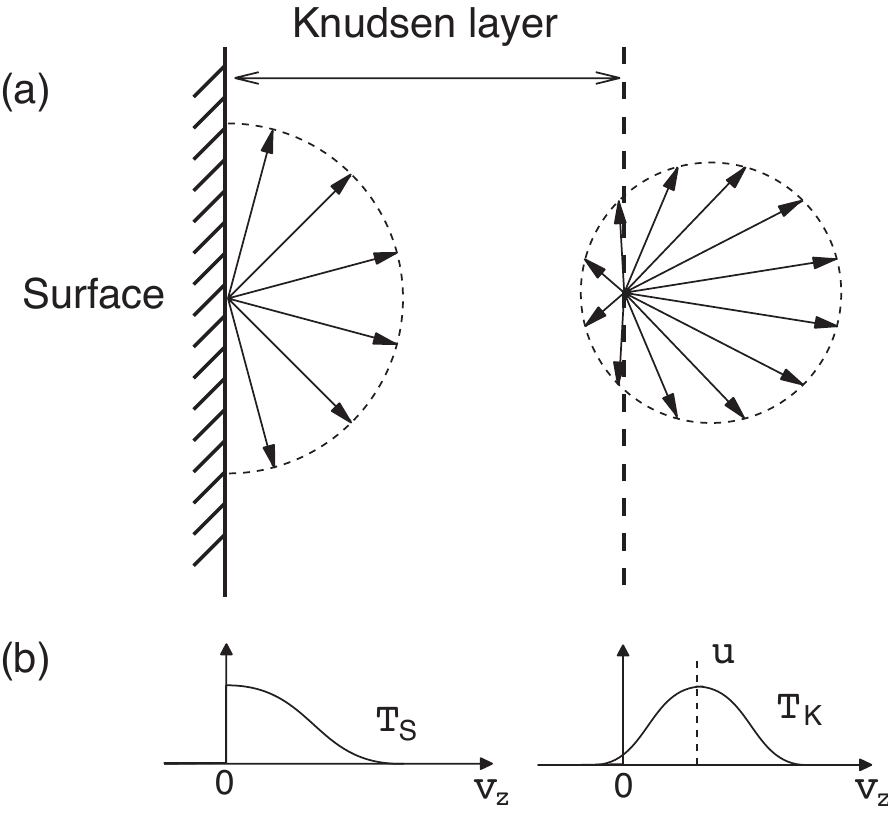} 
\caption{A sketch of Knudsen layer formed above the surface following the laser induced thermal desorption. The desorption flux is assumed to be large enough for the post-desorption collisions to take place and form the Knudsen layer. The arrows denote the velocity vector of desorbed atoms at the surface and at the end of the Knudsen layer. The insets at the bottom shows the schematics of the velocity distribution in $z$ (surface normal) direction at the surface and at the end of the Knudsen layer.\label{KL}}
\end{center}
\end{figure}

Knudsen layer formation theory has been developed in rarefied gas dynamics to model the steady flow of the strong evaporation from the surface \cite{Ytrehus, Cercignani, Davidsson}. As shown in Fig. \ref{KL}, the initial velocity distribution of thermally desorbed species at the surface is well described by a ''half-range'' Maxwell-Boltzmann velocity distribution \cite{Ytrehus}. In the Knudsen layer theory, as a result of intensive post-desorption collisions, the half-range velocity distribution at the surface is thermally equilibrated to a full-range Maxwell-Boltzmann velocity distribution with a stream velocity at some distance from the surface. This thermalization layer is defined as the Knudsen layer. The theory analytically predicts for monoatomic gas that the ratio of the translational temperature at the end of Knudsen layer $T_{\mathrm{K}}$ to the surface temperature $T_{\mathrm{S}}$ and the Mach number of the desorption flow at the end of the Knudsen layer become 0.65 and 1.0, respectively \cite{KellySS1988, KellyNIMB1988}.

Kelly and Dreyfus discussed that the Knudsen layer formation theory may be applicable to the pulsed desorption flow with a large desorption flux. However, it is not straightforward because the pulsed desorption involves complex time evolution of the density and velocity distributions of desorbed atoms \cite{Stein}. Sibold and Urbassek have shown by means of the Monte Carlo simulation of the Boltzmann equation that the pulsed desorption flow at an intense flux is well characterized by the above values predicted by the Knudsen layer formation theory. Although previous experimental studies recognized the collision effect in laser desorption \cite{Taborek, Burgess, Cowin1985, HusslaBBPC1986, HusslaCJP1986}, the formation of the Knudsen layer in laser desorption has been discussed only by theory and simulations \cite{KellySS1988, KellyNIMB1988, Sibold1991}. So far, any experimental confirmation of Knudsen layer formation in laser desorption has not been presented. For an experimental verification of the theory, a systematic observation of the translational temperature and stream velocity of the desorption flow as a function of the desorption flux is strongly required.

\subsection{Nuclear resonant scattering}
With analogy to the absorption of visible light by atoms or infrared light by molecular vibrations, the absorption of high energy x ray by nucleus was considered in 1930s. However, they were not easily observed due to the sharp resonance width ($\sim$ neV) of the nucleus and the relatively large recoil energy at the moment of photon absorption and emission. If one consider a $\gamma$ ray emission at $h\nu=E_{\gamma}$ from a radio isotope and the recoil energy $E_{\mathrm{R}}$ due to the conservation of the momentum of the system, the resultant photon energy $E$ observed from the center of mass system is 
\begin{equation}
E=E_{\gamma}+E_{\mathrm{R}}.
\label{Moss}
\end{equation}
With this relation, the shift of the energy of the $\gamma$ ray is described. Experimentally, one want to observe the absorption of the emitted $\gamma$ ray by the sample. If the loss of the photon energy due to the second term of the right hand side of Eq. (\ref{Moss}) is small compared to the resonance width of the nucleus, the resonant absorption of the $\gamma$ ray should be observed. Unfortunately, based on the conservation of the momentum the recoil energy can be expressed as
\begin{equation}
E_{\mathrm{R}}=\frac{1}{2m}\left(\frac{E_{\gamma}}{c}\right)^{2},
\label{Recoil}
\end{equation}
where $m$ and $c$ are mass of the nucleus and the speed of light. $E_{\mathrm{R}}$ is proportional to the square of $E_{\gamma}$. If one substitutes $m$ and $E_{\gamma}$ of Eq. (\ref{Recoil}) with the mass of $^{57}$Fe and 14.4 keV, one gets $E_{\mathrm{R}}= 2.0$ meV. The resonance width of $^{57}$Fe is 5.0 neV. Therefore, with $\gamma$ ray emission with a recoil of the atom, the resonant absorption of the $\gamma$ is not observed because the resonance is not obtained, whereas the resonant absorption of visible or infrared light can be observed because their recoil energy $E_{\mathrm{R}}$ is sub nano eV and in a order of pico eV, respectively. It had been speculated that with the thermal energy, the lost energy can be compensated. In this way, the absorption of $\gamma$ ray at high temperature was actually observed. This method is based on the doppler effect due to the thermal energy. This indicates that with a higher temperature one can observe a $\gamma$ ray adsorption with a better efficiency.

Nevertheless, M\"{o}ssbauer found that the efficiency of the resonant $\gamma$ ray absorption by $^{191}$Ir is greatly increased by decreasing the temperature of the sample and $\gamma$ ray source \cite{Mossbauer}. It was readily confirmed by other groups \cite{Craig, Lee}. The phenomenon was explained by M\"ossbauer with a recoilless emission and absorption of $\gamma$ ray with a recoilless fraction or a Lamb-M\"ossbauer factor $f_{\mathrm{LM}}$. In a solid, atoms are bound at a lattice site. The vibrational energy is quantized in a unit of $\hbar\omega$. In order for the total energy to be conserved, it holds that
\begin{equation}
E_{\mathrm{R}}=(1-f_{\mathrm{LM}})\hbar\omega.
\label{Recoil2}
\end{equation}
With Debye model $f_{\mathrm{LM}}$ can be expressed as \cite{Sano}
\begin{align}
&\text{At low temperature, } f_{\mathrm{LM}}=A \exp\left(-\frac{3E_{\mathrm{R}}}{2k\Theta_{\mathrm{D}}}\right), \\
&\text{At high temperature, } f_{\mathrm{LM}}=A \exp\left(-\frac{6E_{\mathrm{R}}}{2k\Theta_{\mathrm{D}}^{2}}T\right),
\label{Recoil3}
\end{align}
where $\Theta_{\mathrm{D}}$ is the debye temperature. Those equations well explained the temperature dependence of $f_{\mathrm{LM}}$. Physically, this means that the total energy in $\gamma$ ray emission and absorption is conserved by the excitation of the vibration of atoms in the lattice with a probability $(1-f_{\mathrm{LM}})$. The conservation of momentum is satisfied because the momentum is transferred to the whole material, which results in a very small recoil velocity. This is called the M\"ossbauer effect, which had a significant impact on a large field of physics because it allows us to investigate the nuclear energy level very precisely. M\"ossbauser spectroscopy is significant because it allows us to investigate the hyperfine structure of the nucleus which is determined by a very local electronic and magnetic structure surrounding the atom involved. M\"ossbauser specterscopy has been utilized to investigate the chemical properties of $^{57}$Fe in 
various materials, confirmation of the gravitational red shift or the uncertainty principle.

Although M\"ossbauser spectroscopy has been very widely used in various fields of physics and chemistry, the materials have been limited to those which have convenient radio isotopes. Furthermore, the method posseses a poor sensitivity to the solid surfaces because the signal is too weak compared to those from the bulk. It had been speculated that with a synchrotron radiation one can conduct a similar experiment. In 1990s, the method called nuclear resonant scattering (NRS) of synchrotron radiation was established. The beauty of NRS is that one can make use of the characteristics of the synchrotron radiation which is not obtained using a radio isotope. (1) The selectivity of the wavelength, (2) the brilliance of the light source, (3) High directivity, (4) the polarization of the light source, (5) pulse source.

In the present work, we focus on the physisorption systems. With a conventional M\"ossbauer spectroscopy, the radio isotopes for $^{83}$Kr and $^{129}$Xe could be obtained but they are not convenient. Their life time is short and thus the experiments has to be done in close to the nuclear reactor. On the contrary, synchrotron radiation of naturally white, we are able to use the x ray to excite $^{83}$Kr much easier. Furthermore, its high directivity and the significant brilliance are great features when we consider a measurement of the solid surface. With a glancing angle regime, the sensitivity to the solid surface is expected to be enhanced. This is a particular feature to the NRS. We further make use of the polarization of the synchrotron radiation to select the specific excitation lines. $^{83}$Kr possess a 11 transitions. The lesser the number of the transitions, the easier the analysis of the spectrum. With the selection rule regarding the polarization of the incident light, we discuss that we successfully reduced the number of transitions from 11 to 4.

\begin{figure}
\begin{center}
\includegraphics[scale=.6, clip]{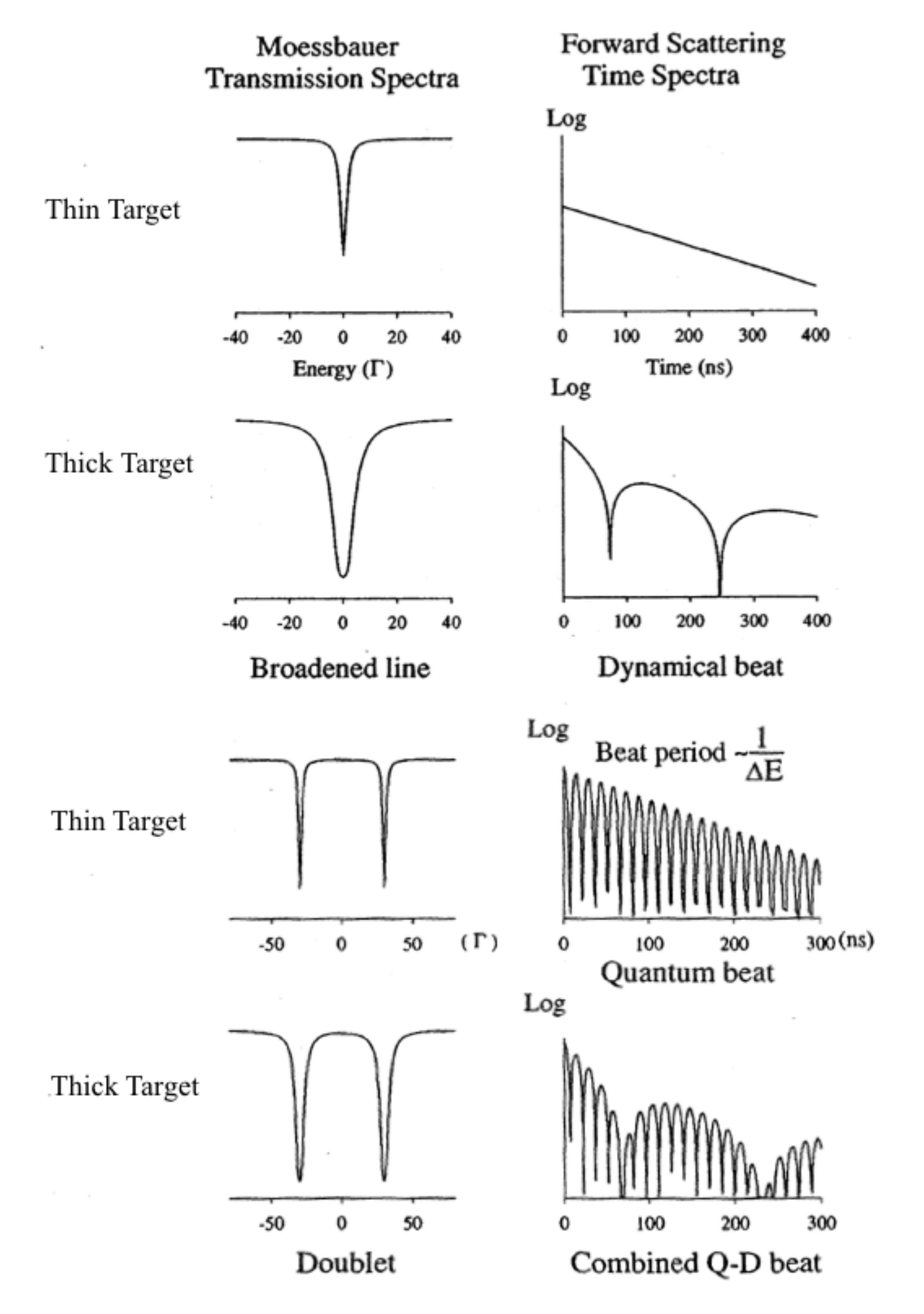} 
\caption{M\"ossbauer transmission spectra (left column) and the nuclear resonant scattering spectra in the time domain (right column). The hyperfine splitting in the energy spectra results in a quantum beat structure. The effect of sample thickness results in the dynamical beat structure. In general, the combination of the both structure are observed. The image is adopted from Ref. \cite{Smirnov1999}. \label{MossNRS}}
\end{center}
\end{figure}

In the conventional M\"ossbauer spectroscopy, the energy spectrum of the nuclear levels are obtained. In the NRS, on the other hand, the time spectrum of the nuclear levels are obtained by a fast time resolved measurement. The time spectrum is related to the energy spectrum with a Fourier transform. The example of the hyperfine structure observed in the conventional M\"ossbauer spectroscopy and the NRS is shown in Fig. \ref{MossNRS}. The single peak in the energy spectrum is transformed into a single exponential decay in the time domain. The double peak due to the hyperfine splitting in the energy spectrum is observed as an oscillation in the time spectrum which is called a quantum beat \cite{Smirnov1996, Smirnov1999}.

The typical representation of the time spectrum of NRS with the two level splitting is \cite{Trammell1, Trammell2, Gerdau}
\begin{equation}
I_{\mathrm{fs}}(t)=C\cdot\exp\left(-t/t_{0}\right)\left[1+\cos\left\{\left(\omega_{1}-\omega_{2}\right)t\right\}\right].
\label{qbeat}
\end{equation}
The beat frequency is understood as the intra-interference between the slightly split energy levels. The larger the hyperfine splitting, the larger the quantum beat frequency becomes. Typically, the hyperfine splitting of neV results in the ns quantum beat structures.

For the simplicity of the analysis, it greatly helps to reduce the number of transitions. This is realized by making use of the polarization dependence as schematically shown in Fig. \ref{POL}. The sample surface in the present study is TiO$_{2}$(110) surface. Ti and O in TiO$_{2}$ is roughly in charged states of $+4$ and $-2$, respectively. The surface used was a Ti riched surface where all the topmost atoms are Ti. Therefore, it is expected that a great electric field and its gradient is present at the surface especially in the surface normal direction rather than the surface lateral direction. The principle axis of the electric field gradient is expected in the surface normal direction as seen in Fig. \ref{POL}. In this case, the quantization axis of the quadrupole splitting of $^{83}$Kr nuclei is also the surface normal direction. The optical transition from the ground state $I=9/2$ to the first excited state $I=7/2$ is M1 transition \cite{Ruby1963}. Therefore, the selection rule for the magnetic sub level is $\Delta m=0, \pm1$. In the M1 transition, the $\Delta m =0$ transition corresponds to the linear magnetic dipole oscillator along the quantization axis, whereas $\Delta m=+1$ and $\Delta m=-1$ transitions correspond to the right-handed circular oscillator and left-handed circular oscillator with respect to the quantization axis, respectively \cite{Grunsteudel, Sladecek, Partykajankowska}. If the magnetic oscillator is parallel to the quantization axis, only $\Delta m =0$ transition is induced. If the magnetic oscillator has a component in the direction perpendicular to the surface normal direction, the $\Delta m=\pm1$ transitions are excited.

\section{Objective of the study}
There are following objectives in the present thesis.
\begin{enumerate}
\item  I report an experimental study of LITD and nonthermal PSD of Xe from a Au(001) surface at photon energies of 6.4 and 2.3 eV. At a high laser fluence, Xe desorption was thermally induced at both photon energies, which is in good agreement with theoretical calculations. At a low laser fluence, on the other hand, Xe desorption was induced nonthermally by 6.4 eV photons as a one-photon process, whereas little desorption was observed with 2.3 eV photons. We argue that the nonthermal PSD proceeds with a transient formation of Xe$^{-}$ as a result of the photoexcitation of substrate conduction electrons. A classical model calculation of Xe desorption reproduces both the experimentally observed TOF and nonthermal PSD cross section, assuming a value of the Xe$^{-}$ lifetime to be $\sim$15 fs.

\item  I investigated the laser induced thermal desorption (LITD) of Xe from an Au(001) surface by means of the time-of-flight (TOF) measurement as a function of the wide range of desorption flux by varying the initial Xe coverage $\Theta$. We found that at $\Theta$ close to 0 ML, the TOF was well analyzed by a Maxwell-Boltzmann velocity distribution. Hence, the desorption at $\Theta$ close to 0 ML is rationalized by the thermal desorption followed by the collision-free flow. At $\Theta$ close to monolayer, we observed that the peak positions of the TOF spectra shift towards smaller values. Assuring that the desorption is only thermally activated, we regard this modification of the TOF as the manifestation of the collision effect. At larger $\Theta$, the peak positions of the TOF spectra become constant. We deduced the Mach number $M$ of the desorption flow to be 0.96 at large $\Theta$ under the assumption that $T_{\mathrm{K}}/T_{\mathrm{S}}=0.65$. The obtained value of $M$ and the saturating behavior of u at $\Theta > 4$ ML both well coincide with the previously reported values by the Knudsen layer theory and simulations. The facts suggest the formation of the Knudsen layer in LITD at large $\Theta$. Furthermore, we tentatively estimated the Knudsen number of the initial desorption flow.

\item I developed an UHV chamber compatible with the nuclear resonant scattering measurements in SPring-8 BL09XU, which was used to observe the nuclear resonant scattering of synchrotron radiation by multi-layer and monolayer $^{83}$Kr fabricated on a clean single crystal solid surface for the first time. The possibility of the observation of the quadrupole splitting of  $^{83}$Kr nuclei was discussed.
\end{enumerate}



\chapter{Experiment}

In this chapter, two experimental setups are presented, both of which are newly developed for the present study. One is for the laser induced desorption experiment and is equipped with a time-of-flight (TOF) measurement system for the laser desorbed atoms as schematically drawn in Fig. \ref{exp1}. The other is for the nuclear resonant scattering (NRS) experiment. For this experiment, I installed an ultra-high vacuum (UHV) chamber in a beam line of a synchrotron radiation facility called SPring-8 BL09XU. The whole experimental setup for the NRS consists of the synchrotron, monochromator, the sample enclosed in the UHV chamber, fast photon detector and the signal collection circuit as schematically drawn in Fig. \ref{exp2}. Detailed descriptions of the experimental instruments follows.

\section{Laser induced desorption}
\begin{figure}
\begin{center}
\includegraphics[scale=1., clip]{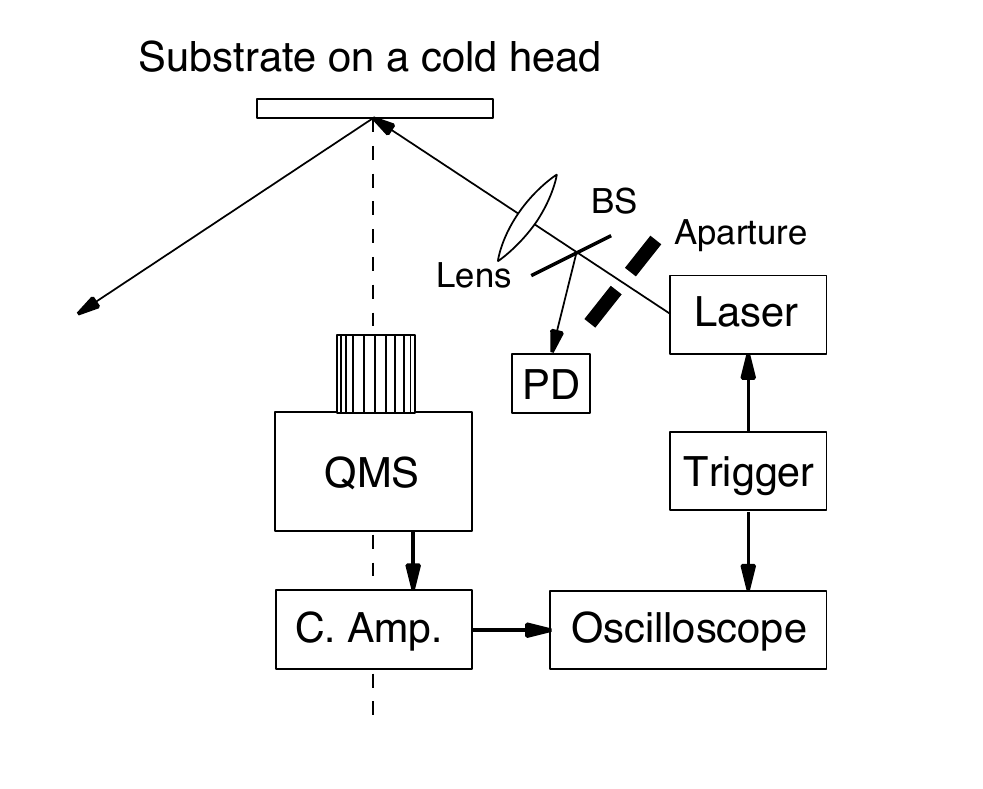} 
\caption{A schematic drawing of the experimental setup for the time-of-flight measurement of Xe following the laser induced thermal desorption from Au surfaces. C. Amp., PD and BS denote fast current amplifier, photo diode and beam splitter, respectively.\label{exp1}}
\end{center}
\end{figure}

Figure \ref{exp1} shows a schematic drawing of the whole setup of the laser induced desorption system. The system consists of the sample and a quadrupole mass spectrometer (QMS) enclosed in a UHV chamber, pulse laser source, optics, fast current amplifier and an oscilloscope synchronized with the pulse laser. In the following, I explain those parts in order.

\begin{figure}
\begin{center}
\includegraphics[scale=1.2, clip]{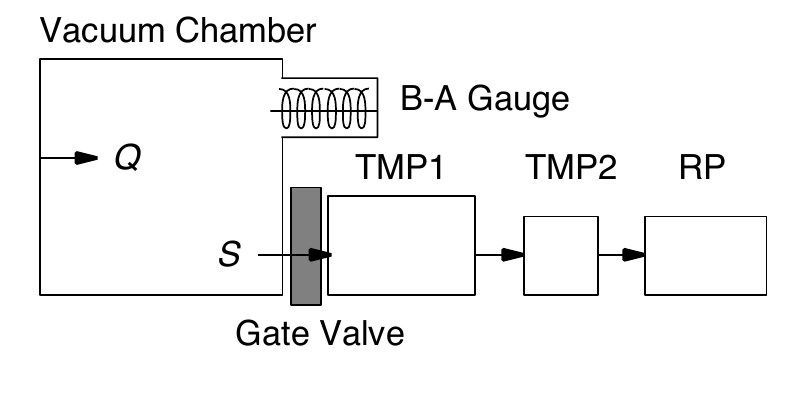} 
\caption{A schematic drawing of the evacuation system of the UHV chamber for the laser desorption experiment.\label{exp4}}
\end{center}
\end{figure}

The base pressure of the UHV chamber for the laser desorption is at $2.0\times10^{-8}$ Pa. The evacuation system of the UHV chamber is schematically shown in Fig. \ref{exp4}. The UHV chamber is evacuated with two turbo molecular pumps (TMP1 and TMP2) and a rotary pump (RP) in a series circuit. The main TMP and sub TMP are Pfeiffer TMU 261 with the pumping speed $S = 220$ l/s and Mitsubishi PT-50 with $S = 50$ l/s, respectively. The rotary pump is BOC Edwards RV-5 with $S = 1.4$ l/s. The total pressure of the main UHV chamber was measured with a B-A gauge controlled by a digital gauge controller ULVAC GI-N8. The gas line is evacuated by the TMP2. Ar, Kr and Xe gas are introduced into the UHV chamber via a variable leak valve. The total pressure was measured with a MKS Baratron with the range of ambient pressure to $\sim1$ Pa. In order to get the chamber to UHV, one needed to bake the whole chamber at about 420 - 570 K over night. This was done with a number of tape heaters.

As a sample, an atomically flat Au(001) surface was used in the laser desorption experiment. In order to prepare the specimen, first, a gold disc was cut from a single crystal rod orienting to the [001] direction. The cross section of the disc was a circle with $\phi=$ 6 mm. It was 1.5 mm thick. The orientation of the [001] direction was confirmed by x ray diffraction pattern with an accuracy of about $0.5^{\circ}$. The x ray diffraction pattern were taken at Edagawa laboratory. The diffraction pattern were not visible right after the disc as cut out from the single crystal rod. The diffraction spots became clear after the sample was chemically etched with Aqua regia which was prepared by mixing concentrated nitric acid and concentrated hydrochloric acid with a volume ratio of 1 to 3. After fixing the surface orientation of the sample disc with a goniometer, it was mechanically polished on an automatic lapping and polishing machine Musashino Denshi MA-200. The polishing cloth and polishing particles were Nylon polishing cloth by Buehler and Alumina particles with diameters of 1.0, 0.3 and 0.05 $\mu$m called Buehler Micropolish II, respectively. The flatness of the disc after each polish was checked by an optical stereo microscope. After the disc was polished with the Alumina particles of 0.3 $\mu$m, it comes to get mirror reflection to the naked eyes. Further polishing with Alumina particles of 0.05 $\mu$m got things worse, surprisingly. The disc get several scratches visible to the naked eyes after polished with the Alumina particles with 0.05 $\mu$m. I suspect that the quality of the product is not as good as it should be. Therefore, I finished polishing process with the Alumina particles of 0.3 $\mu$m.

\begin{figure}
\begin{center}
\includegraphics[scale=1.1, clip]{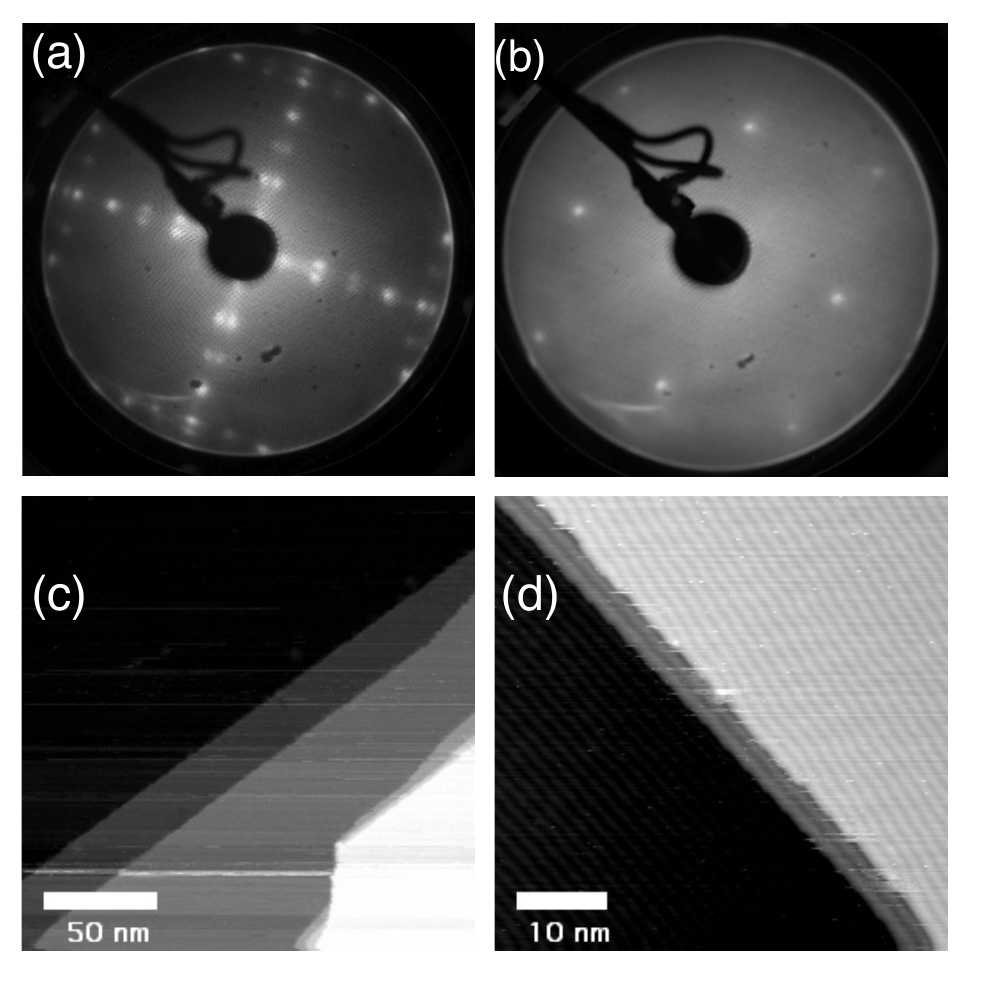} 
\caption{(a) Low energy electron diffraction pattern of a Au(001) surface after cleaning process including several cycles of Ar$^{+}$ sputtering and annealing at 700 K in the ultra-high vacuum conditions. The electron energy was at 48 eV. The image shows a clear super structure of (5$\times$20) reconstruction of topmost layer, which is consistent with previous reports \cite{Vanhove1, Vanhove2}. (b) Low energy electron diffraction pattern of the same surface observed in the UHV after it was exposed to the ambient air for about 10 min. The electron energy was 60 eV in this image. The image shows a ($1\times1$) pattern with a relatively larger background intensity compared to (a). (c) and (d) are STM images taken at RT showing the herringbone structure.\label{exp3}}
\end{center}
\end{figure}

An atomically flat surface of Au(001) was prepared in a RT-STM system by Omicron. The crystallization was checked by low energy electron diffraction (LEED) and scanning tunneling microscopy (STM) at the room temperature. The chemical cleanliness was checked with the Auger electron spectroscopy (AES). The optimized cleaning process was then repeated in the UHV chamber for the laser desorption, although the cleanliness was not checked in situ, for this chamber is not equipped with LEED. Following the previous reports \cite{Vanhove1, Vanhove2}, the sample was cleaned with several cycles of annealing at 670 K and Ar$^{+}$ sputtering at 500 eV. After these cleaning process, a clear LEED pattern was observed as shown in Fig. \ref{exp3} (a). The image is in good agreement with the previous report \cite{Vanhove1, Vanhove2, Figuera, Meyer}, indicating that the Au(001) surface was reconstructed with ($5\times20$) pattern. The ideal surface of the Au(001) surface is square lattice, whereas the ($5\times20$) pattern can be interpreted in such a way that the topmost layer is compressed into a close-packed hexagonal lattice. Thus, we ideally can regard this surfaces to possess a (111) like structure rather than (001) surfaces. This is also confirmed by a STM observation of this surface. I note that the annealing temperature is the best at 670 K.

In order to optimize the annealing temperature, I varied the annealing temperature. With increasing annealing temperature up to 770 K, the sharpness of the LEED spot observed at room temperature gets worse. With decreasing annealing temperature down to 570 K, the sharpness of the LEED spot also got worse. Therefore, it is concluded that the best annealing temperature is around 670 K. It took only 5 to 10 min to get the LEED spot sharp. Furthermore, I checked the resistivity of this clean surface to the ambient air. After obtaining the sharp LEED spot, I take out the Au sample from the UHV chamber via the load lock. Then, after the exposure of the sample to the ambient air for about 10 min, I got the Au sample into the UHV chamber again. Soon after that, I checked the cleanliness of the surface by observing the LEED pattern. The LEED pattern could be observed without any treatment. The observed LEED image showed ($1\times1$) pattern as shown in Fig. \ref{exp3} (b). This may have resulted from the fact that the ($5\times20$) reconstruction of the topmost Au layer was lifted or that the some adsorbed molecule formed the periodic lattice of ($1\times1$) pattern. I annealed the Au surface at 670 K for 5 min and observed the LEED pattern again. The LEED image showed a clean ($5\times20$) reconstructed pattern. This means that with the annealing 670 K the Au(001) surface restores the clean reconstructed surface even after it was exposed to the ambient air. This result shows the significant nobleness of this surface as compared to other metal surfaces such as Pt, Pd or Ir, for with those surfaces does not retain a clean surface such easily. In this way, the cleaning process was successfully optimized.

\begin{figure}
\begin{center}
\includegraphics[scale=.8, clip]{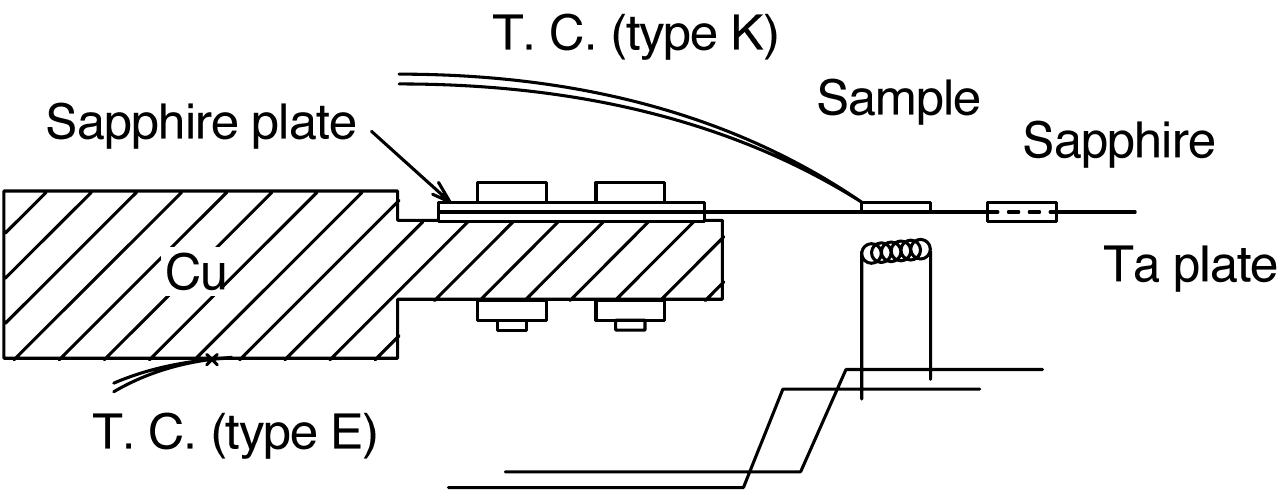} 
\caption{A schematics of the sample holder. \label{exp11}}
\end{center}
\end{figure}

The Au disc was mounted on a sample manipulator as shown in Fig. \ref{exp11}, which was then installed into the UHV chamber for the laser desorption experiment. The sample manipulator was prepared so that the Au surface could be annealed and cooled. The sample manipulator consists of the closed-cycle He-compression type refrigerator IWATANI Cryo mini D310 and the copper block mounted at the end of the refrigerator. The Au disc was mounted on a Ta plate with the thickness of about 0.1 mm. The Au disc was held by two Ta wires with $\phi=0.5$ mm and was directly spot welded to those wires. The Ta wires were also spot welded to the Ta plate. In this way, the thermal contact of the Au disc to the Ta plate was assured. It was not easy to spot weld the Au disc to the Ta wire. Thus, I increased the power of the spot welder to do that. It was successful, although the bond was not as strong as the bond formed between Ta and Ta or Ta and W. I did this because it is of some importance to make sure that the thermal contact between Au disc and the Ta plate.

The Ta plate was electrically floated from the Cu block on which the Ta plate was mounted. The Ta plate was electrically floated so that the Au sample could be heated by electron bombardment. Between the Cu block and the Ta plate, a single crystal sapphire spacer was inserted so that the Ta plate should be electrically floated and should gain the most of the thermal contact with the Cu block. Right behind the Ta plate, I placed a W wire of $\phi=0.3$ mm as a heater. The W wire was used both as a radiative heater and as an electron source for the electron bombardment. An area of the Ta plate right behind the Au disc was cut out in a circle of $\phi=5$ mm so that the electron could bombard the disc directly. For the measurement of the temperature of the sample, I directly spot welded the thermocouple of type K (chromel-alumel) to the side of the Au disc. This welding was also tough thanks to the good thermal conductivity of gold. This is worth doing because it is always best to measure the temperature of the sample directly. In the case of a metal sample, little temperature distribution within the sample disc is expected. At the sample, the lowest temperature reached was 20 K. With a radiation shield, the sample temperature should have got to the same temperature as the cyrohead. The temperature at the cryohead was monitored by another thermocouple of type E (chromel-constantan), where the lowest temperature observed was 10 K.

For the heating of the sample, two regulated DC power supplies were used. KIKUSUI PAS40-9 was used for the current supply to the W wire up to 7 A. KIKUSUI PAS500-1.2 was used to accelerate the thermal electrons emitted from the hot W filament for the electron bombardment up to 500 V. At a  typical heating condition of 7 A and 10 V, the sample temperature gets to 600 K with the refrigerator working. With the sample bias of $-500$ V, the emission current gets to 0.2 or 0.3 A. The emission current was very sensitive to the current supplied to the filament at around 7 A. It is because the temperature of the filament $T$ is determined by the current supplied to the filament and because the yield ($I_{\mathrm{EM}}$) of the thermal emission of the electrons obeys the form of $I_{\mathrm{EM}}\propto\exp(-E_{0}/kT)$, where $E_{0}$ is the work function of the filament. The highest temperature of the sample was about 1000 K with the refrigerator working. 

The physisorbed layers of Xe and Kr were prepared by backfilling the chamber. As mentioned above, the clean surface of Au(001) is reconstructed to a close-packed hexagonal structure. The density of atoms at the topmost layer is then  about $1.4\times10^{19}$ atoms/m$^{2}$. The Van der Waals radius of Xe and Kr are 2.2 and 2.0 \AA, whereas the half of the lattice constant of Au is 1.4 \AA. Therefore, neither Xe nor Kr can form a ($1\times1$) structure. Although there has been no reports on the adsorption structure of Xe and Kr on Au(001), on many other metal (111) surfaces it has been reported that Xe and Kr form a ($\sqrt{3}\times\sqrt{3}$) structure \cite{Bruch}. Therefore, I assumed that the first layer forms a ($\sqrt{3}\times\sqrt{3}$) structure and estimated the density of the first physisrobed layer to be 1/3 of the density of the topmost layer of the substrate. In this study, I defined the monolayer saturation density $\Theta_{\mathrm{S}}$ to be $4.5\times10^{18}$ atoms/m$^{2}$.

\begin{figure}
\begin{center}
\includegraphics[scale=.5, clip]{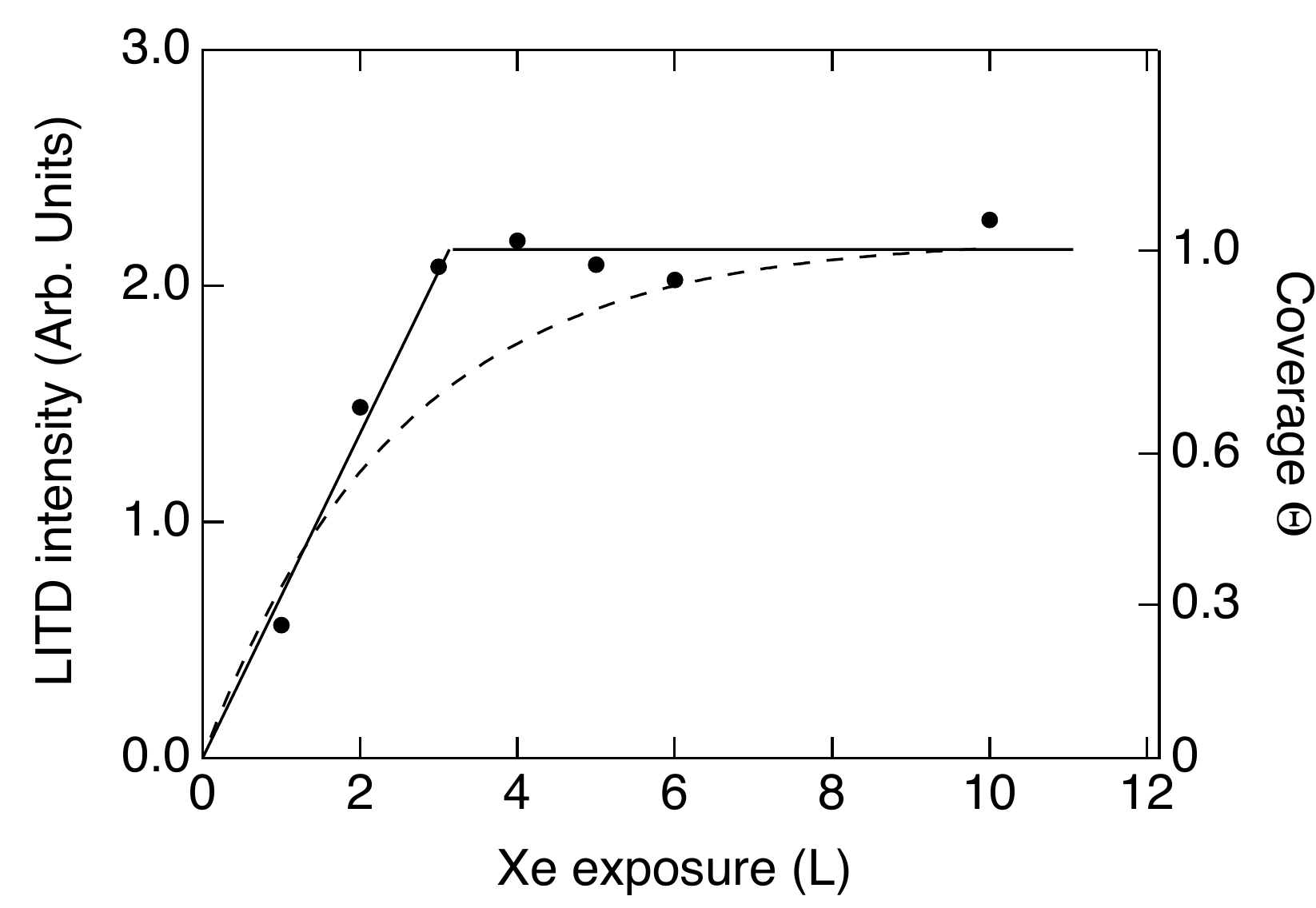} 
\caption{Laser induced thermal desorption intensity as a function of initial Xe exposure. The substrate was Au(001) at 80 K, where only monolayer can form. The solid line is the guide for the eyes. The dashed line is the theoretical result based on the Langmuir type adsorption. The deviation means that the sticking probability $s$ is constant such as $S(\Theta)=c$ rather than $S(\Theta)=c(\Theta/\Theta_{\mathrm{S}}-1)$ during the first layer growth.\label{exp7}}
\end{center}
\end{figure}

The coverages of Xe and Kr layers were determined in the following way. The yield of exposure of Xe and Kr is measured in Langmuir (L), where 1 L corresponds to ($1\times10^{-4}$ torr) $\times$ (100 s). The time-evolution of coverage $\Theta$ (atoms/m$^{2}$) of adsorbed atoms can be expressed as 
\begin{equation}
\Theta=\int_{0}^{t}S(\Theta)\Gamma dt,
\end{equation}
where $S(\Theta)$ is the sticking probability and $\Gamma$ is the impinging rate of gas atoms. The impinging rate is expressed as
\begin{equation}
\Gamma=\frac{p}{\sqrt{2\pi mkT}},
\end{equation}
where $p$, $m$, $k$ and $T$ are pressure, mass of the gas atom, Boltzmann constant and the temperature of the gas. Given $S(\Theta)$ is constant at unity, the monolayer completion is calculated to occur at 2.8 L. In the Langmuir type adsorption, $s$ is expressed as $S(\Theta)=c(\Theta/\Theta_{\mathrm{S}}-1)$, where $c$ and $(\Theta/\Theta_{\mathrm{S}}-1)$ are the initial sticking probability and the repulsion factor. I experimentally observed the Xe coverage $\Theta$ on a clean Au(001) as a function of Xe exposure by measuring the laser induced thermal desorption yield, the detail of which will be described later. On Ag and Pt surfaces, only monolayer can be formed at 80 K. Thus, I expect that on Au(001) at 80 K, only monolayer is formed. The result is shown in Fig. \ref{exp7}. The result shows that $\Theta$ increases almost linearly with Xe exposure at 0 to 3 L. Above 3 L, the LITD yield become constant. The solid lines are drawn as the guide for the eyes. Figure \ref{exp7} also shows the dashed line, which is the theoretical line based on the Langmuir type adsorption (i.e. $S(\Theta)=c(\Theta/\Theta_{\mathrm{S}}-1)$). The theoretical line is not in good agreement with the experimental result. The fact indicates that the sticking probability at $0<\Theta<1$ ML is not dependent on the coverage $\Theta$ described simply as $s=c$ rather than $S(\Theta)=c(\Theta/\Theta_{\mathrm{S}}-1)$.

As a pulse laser source, an ArF excimer laser LAMBDA PHYSIK OPTex and a Nd: YAG (yttrium aluminum garnet) laser Spectra-Physics Quanta-Ray DCR-11 were used. The photon energy and the pulse duration were (6.4 eV, 8 ns) and (2.3 eV, 7 ns), respectively. As for the Nd:YAG laser, only the second harmonics of the fundamental IR of 1064 nm was used in the experiment. With the use of the third harmonic, the photons at the energy of 3.4 eV was also available, although it was not used. The maximum power of the excimer laser at the exit of the laser was at around 14 mJ per pulse. The UV laser seemed to be slightly absorbed by the ambient air. I guess it was the O$_{2}$ that absorbed the UV.

The UV laser pulse is not visible to the naked eyes. Thus, they were guided to the UHV chamber with a help of He-Ne continuous laser (532 nm). The laser path was first established with the He-Ne laser. After that the UV laser pulse was tuned and guided to follow the same path. The position of the UV laser is easily seen by disturbing the light path with any paper. I preferred ordinary copy paper. The UV light was easily checked in the air. However, the position of the UV laser spot on the sample surface suspended inside the vacuum chamber is not easy to see. The UV laser does not reveal itself on the sample surface unlike the visible lasers do whose location is visible by their diffused scattering even on mirror-reflective surfaces. In order to make the position of the UV laser spot on the sample in the vacuum chamber, I placed a sapphire block next to the Au sample. I first irradiate the sapphire surface by the UV laser pulse to make sure it is there and that the UV laser pulse is exactly following He-Ne laser. The fluorescence of the UV laser by the sapphire block was clearly seen as a bright violet light spot. The sample manipulator slides on a linear guide. The sample is located the position so that with it can be moved to the place where the UV laser was irradiating the sapphire with a slide of 2 cm.

The incident angle of the incoming laser light on the sample surface is about 25$^{\circ}$ from the surface normal. The incident laser light gets in the UHV chamber via a quartz window, which is transparent for the UV light. The reflected laser light escapes the UHV chamber through another quartz window. The UV and visible laser were both focused with a quartz lens of which the focal length was 30 cm. The sample surface needed to be off focus if otherwise the surface is greatly damaged by the laser as extensively as it can be seen by the eyes. The reflectivity of the UV laser of 6.4 eV and visible laser of 2.3 eV on the gold surface was calculated using the Fresnel equation and deduced to be 0.2 and 0.8, respectively.

\begin{figure}
\begin{center}
\includegraphics[scale=.6, clip]{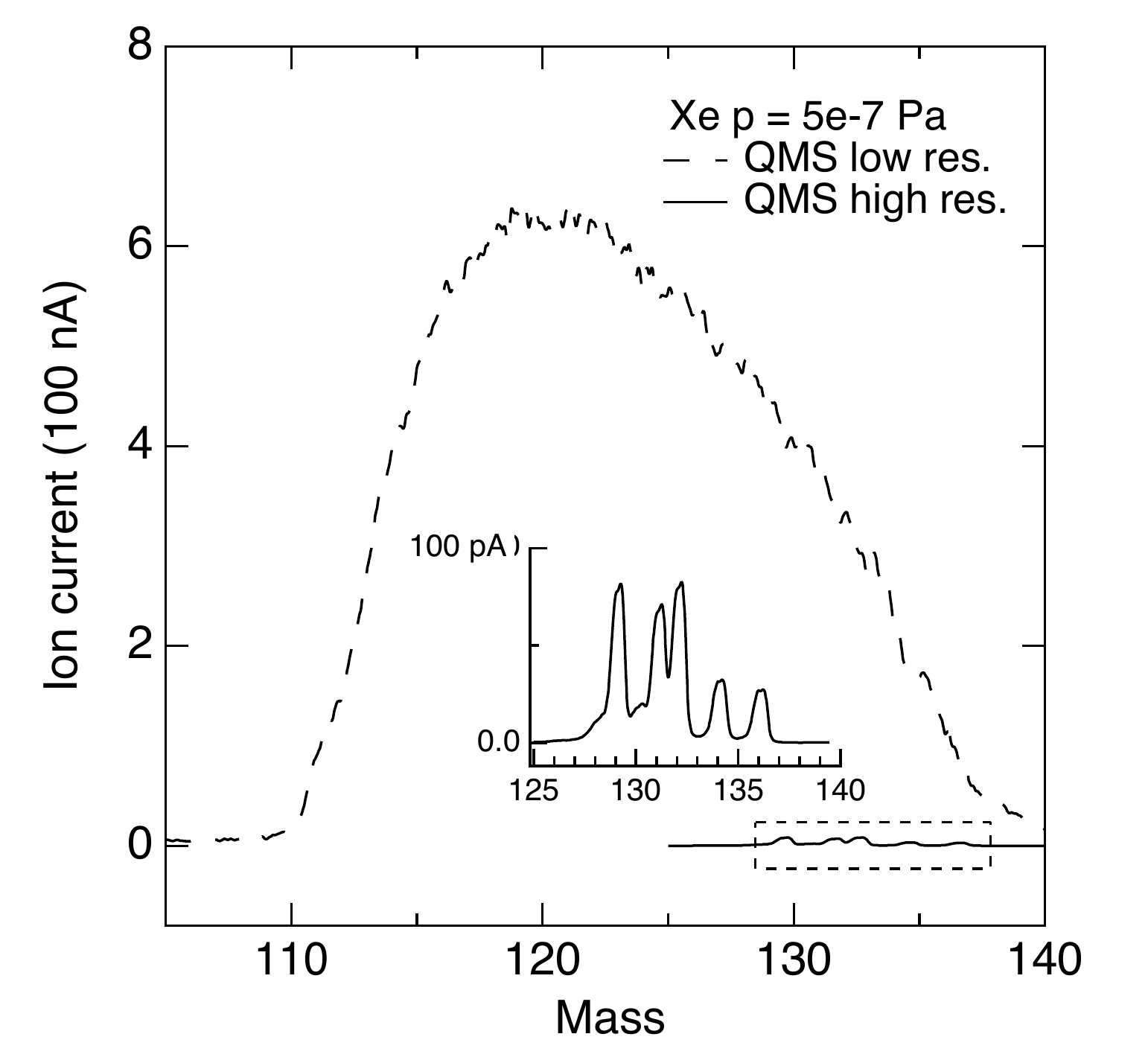} 
\caption{Mass spectrum of Xe gas at $p=5\times10^{-7}$ Pa with the QMS tuned to high mass resolution and with the QMS tuned to a low mass resolution. The inset shows the magnification of the area wrapped by the dashed rectangular.\label{exp9}}
\end{center}
\end{figure}

As a detector for the laser desorbed species, a QMS is placed in surface normal direction. The flight distance of the desorbed atoms from the surface and the ionization chamber of the QMS was 10 cm. The QMS consists of the analyzer called Balzars QMA 125 90$^{\circ}$ SEM type, the power supply called Balzars QME 125 and the controller called Balzars QMG 112A. The ion current was magnified with the SEM inside the QMA 125. When the mass spectrum of the residual gas was measured, the signal current from the SEM was further amplified with current amplifier EP 422 and fed to the computer. On the other hand, in the laser induced desorption experiment the EP 422 needed to be replaced by a fast current amplifier called Keithley 427, the output of which is fed to the oscilloscope called Tektronix TDS-620B. This alternation concerns the time response of the electronics, which will be mentioned a little later. The mass spectrum of Xe is measured as a mass spectrum of the QMS as shown in Fig. \ref{exp9}. Natural abundance of Xe isotopes were is reported to be ($^{129}$Xe, 26 \%), ($^{131}$Xe, 21 \%), ($^{132}$Xe, 27 \%), ($^{134}$Xe, 10 \%) and ($^{136}$Xe, 9 \%). The experimental observation is in good agreement with the reported values.

Generally it can be said that the sensitivity of the QMS to the massive atoms or molecules is relatively small compared to that to the light atoms or molecules. This is due to the time required for an ionized particle to pass through the mass pole. Thus, the sensitivity of QMS to massive atoms like Xe is small. TOF measurement in laser desorption experiment requires high sensitivity. In order to detect the desorption yield as a function of time, a good time response of the measurement electronics is required. To achieve a good time response of the electronics, the measured current or voltage has to be large. This also can be generally said that the selectivity and the sensitivity of a detector has a relation of trade off with each other. Hence, I further tuned the mass resolution of the QMS to the lowest achieved. The result is also shown in Fig. \ref{exp9}.

For the TOF measurement with a QMS, the use of a fast current amplifier such as Keithley 427 is of some importance. The thermal velocity of atoms are in the same order of the sound speed, which is a few hundred m/s. The TOF distance from the sample to the detector is usually a few to a few tens of cm. Thus, the typical TOF is in the order of several microns to several milli seconds. The electronics has to have this range of time response for measurements of the TOF spectra. The typical ion current output from the SEM of the QMS is in the order of pico A as shown in the inset of Fig. \ref{exp9}. Ordinary picoammeter such as Keithley model 617 or others have a poor time response of about a few seconds during the measurement of pico to nano A. The fast current amplifier (Keithley 427) has the time response of about a few micron seconds during measuring a few nano A (i.e. amplification of 10$^{6}$ V/A). The smaller the ion current, the time response gets longer. Thus, it is definitely needed in the TOF measurements that the signal current supplied from QMS/SEM to the fast current amplifier is in the order of a few nano A or larger. Signals weaker than a few nano A is not able to amplified enough to observed by these electronics. The resultant signal is in the order of a few milli Volts in order for them to be recorded by A/D converter of an oscilloscope or computer.

\begin{figure}
\begin{center}
\includegraphics[scale=.6, clip]{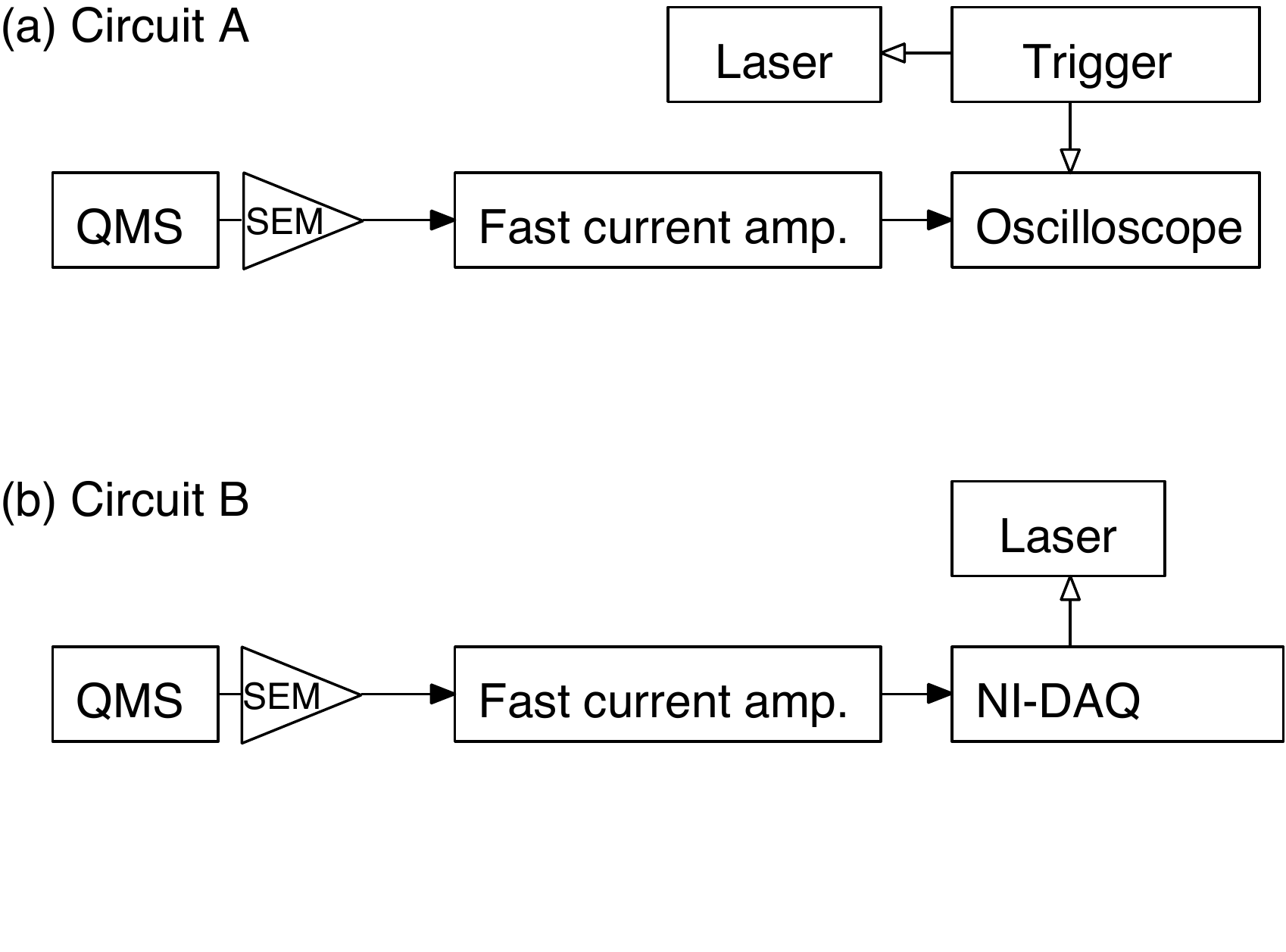}
\includegraphics[scale=.45, clip]{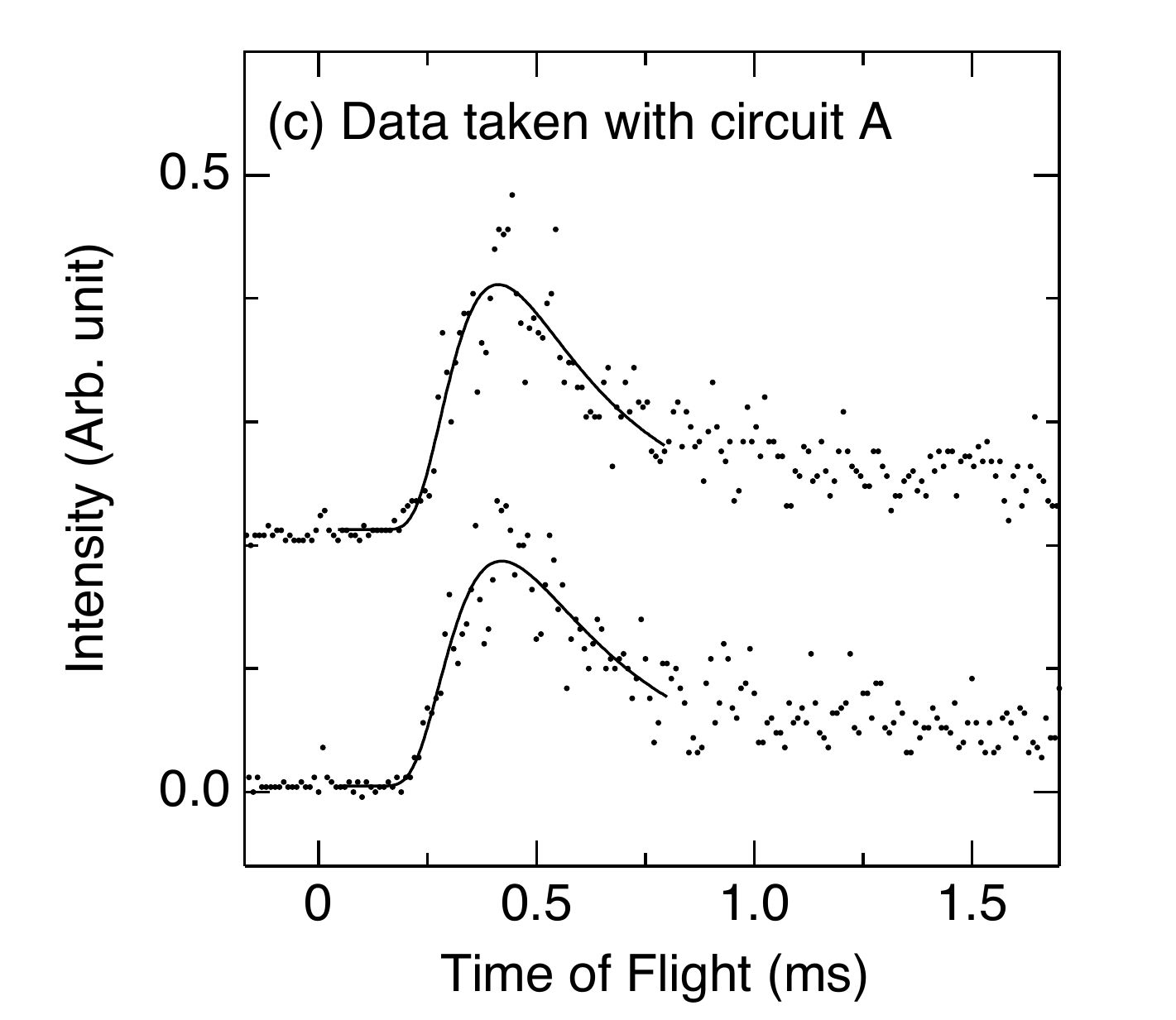}
\includegraphics[scale=.45, clip]{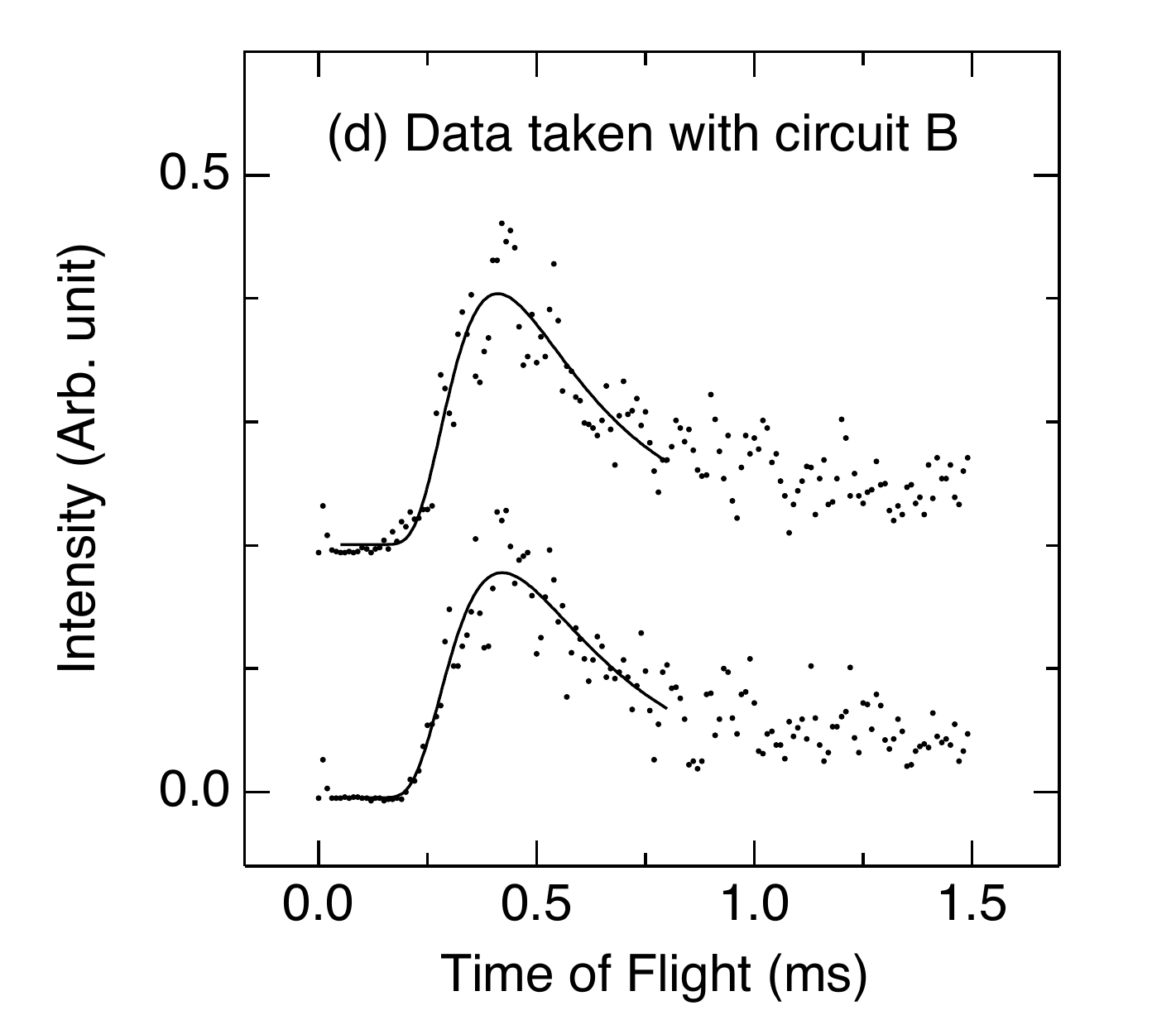} 
\caption{(a) A data acquisition circuit for laser desorption experiment using an oscilloscope as a recorder. (b) Another data acquisition circuit for the same objective using an A/D converter as a recorder and a trigger at the same time. (c) Time of flight spectrum of Xe from Au(001) following laser irradiation. This data were recorded with the circuit A. (d) Same data as shown in (c) recorded at the same time with circuit B. \label{exp8}}
\end{center}
\end{figure}

I established two kinds of electronics for the TOF measurement with QMS as shown in Fig. \ref{exp8} (a) and (b). I first established the circuit A for laser desorption experiment. Later, for the simplicity of the measurement and data handling, I established circuit B. The circuit B was as good as circuit A used in the laser induced desorption with a single pulse, although it is not sophisticated enough for the experiment with multiple pulses. Both circuit A and B amplify the ion current from the QMS with a fast current amplifier Keithley 427 at a gain of 10$^{6}$ V/A where the time response is $\sim3$ $\mu$s. The output of Keithley 427 is a voltage of a few mV which shows a time evolution at $\mu$s order. In the circuit A, the output signal is fed to a digital oscilloscope (Tektronix TDS-620B) which is synchronized with the laser by a trigger pulse which is generated by a pulse generator Stanford Research DG535. The time response of the oscilloscope is sub nano second in the fastest regime. Thus, the time response of the whole electronics is determined by the fast current amplifier as well as the sensitivity. Circuit B has an A/D converter National Instruments NI-USB-6363 connected to a computer in stead of an oscilloscope and a pulse generator. The USB-6363 is able to record the signal voltage at a speed of sub micron seconds, which is not as fast as the oscilloscope but is fast enough to record the TOF spectra. The USB-6363 is also equipped with a digital output, which is used as a pulse source in this circuit.

Two TOF spectra of Xe from Au(001) following laser irradiations were recorded with circuit A and circuit B at the same time as shown in Fig. \ref{exp8} (c) and (d). Initially prepared Xe coverage on Au(001) was 1 ML. Keithley 427 has two output channels which were thus able to be fed to the oscilloscope and NI-USB at the same time. The time resolution of two recorders were both set to 2 $\mu s$. The two TOF spectra shown in Fig. \ref{exp8} (c) is almost identical to those in Fig. \ref{exp8} (d), although they do not look precisely the same with each other. The slight difference seen with a close look at them may resulted from the different clock timing of each A/D converters, although the input signals were identical. Anyway, it was confirmed that those two electronic circuits were effectively the same. In the experiment, both circuits were used considering the conditions of measurements. The circuit A was used both in laser induced thermal desorption experiment and non-thermal photon stimulated desorption experiment, whereas the circuit B was used only in the laser induced thermal desorption experiment.

With some analyses of the observed TOF, several parameters such as translational temperature of the desorbing particles and its desorption yield could be deduced. The analysis is based on an ideal assumption. The assumption is that as a result of the laser irradiation on the surface, it produces a thermalized gas at the surface. Some of the gas atoms fly to the detector in surface normal. Another assumption is that the gas disappears in a so short time period as a few nano seconds. This time period can be regarded as almost a Delta function compared with the TOF.  What I observed in the experiment are shown for instance in Fig. \ref{exp8} (c) and (d). The velocity distribution of the gas at a temperature of $T$ at the surface is expressed using a Maxwell-Boltzmann velocity distribution 
\begin{equation}
f(\bm{v}, z=0)d\bm{v}=A\exp\left(-\frac{m\bm{v}^{2}}{2kT}\right)d\bm{v},\label{mb1}
\end{equation}
where $\bm{v}$, $z$, $A$, $m$ and $k$ are velocity vector in Cartesian coordinate $\bm{v}=(v_{x}, v_{y}, v_{z})$, distance from the surface, normalization factor, mass of a atom or molecule and Boltzmann constant, respectively. Equation (\ref{mb1}) can be transformed into polar coordinate with the relations $(v_{x}, v_{y}, v_{z})=(v \cos\theta \cos\phi, v \cos\theta \sin\phi, v \sin\theta)$, which yields
\begin{equation}
f(v, \theta, \phi, z=0)dv d\theta d\phi=A\exp\left(-\frac{mv^{2}}{2kT}\right) v^{2}\sin\theta dv d\theta d\phi.\label{mb2}
\end{equation}

In the present geometry, the origin of the polar coordinate is fixed at the irradiation spot. The detector is located at a distance of 10 cm from the surface plane in the surface normal direction. The irradiation area is in an order of 1 - 10 mm$^{2}$. Thus, the coordinate of the detector is ($\theta\sim0$, $\phi$, $z$ = 10 cm). The velocity distribution of atoms directing in the direction to the detector in the present geometry can be expressed as 
\begin{equation}
f(v, z=0)dv=A' v^{2}\exp\left(-\frac{mv^{2}}{2kT}\right) dv,\label{mb3}
\end{equation}
where $A'$ is a normalization factor.

I first consider an ideal case where the desorption occurs in a flash. In this case, the desorption flux can be expressed as $F(t)=\delta(t-t_{0})$, where $t_{0}$ is the instance of pulse laser irradiation. The relation between time of flight $t-t_{0}$, the flight distance $L$ and the velocity of an atom $v$ is 

\begin{equation}
t=L/v+t_{0}.
\label{mb8}
\end{equation}

If $t_{0}$ is taken to be 0, Eq. (\ref{mb8}) becomes $t=L/v$. $f(v)dv$ is proportional to the number of atoms having the velocity between $v$ and $v+dv$. These atoms are considered to reach the detector at the time between $t$ and $t+dt$ where $t=L/v$ holds. Therefore, it is reasonable to consider a time-of-flight function $f_{\mathrm{F}}(t)dt$ which is related to $f(v)dv$ as 
\begin{equation}
f_{\mathrm{F}}(t)dt=f(v)dv.
\label{mb4}
\end{equation}
With $v$ being replaced with $L/t$, Eq. (\ref{mb4}) is transformed into,
\begin{equation}
f_{\mathrm{F}}(t)dt=A'' t^{-4}\exp\left\{-\frac{m}{2kT}\left(\frac{t}{L}\right)^{2}\right\}dt,
\label{mb5}
\end{equation}
where $A''$ is a normalization factor. Equation (\ref{mb4}) is directly applied to the experimentally obtained TOF spectra using a flux sensitive detector.

In the case where a density sensitive detector is used, the TOF spectra $f_{\mathrm{D}}(t)dt$ should look a little different from Eq. (\ref{mb5}) as
\begin{equation}
f_{\mathrm{D}}(t)dt=\frac{f_{\mathrm{F}}(t)}{v}dt,
\label{mb6}
\end{equation}
with a factor of $1/v$. The density sensitive detector sense the particle density rather than particle flux. The density is proportional to the particle flux divided by the particle velocity in case the particles flies in one direction. It readily follows from Eq. (\ref{mb5}) and Eq. (\ref{mb6}) that
\begin{equation}
f_{\mathrm{D}}(t)dt=A''' t^{-3}\exp\left\{-\frac{m}{2kT}\left(\frac{t}{L}\right)^{2}\right\}dt,
\label{mb7}
\end{equation}
where $A'''$ is the normalization factor and the only difference from Eq. (\ref{mb5}) is the exponents of the first $t$ on the right hand side. The detectors have each principles of detection which are either classified into density sensitive detections, flux sensitive detections or else. The QMS as used in the present study principally detects the density of the gas. Therefore, Eq. (\ref{mb7}) is the most frequently used form in the analysis of the TOF spectra of the present study.

\begin{figure}
\begin{center}
\includegraphics[scale=.7, clip]{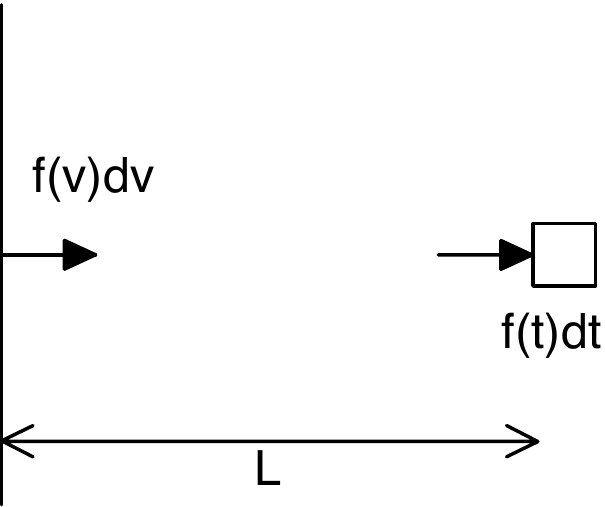} 
\caption{A schematics of desorption from the surface and the TOF measurement.\label{exp12}}
\end{center}
\end{figure}


\clearpage
\section{Nuclear resonant scattering}
The purpose of the experiment was to investigate the hyperfine structure of the nucleus of $^{83}$Kr physisorbed on solid surfaces. It was speculated that through hyperfine structure of $^{83}$Kr I could acquire the physical information on the electronic structure of physisorbed Kr. In order to realize the measurement, nuclear resonant scattering (NRS) was performed at SPring-8 BL09XU, Japan. Before this experiment, NRS of gas atoms at a well-defined surface was not performed anywhere. For the experiment I establish and installed a UHV chamber into an experiment hatch in BL09XU which can hold a well defined surface and position the sample precisely against the synchrotron radiation at the same time.

Present experiments at SPring-8 were performed from late 2011 to late 2012 under the approval of 4 proposals below by the SPring-8 proposal review committee. The proposals were and a general proposal of which the project leader was Prof. T. Okano and three budding researchers support proposals of which the project leader was A. I.

\begin{enumerate}
\item(2011B1188) Simultaneous measurement of nuclear resonant x ray scattering and conversion electron emission from physisorbed krypton layer.
\item(2011B1695) Nano-second observation of 2-dimensional phase transition and surface diffusion of the adsorbed monolayer
\item(2012A1560) M\"{o}ssbauer spectroscopy of physisorbed krypton mono-layer on a solid surface: As a probe of the electric field gradient and dynamics on a solid surface
\item(2012B1699) Measurement of electric field gradient on solid surfaces using M\"{o}ssbauer spectroscopy of physisorbed Kr layer: Metal surfaces and amorphous solid water surfaces
\end{enumerate}

\begin{figure}
\begin{center}
\includegraphics[scale=.4, clip]{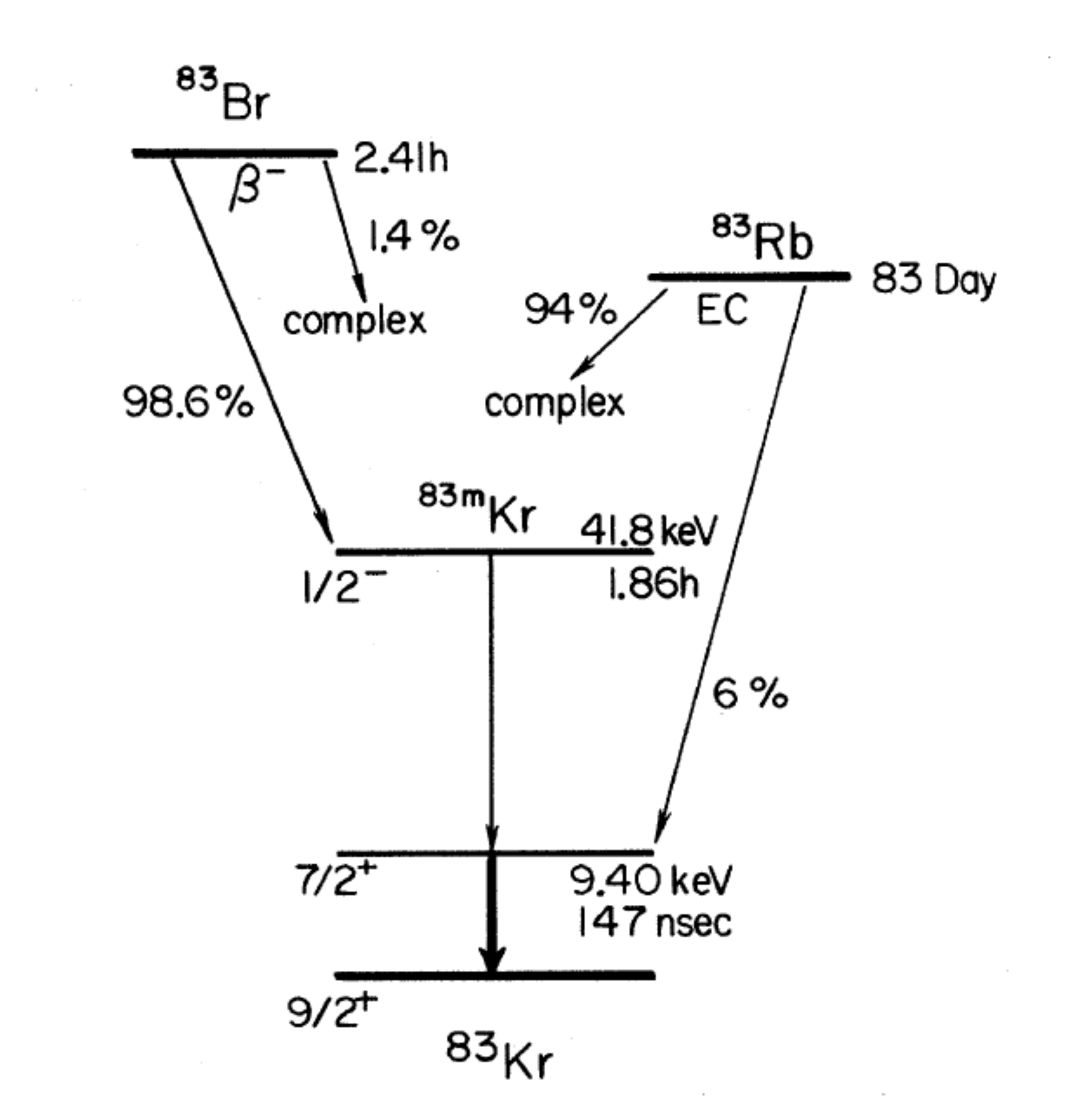} 
\caption{A schematics of decay scheme of $^{83}$Kr using radio isotope of $^{83}$Br or $^{83}$Rb. The image is adopted from Ref. \cite{Kolk1}.\label{decay}}
\end{center}
\end{figure}

\subsection{Nuclear energy levels of $^{83}$Kr} 
At the ground state, the $^{83}$Kr nucleus posseses a nuclear spin of 9/2 \cite{Ruby1963, Ruby1966}. With radio isotopes of $^{83}$Br or $^{83}$Rb, one can observe a $\gamma$ decay of an excited $^{83}$Kr to the ground state with a $\gamma$ ray emission at 9.4 keV. With conventional time to pulse height technique, Ruby and coworkers measured the natural half-life of the first excited state of $^{83}$Kr to be 147 ns. The parity of the both states are + (even). A schematic diagram is shown in Fig. \ref{decay}.

\begin{figure}
\begin{center}
\includegraphics[scale=.2, clip]{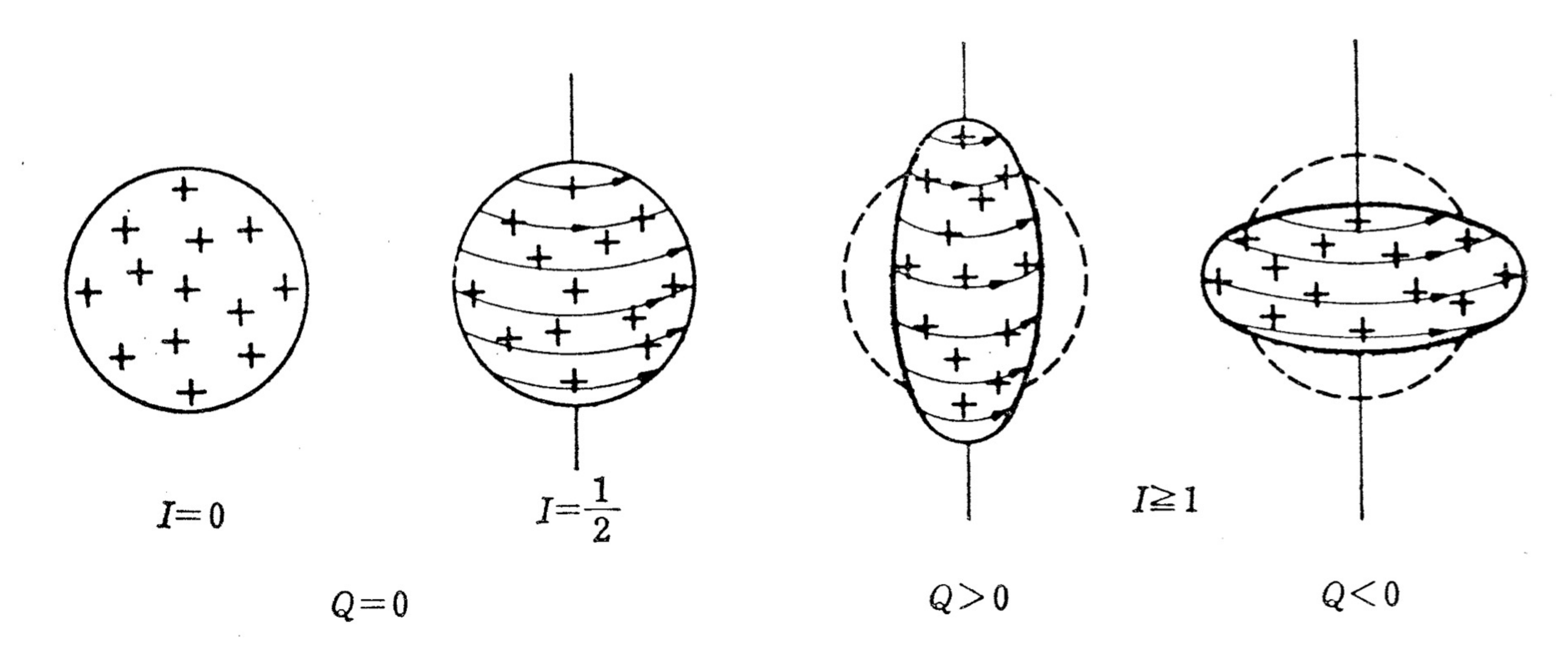} 
\caption{Schematic drawings of the distribution of nuclear charge and the nuclear spin $\bm{I}$. With $\bm{I}\geq1$ the charge distribution become non-spherical and possess a quadrupole moment $\bm{Q}>0$ or $\bm{Q}<0$. The image is adopted from Ref. \cite{Sano}.\label{nucleus}}
\end{center}
\end{figure}

Nuclei consist of protons and neutrons. The whole nuclei possess a nuclear spin $\bm{I}$. The nuclear spin $\bm{I}$ is expressed as $\bm{I}=\bm{S}+\bm{L}$, where $\bm{S}$ and $\bm{L}$ are total spin of the nucleon and total orbital angular momentum of the nucleon, respectively. Generally, the nuclei are spherical with $\bm{I}=0$ and 1/2, whereas they possess magnetic moment with $\bm{I}>1/2$. The charge distribution of the nuclei with $\bm{I}>1/2$ are no longer spherical, which is schematically shown in Fig. \ref{nucleus}.

The interaction between the charge and the current of the nucleus and the surrounding electric field is described as 
\begin{equation}
E=\int\rho(\bm{r})V(\bm{r})d\bm{r},\label{HE1}
\end{equation}
where $\rho(\bm{r})$, $V(\bm{r})$ are charge distribution of the nucleus and the electric potential at the nucleus. The magnetic interactions are not taken into account here. In the present thesis we focus only on the electric interaction. The scalar potential in Eq. (\ref{HE1}) can be expanded as a Taylor series at the nucleus as
\begin{align}
&V(\bm{r})=V(0)+\sum_{i}x_{i}x_{j}V_{i}+\frac{1}{2}\sum_{i, j}x_{i}x_{j}V_{ij}+\dotsb,\label{HE2}\\
\intertext{where,}
&V_{i}=\left(\frac{\partial V}{\partial x_{i}}\right)_{\bm{r}=0} \text{and  } V_{ij}=\left(\frac{\partial^{2}V}{\partial x_{i}\partial x_{j}}\right)_{\bm{r}=0}.\notag
\end{align}
Equation (\ref{HE2}) is then inserted into Eq. (\ref{HE1}). Then, we get
\begin{equation}
E=V(0)\int\rho d\bm{r}+\sum_{i}V_{i}\int\rho(\bm{r})x_{i}d\bm{r}+\frac{1}{2}\sum_{i, j}V_{i, j}\int\rho(\bm{r})x_{i}x_{j}d\bm{r}+\dotsb,\label{HE3}
\end{equation}
The first term is the direct interaction between the total charge of the nucleus and the electric potential at the nucleus. In this term, the charge is regarded as a point charge at the nucleus. The second term describes the interaction between the electric dipole moment and the electric field. In the case of the nucleus, the nucleus does not have an electric dipole moment because the parity of $\rho(\bm{r})$ is always even. Thus, we can neglect this term. The third term is what we concern in this study. It is called a quadrupole interaction against the electric field gradient. Both quadrupole moment and the electric field gradient are tensor of rank 2.

If we concentrate on the interaction between the quadrupole moment of the nucleus and the electric filed gradient, the energy of the quadrupole interaction reads
\begin{equation}
H_{\mathrm{Q}}=\sum_{i, j}\bm{Q}_{i, j}\left(\frac{\partial E_{j}}{\partial x_{i}}\right).\label{HQ}
\end{equation}
$\bm{Q}$ is the tensor defining the quadrupole charge distribution in the nucleus. Its irreducible components in terms of coordinates $x$, $y$, $z$ are given by
\begin{align}
&Q^{0}=\frac{eQ}{2I(2I-1)}(3I^{2}_{z}-\bm{I}^{2}),\notag\\
&Q^{\pm1}=\frac{eQ}{2I(2I-1)}\frac{\sqrt{6}}{2}[I_{z}(I_{x}\pm iI_{y})+(I_{x}\pm iI_{y})I_{z}],\label{Q2}\\
&Q^{\pm2}=\frac{\sqrt{6}eQ}{4I(2I-1)}(I_{x}\pm i{I}_{y})^{2},\notag
\end{align}
where scalar quadrupole moment $Q$ of the nucleus is defined by
\begin{equation}
eQ=\int\rho(\bm{r})(3x_{i}x_{j}-r^{2})d\bm{r}
\end{equation}
where $\rho_{i}$ is the charge density in a small volume element $d\tau_{i}$ inside the nucleus at a distance $r_{i}$ from the center, and $\theta_{iI}$ is the angle which the radius vector $\bm{r}_{i}$ makes with the nuclear spin axis.

The electric field gradient (EFG) at the nucleus appears in Eq. (\ref{HE2}) and (\ref{HQ}) is also defined by the tensor having 9 components $V_{ij}$ in Cartesian coordinates, where
\begin{equation}
V_{ij}=\frac{\partial E_{j}}{\partial x_{i}}=\frac{\partial^{2}V}{\partial x_{i}\partial x_{j}}.
\label{EFG1}
\end{equation}
We are considering the electric field at nucleus generated by a surrounding electron cloud. The electrons can be considered to be outside the nucleus. Thus, if the Laplace equation gives $V_{xx}+V_{yy}+V_{zz}=0$. In this case $V_{ij}$ is a symmetric traceless tensor. If we take a set of principal axes $X$, $Y$, $Z$ in a way that the EFG in $Z$ direction is the largest, they are expressed as
\begin{align}
&(\nabla\bm{E})_{0}=\frac{1}{2}V_{ZZ}=\frac{1}{2}eq,\notag\\
&(\nabla\bm{E})_{\pm1}=0,\label{EFG2}\\
&(\nabla\bm{E})_{\pm2}=\frac{1}{2\sqrt{6}}(V_{XX}-V_{YY})=\frac{1}{2\sqrt{6}}\eta eq,\notag
\end{align}
where $\eta$ is the asymmetry parameter for the field gradient tensor defined as
\begin{equation}
\eta=\frac{V_{XX}-V_{YY}}{V_{ZZ}}.
\end{equation}

Given that the field gradient is axially symmetric, we can take $\eta=0$. In that case, we only need to consider the EFG in the axial direction $V_{ZZ}$. In our discussion, we take the relation $|V_{XX}|\le|V_{YY}|\le|V_{ZZ}|$. Therefore, we ensure that $0\le\eta\le1$. Our experiment was performed on a TiO$_{2}$(110) surface. We regard the main axis to be the surface normal. Intuitively, (110) surface is not axially symmetric. In my experiment, I consider the physisorbed Kr which is located a few \AA\  above the surface. Because the potential well of the physisorbates are located far from the surface, the symmetric parameter $\eta$ may be considered to be close to 0 in the present case. I discuss this later in the section of DISCUSSION.

As we can see from the Eq. (\ref{Q2}) and Eq. (\ref{EFG2}) under the condition that $\eta=0$, the matrix elements of the Hamiltonian of the quadrupole interactions with regard to the nucleus at the state of $|I, m\rangle$ 
\begin{equation}
\langle I, m'|H_{\mathrm{Q}}|I, m \rangle =\frac{e^{2}qQ}{4I(2I-1)}\left\{3m^{2}-I(I+1)\right\}\delta_{mm'}.
\label{QS1}
\end{equation}
Then, the quadrupole interaction energy as a function of magnetic quantum numbers $E_{\mathrm{Q}m}$ can be written as,
\begin{equation}
E_{\mathrm{Q}m}=\frac{e^{2}qQ}{4I(2I-1)}\left\{3m^{2}-I(I+1)\right\},
\label{QS2}
\end{equation}
where $m=I_{z}$ with $z$ as a quantization axis. With Eq. (\ref{QS2}), the energy levels of the nuclei with the state of $|I, m\rangle$ under the quadrupole interactions are estimated. In the cases where EFG have the non-axial symmetry ($\eta\neq0$), the estimation is a bit more complex, involving solutions of secular equation with the order of $(I+1/2)/2$ \cite{Das}.

\begin{figure}
\begin{center}
\includegraphics[scale=.9, clip]{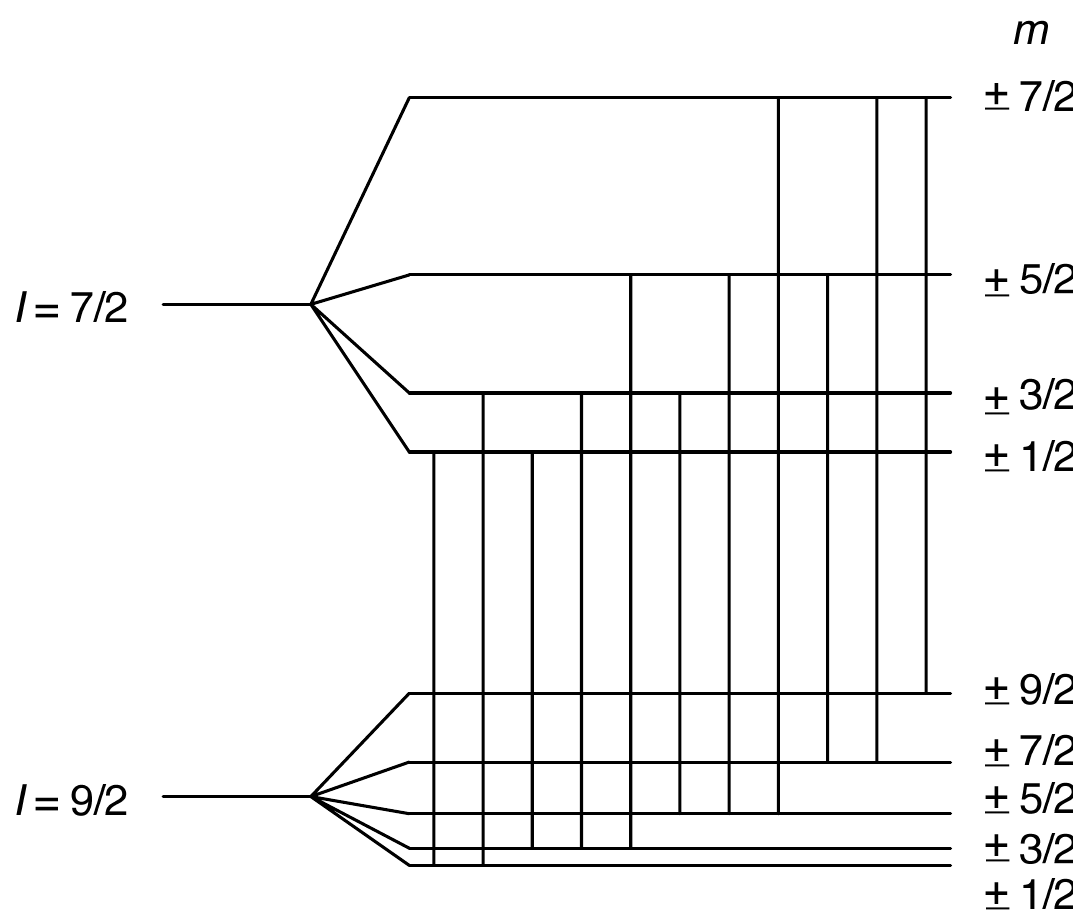} 
\caption{A schematic drawing of nuclear energy levels of $^{83}$Kr at ground and first excited state under electric field gradient. The relative strength of the quadrupole splitting is determined using Eq. (\ref{QS2}) and assuming the ratio of the quadrupole moment $R=Q(7/2)/Q(9/2)=2$ as reported by Kolk and coworkers \cite{Kolk1, Kolk2}. The vertical lines shows a eleven transitions with $\Delta m=0, \pm1$ which are allowed in M1 transition. The transition with $\Delta m=0$ are emphasized with four thick solid lines.\label{eleven}}
\end{center}
\end{figure}

We consider a case of $^{83}$Kr. It possess a nuclear spin $I = 9/2$ at the ground state and it possess the nuclear spin $I=7/2$ at the first excited state. If the nucleus is put in an EFG with axial symmetry where $\eta=0$, the quadrupole splitting of the energy levels are described by Eq. (\ref{QS2}). As can be seen from the Eq. (\ref{QS2}), the energy levels are degenerated such as $E_{\mathrm{Q}\frac{1}{2}}=E_{\mathrm{Q}-\frac{1}{2}}$, $E_{\mathrm{Q}\frac{3}{2}}=E_{\mathrm{Q}-\frac{3}{2}}$ and so forth. Therefore, the ground state of $^{83}$Kr $I=9/2$ are split into 5 levels. The excited state $I=7/2$ is split into 4 levels. If all the transitions are allowed in x ray absorption, 20 transitions are present. By virtue of the selection rule of the M1 transition, only transitions with $\Delta m=0, \pm1$ are allowed. Hence, we expect 11 transitions from the ground state to the excited state in terms of the x ray absorption by the nuclei of $^{83}$Kr as schematically shown in Fig. \ref{eleven}. This number of transitions are expected if the incident x or $\gamma$ ray are unpolarized and the absorber sample is a poly-crystal. The number of transitions are further decreased if the polarization and the crystal axis of the sample is controlled.

\begin{figure}
\begin{center}
\includegraphics[scale=.4, clip]{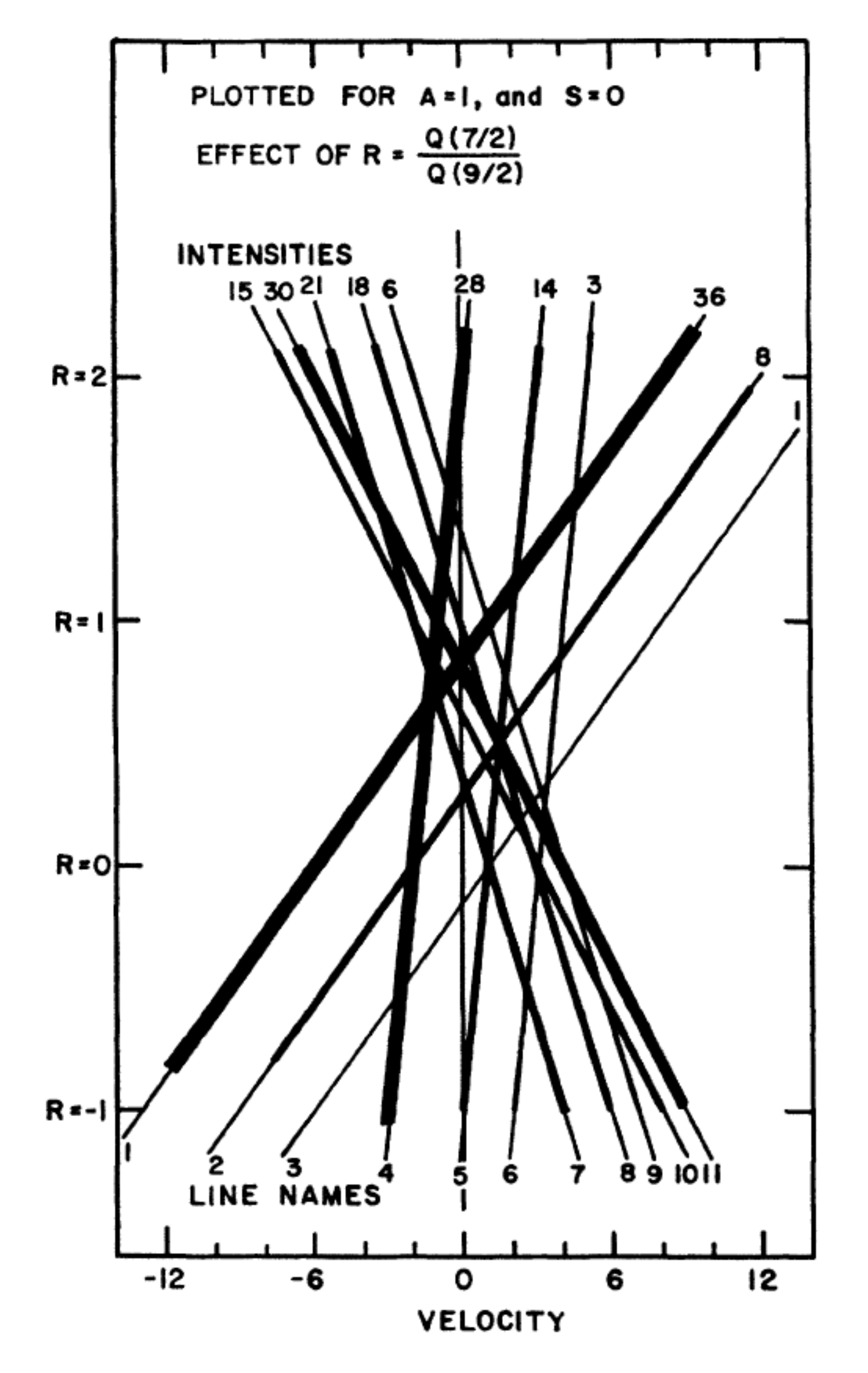} 
\caption{Possible pure quadrupole spectra showing the dependence on the ratio $Q(7/2)/Q(9/2)$. The image is adopted from Ref. \cite{Ruby1966}.\label{RubyLines}}
\end{center}
\end{figure}

\subsection{Experimental setup in SPring-8 BL09XU}
In the SPring-8 BL09XU, there are instruments for the nuclear resonant scattering experiment, including the monochromator of the incident light for various M\"ossbauer isotopes and the fast electronics for the measurement of the time spectra. They are developed and maintained by Dr. Y. Yoda. What we did here was to installed a UHV chamber with which the well-defined monolayer $^{83}$Kr on solid surfaces can be prepared. The objective of the experiment was to see if it is possible to obtained the hyperfine structure in the time spectra of NRS of $^{83}$Kr at the solid surface. Therefore, as a plausible surface, the TiO$_{2}$(110) surface was used. This surface is relatively resistive against contaminations compared with metal surfaces. The surface was once cleaned in another UHV chamber by annealing at 1000 K. Then it was installed in the UHV chamber in SPring-8 BL09XU.

\begin{figure}
\begin{center}
\includegraphics[scale=.9, clip]{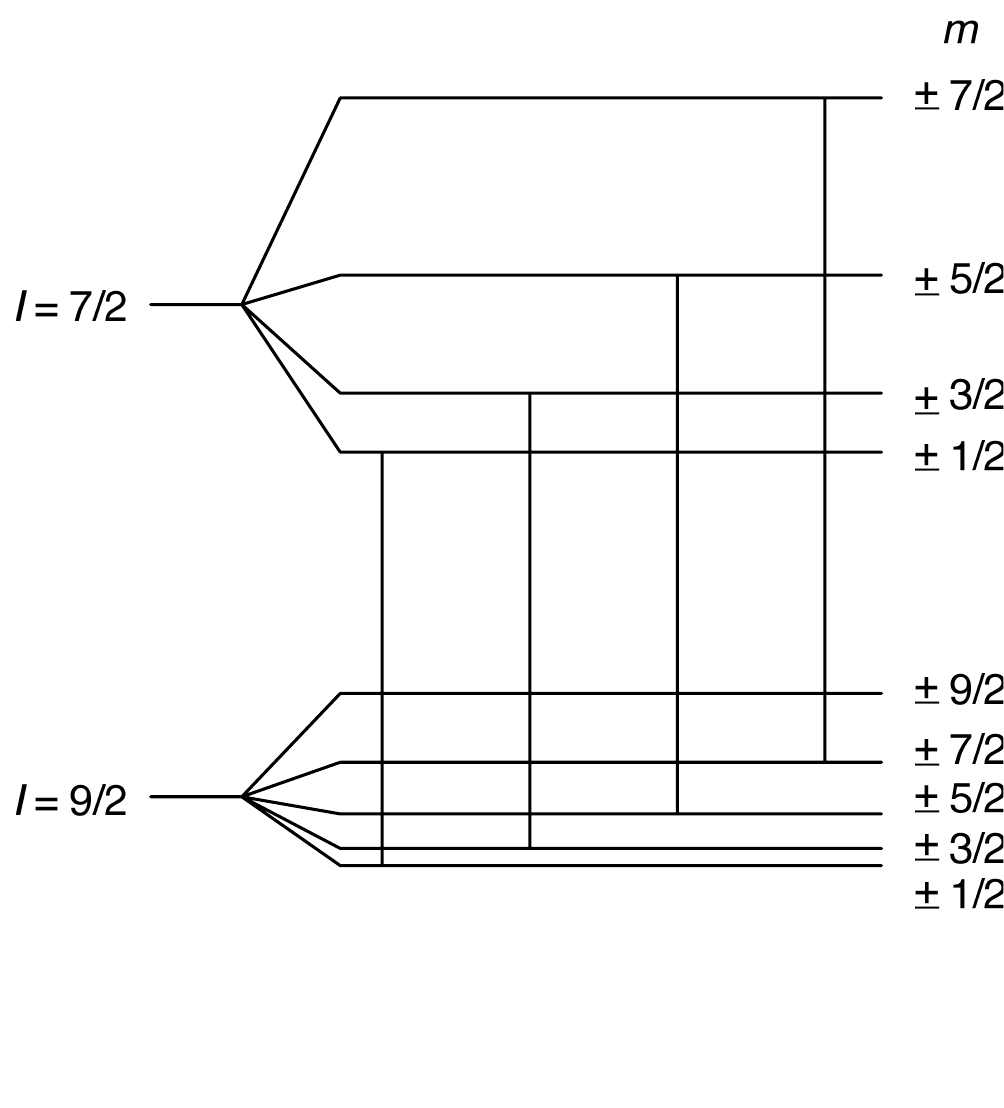}\\ 
\includegraphics[scale=.8, clip]{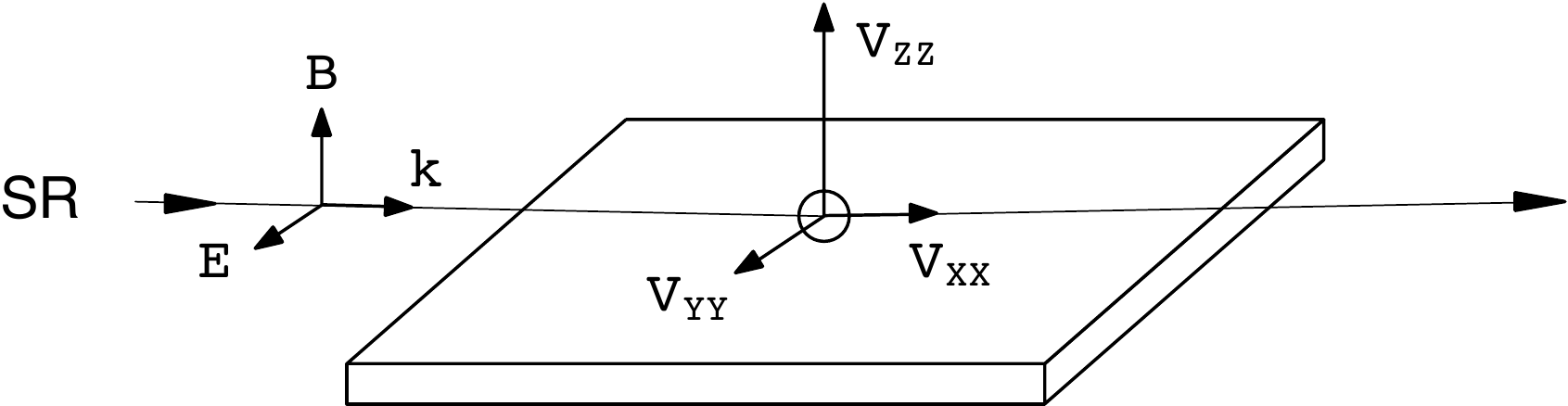}
\caption{A schematic drawing of experimental geometry of the NRS of $^{83}$Kr at TiO$_{2}$(110). The polarization of the synchrotron radiation (SR) with regard to the sample surface is emphasized. Looked from a physisorbate which is sits in a potential well whose bottom is a few \AA \ away from the surface, generally speaking the principal axis of the electric field gradient is directing a surface normal direction. A striking feature of the synchrotron radiation is that they are linearly polarized. In M1 transition, if the polarization of the magnetic vector is parallel to the quantization axis ($z$ in this case) of the sample, only $\Delta m=0$ transitions are excited. On the other hand, if the magnetic vector is polarized perpendicular to the quantization axis ($z$ in this case), only $\Delta m=\pm1$ transitions are excited. \label{POL}}
\end{center}
\end{figure}

As shown in Fig. \ref{POL}, in the present study the polarization is parallel to the surface normal direction. In this configuration, it is expected only 4 $\Delta m=0$ transitions in all the M1 transitions are allowed. It is thus expected that a simple quantum beat structure appears in the time spectrum of NRS of $^{83}$Kr at the TiO$_{2}$(110) surfaces.

\begin{figure}
\begin{center}
\includegraphics[scale=1.5, clip]{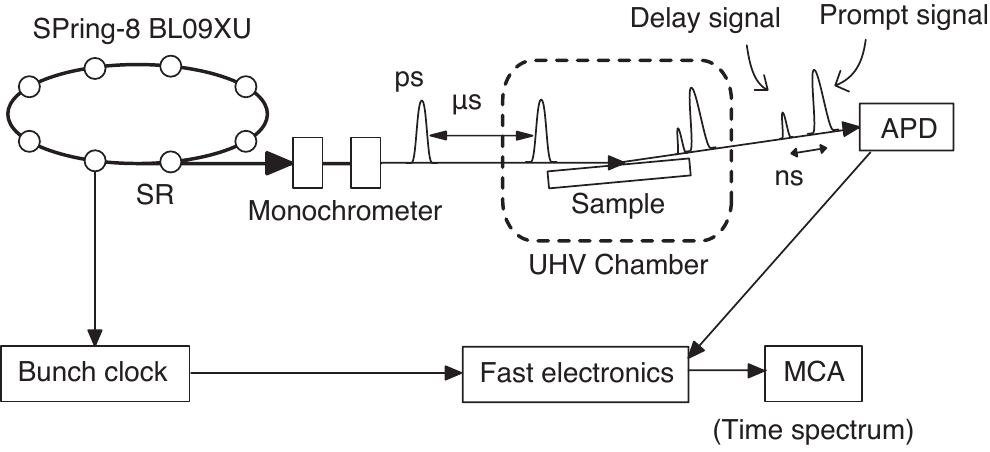} 
\caption{A schematic drawing of the experimental setup for the nuclear resonant scattering from $^{83}$Kr at surfaces. Synchrotron radiation (SR) is monochromized with two monochromators. The SR is then incident on a sample surface with a glancing angle. The mirror reflection is detected with an Avalanche photodiode (APD), which is a fast detector for x ray. \label{exp2}}
\end{center}
\end{figure}

A schematic drawing of the experimental setup for the NRS experiment is shown in Fig. \ref{exp2}. The synchrotron radiation was generated at the undulator installed in the beam line. The brilliance of the initial synchrotron radiation is about 10$^{\mathrm{13-15}}$ photons/s with the energy width of a few eV. The photon energy is about 9.4 keV for the excitation of $^{83}$Kr from $I=9/2$ to $I=7/2$. The excitation involves the M1 transition where the parity is unchanged. In the present case the parity of the ground and first excited state of both + (even). The natural lifetime of the excited $^{83}$Kr is reported to be 212 ns which corresponds to 3 neV in the energy domain \cite{Ruby1963}. The incident light is further monochromized with two monochromators to an energy resolution of a few meV to a few tens of meV. The incident light is generated by the electron bunch in the synchrotron. Therefore, they are pulsed in the time scale of a few pico seconds. The time interval between the pulsed light is determined by the bunch mode of the synchrotron. It is not possible to measure a time spectra which is longer than the bunch period. Therefore, in the measurement of the time spectra, one needs to use a bunch mode which is relatively longer than the natural life time or quantum beat frequency. In the present study, we chosed the bunch modes in SPring-8 which is about 300 ns and 600 ns. 

The pulsed synchrotron radiation is incident on the sample surface in the UHV chamber with a glancing angle regime through a Be window. This is realized with a UHV type goniomator on which a sample manipulator is mounted. This is developed by Kawauchi \textit{et al.} \cite{KawauchiVac}. With this UHV reflectometer, the sample was precisely angled against the synchrotron radiation from -2$^{\circ}$ to +2$^{\circ}$ with an accuracy of $\pm1/1000^{\circ}$. The mirror reflected x ray on the sample surface gets out of the UHV chamber via another Be window and incident on the avalanche photo-diode (APD) which posseses eight detectors for an efficient detection of the resonant signal. The measurements are typically a repetition of the pulse irradiation of x ray. The time duration of the repetition is about $\sim1$ micron seconds. The pulse duration is about a few pico seconds order. The delay signal is in the order of a few or a few tens of nano seconds. The electronics has to be as fast as sub nano seconds which is realized by a pulse count system. The pulse and the electronics are synchronized with a timing signal generated in and sent from the beam line.

The UHV chamber consists of the main chamber which is evacuated by a main turbo molecular pump which is evacuated by a pumping station. The sample manipulator is identical to the one used for the laser desorption experiment. In this case, a little modification of the sample mounter is conducted. In Fig. \ref{exp11} the sample is electrically floated from the ground. In the NRS experiment, this spacer is removed for a better thermal contact with the refrigerator. With this treatment, the sample is stably cooled at 19 - 20 K. The base pressure is at $2.0\times10^{-8}$ Pa. The sample gas are natural Kr and isotope enriched $^{83}$Kr gas. Natural abundance of $^{83}$Kr is 10 \%. The isotope enriched gas contains $^{83}$Kr with a ratio of 70 \%. Both gas were introduced into the UHV chamber via a variable leak valve. By controlling the exposure time and pressure, the sample thickness was varied.

\chapter{Results}
In this chapter, experimental results of laser desorption of Xe and nuclear resonant scattering are presented. In the first section, the results of laser desorption experiments are presented. All these experiments were carried out using the setup developed for the laser desorption experience as shown in Fig. \ref{exp1}. The desorption of Xe was investigated by the time-of-flight measurement, which showed a wavelength dependence, fluence dependence and the desorption flux dependence. In the next section, the experimental results obtained in SPring-8 BL09XU are presented. The experiment starts from the x ray reflectivity measurement. The resonance of $^{83}$Kr was successfully obtained for the first time in SPring-8 BL09XU. Furthermore, the determination of the coverage of Kr was achieved using the NRS intensity as a probe. In the end, the measurement of the time spectra of NRS from $^{83}$Kr at the solid surface was presented.

\section{Laser desorption of Xe from Au(001)}
In this section, the experimental results of laser induced desorption of Xe from a Au(001) surface are presented. The whole results are divided into three parts. First, the laser induced desorption of Xe in the condition where the coverage of Xe $\Theta$ kept constant at 1 ML is presented. The parameter here are the wavelength of the incident laser pulse at 2.3 and 6.4 eV, and the incident laser fluence. The result shows a striking dependence both on the wavelength and the fluence of the incident laser pulse. Secondly, the results of laser induced desorption under the condition where only laser induced thermal desorption is operative are presented. Here, the parameter is $\Theta$, while the laser fluence or photon energy was kept constant.

\begin{figure}
\begin{center}
\includegraphics[scale=.7, clip]{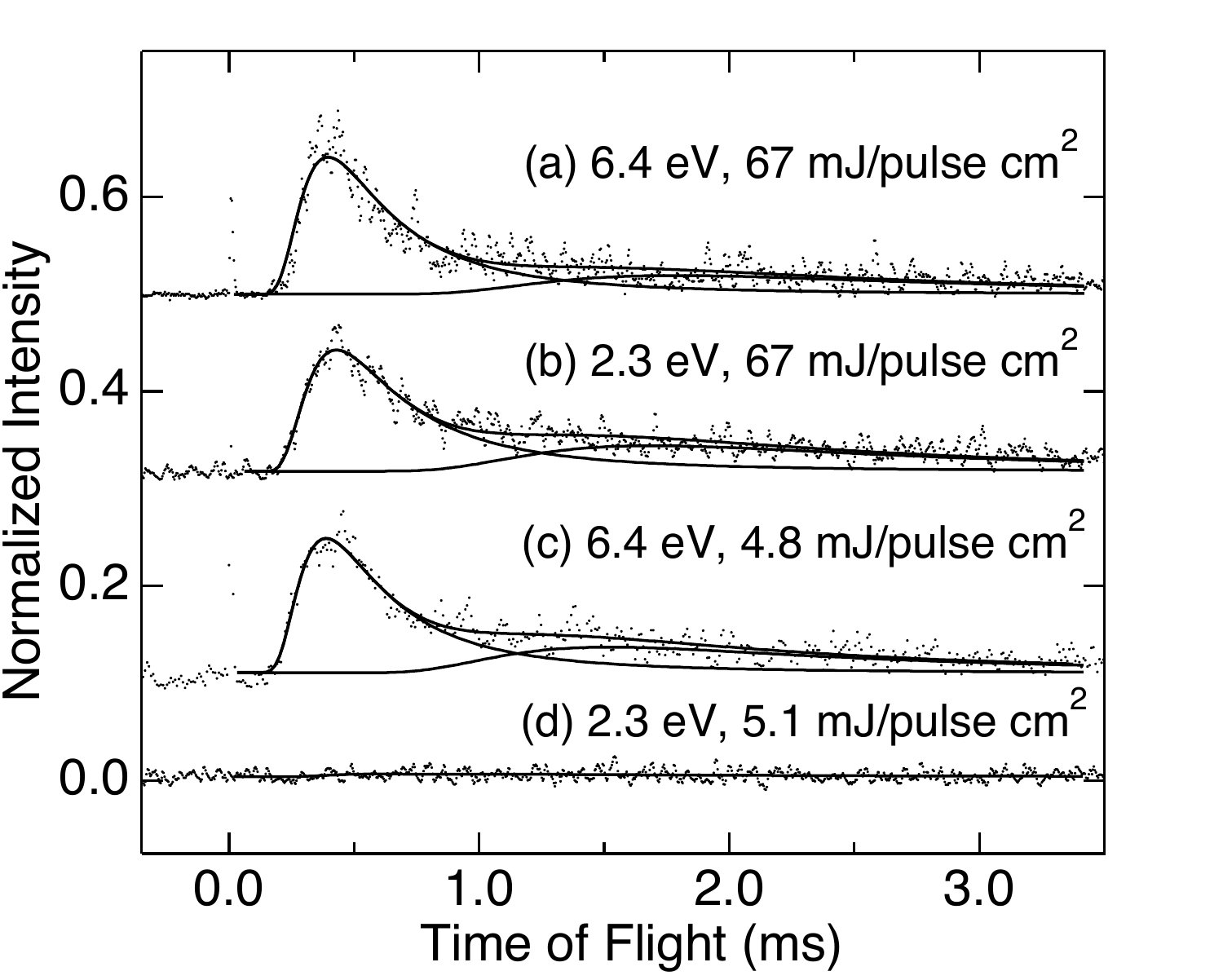} 
\caption{Time-of-flight spectra of desorbing Xe atoms from Au (001) following pulse laser irradiation. On each spectra, the photon energy and the absorbed energy by the sample are denoted. Data of (a), (b) and (c), (d) were recorded with a single pulse and the accumulation of over 120 pulses, respectively.\label{tof}}
\end{center}
\end{figure}

\subsection{Fluence dependence}
The time-of-flight (TOF) of desorbing Xe atoms upon laser irradiation was recorded at a wide range of laser pulse energy absorbed by the sample ($I_{\mathrm{L}}$) for both 6.4 and 2.3 eV photons. $I_{\mathrm{L}}$ was estimated by taking account of the reflectivity on Au (0.8 for 2.3 eV and 0.2 for 6.4 eV). Figure~\ref{tof} shows typical TOF results. The data reveal a maximum at a TOF of 400 $\mu$s with a tailing feature in the long TOF region. TOF was recorded with only one pulse for both 6.4 and 2.3 eV photons with $I_{\mathrm{L}}>$ 10 mJ/pulse cm$^{2}$ [Figs.~\ref{tof} (a) and \ref{tof} (b)]. With $I_{\mathrm{L}}<$ 10 mJ/pulse cm$^{2}$, on the other hand, TOF was recorded by accumulating over 120 data [Figs.~\ref{tof} (c) and \ref{tof} (d)] because the Xe desorption yield was small. Whereas a substantial desorption yield was observed with 6.4 eV photons as shown in Fig.~\ref{tof} (c), no significant signal was recorded with 2.3 eV photons as shown in Fig.~\ref{tof} (d). Solid curves in Fig.~\ref{tof} are fits to the data with a sum of two Maxwell-Boltzmann (MB) velocity distributions described as 
\begin{equation}
f(v)=A_{1}v^{2}\exp\left(-\frac{mv^{2}}{2kT_{\mathrm{D1}}}\right)+A_{2}v^{2}\exp\left(-\frac{mv^{2}}{2kT_{\mathrm{D2}}}\right),
\end{equation}
where $m$ and $k$ are mass of a Xe atom and the Boltzmann constant, respectively, and $A_{i}$ and the translational temperature $T_{\mathrm{D}i}$ are fitting parameters. In the analysis, the form is converted to the flux weighted form. In the following, we discuss only the fast component of the TOF.

\begin{figure}
\begin{center}
\includegraphics[scale=.6, clip]{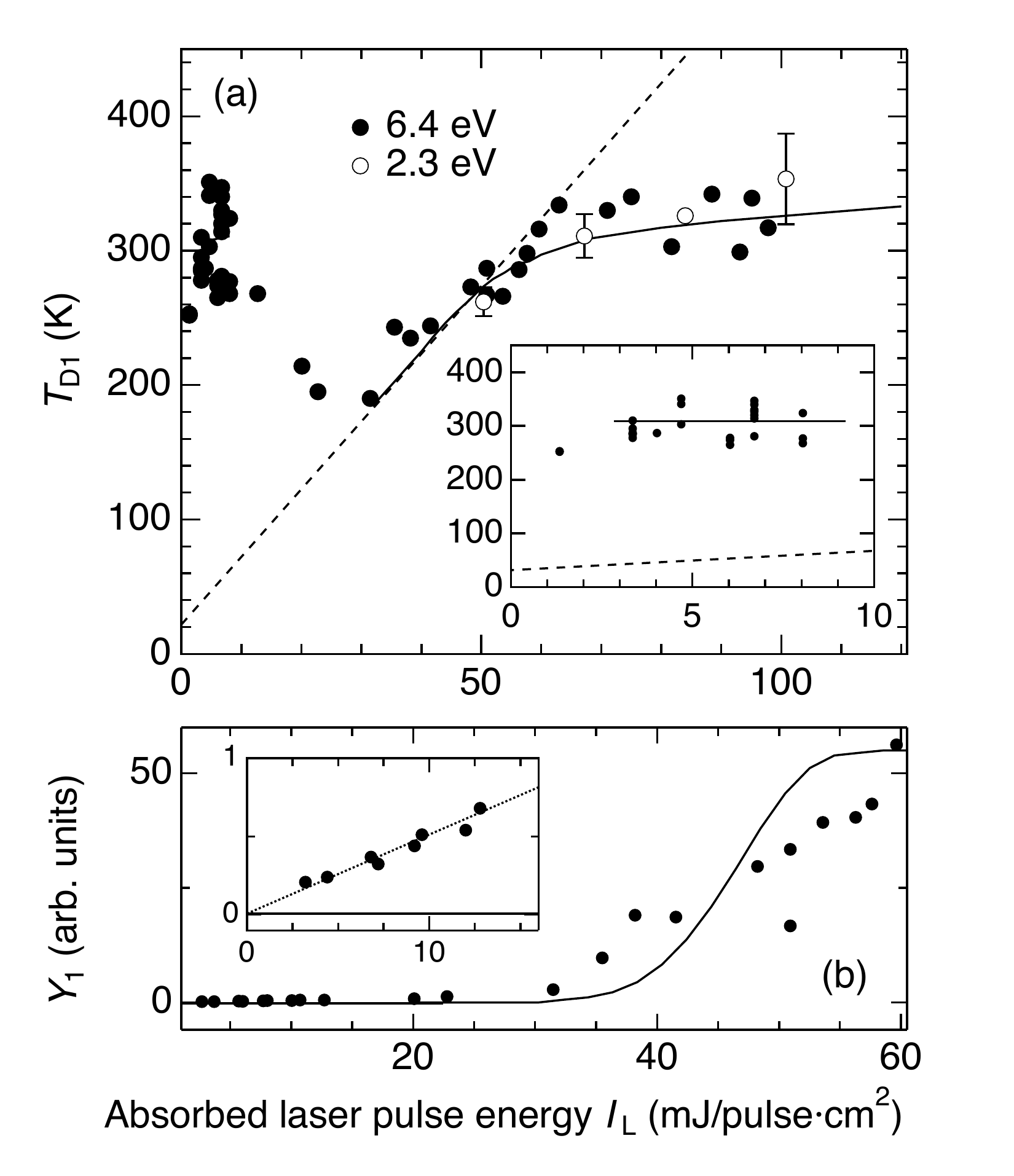} 
\caption{(a) Translational temperature ($T_{\mathrm{D}1}$) of Xe atoms desorbed from Au(001) following laser irradiation as a function of absorbed laser pulse energy ($I_{\mathrm{L}}$). Solid and dashed lines are calculated results of the surface temperature at the maximum desorption rate ($T_{\mathrm{DM}}$) and the maximum surface temperature ($T_{\mathrm{SM}}$), respectively. The inset is a magnification of the small $I_{\mathrm{L}}$ region. (b) Desorption yield of Xe atoms ($Y_{\mathrm{1}}$) following the laser irradiations with 6.4 eV photons as a function of $I_{\mathrm{L}}$. Calculated results of $Y_{\mathrm{1}}$ are shown by the solid curve. The inset is a magnification of the small $I_{\mathrm{L}}$ region. \label{td}}
\end{center}
\end{figure}

\begin{figure}
\begin{center}
\includegraphics[scale=.6, clip]{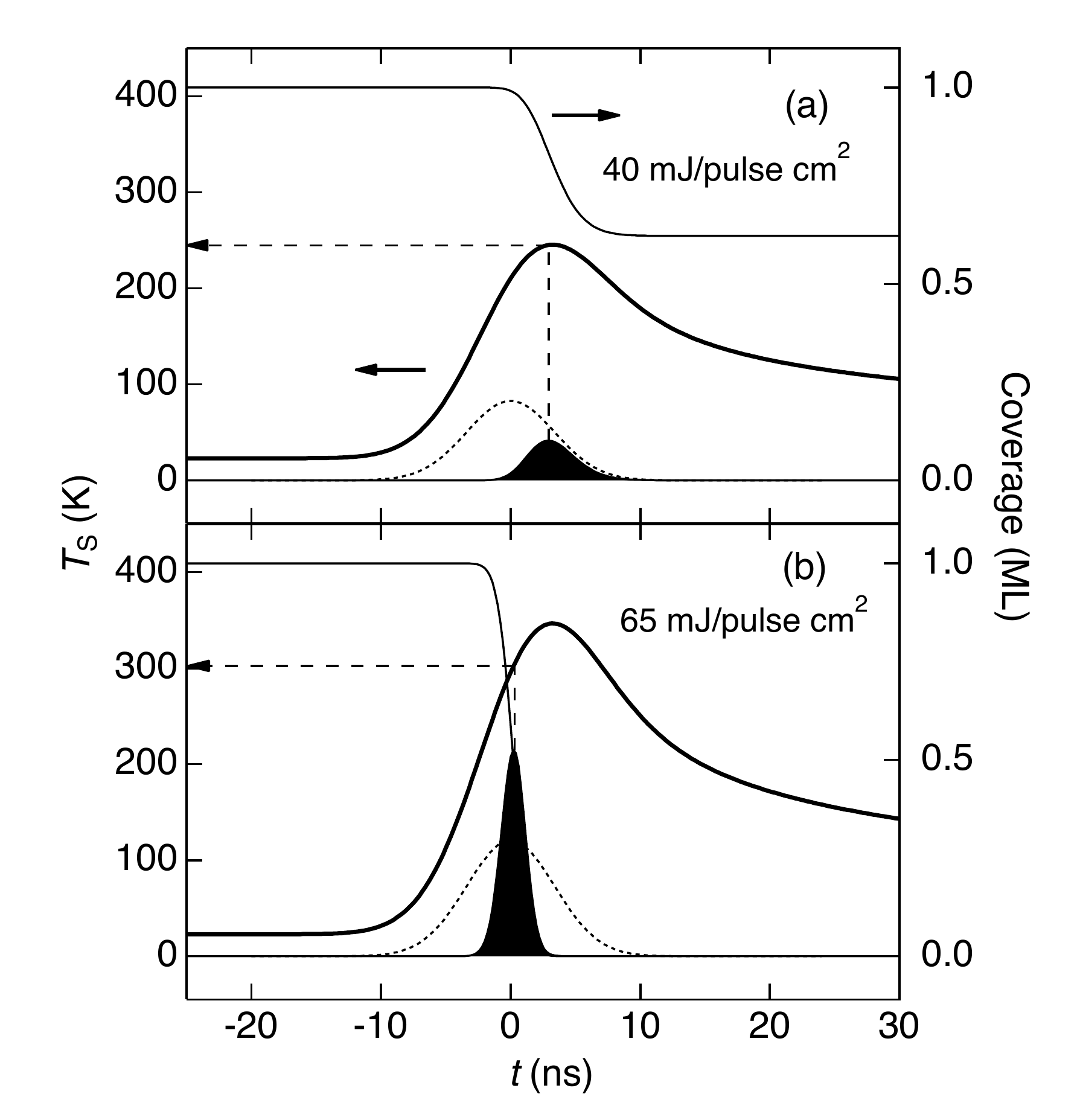} 
\caption{Time evolution of surface temperature $T_{\mathrm{S}}$, surface coverage $\Theta$, and desorption rate following the laser irradiation at (a) 40 mJ/pulse cm$^{2}$ and (b) 65 mJ/pulse cm$^{2}$, respectively. \label{cal}}
\end{center}
\end{figure}

Figures~\ref{td} (a) and \ref{td} (b) show $T_{\mathrm{D}1}$ and the Xe desorption yield of the first component ($Y_{\mathrm{1}}$) plotted as a function of $I_{\mathrm{L}}$, respectively. We first focus on the region of $I_{\mathrm{L}}>32$ mJ/pulse cm$^{2}$. In this region, $T_{\mathrm{D}1}$ increases with increasing $I_{\mathrm{L}}$ from about 200 K at 32 mJ/pulse cm$^{2}$ and saturates at about 300 K, for both 6.4 and 2.3 eV photons. In Fig~\ref{td} (b), $Y_{\mathrm{1}}$ shows a sharp increase at 32 mJ/pulse cm$^{2}$.  The behavior of $T_{\mathrm{D}1}$ and $Y_{\mathrm{1}}$ indicates that the Xe desorption is thermally activated with $I_{\mathrm{L}}$ $>32$ mJ/pulse cm$^{2}$ \cite{Wedler, Hussla}. For the quantitative analysis, we carried out numerical calculations of the surface temperature ($T_{\mathrm{S}}$) during laser irradiation on the basis of the one-dimensional heat conduction equation \cite{Hicks}. Subsequently, the time evolution of the Xe coverage and the Xe desorption rate (i.e., LITD) was deduced by employing the first-order desorption kinetics assuming the activation energy for desorption of Xe from Au(001) to be 240 meV \cite{Mcelhiney}. By the calculations above, we obtained the surface temperature at maximum Xe desorption rate ($T_{\mathrm{DM}}$) and the maximum surface temperature ($T_{\mathrm{SM}}$) following laser pulse irradiation with $I_{\mathrm{L}}$. The calculated results are shown in Fig. \ref{cal}.

$T_{\mathrm{DM}}$ and $T_{\mathrm{SM}}$ deduced from the calculation are depicted as a function of $I_{\mathrm{L}}$ in Fig.~\ref{td} (a) as dashed and solid curves, respectively. $T_{\mathrm{D1}}$ obtained by the experimental results is in good agreement with $T_{\mathrm{DM}}$, indicating that LITD of Xe is dominant with $I_{\mathrm{L}}>32$ mJ/pulse cm$^{2}$. $T_{\mathrm{DM}}$ deviates from $T_{\mathrm{SM}}$ with $I_{\mathrm{L}}>50$ mJ/pulse cm$^{2}$ because desorption occurs before the surface temperature reaches its maximum \cite{Wedler}. The solid line in Fig.~\ref{td} (b) shows a calculated result of the $Y_{\mathrm{1}}$ by LITD as a function of $I_{\mathrm{L}}$. The calculated $Y_{\mathrm{1}}$ sharply increases at 35 mJ/pulse m$^{2}$ and saturates above 55 mJ/pulse cm$^{2}$. This is in good agreement with the experimental data that the desorption yield exhibits a steep increase at 32 mJ/pulse cm$^{2}$. This thresholdlike behavior is typical of LITD. However, the experimental data in Fig.~\ref{td} (b) monotonously increases in contrast to the saturating behavior of the calculated curve, which may be caused by either spatial inhomogeneity of the desorption laser intensity \cite{Koehler} or coverage dependence of the activation energy for desorption due to the attractive interactions between the adsorbates \cite{Wedler}.

We turn next to the region of $I_{\mathrm{L}}<$ 24 mJ/pulse cm$^{2}$, where desorption of Xe is observed only at 6.4 eV. As can be seen in Fig.~\ref{td} (a), $T_{\mathrm{D}1}$ significantly deviates from the calculated result of $T_{\mathrm{SM}}$ in this region of $I_{\mathrm{L}}$. The inset in Fig.~\ref{td} (a) shows a magnification of $T_{\mathrm{D}1}$ with $I_{\mathrm{L}}<10$ mJ/pulse cm$^{2}$. In this region, $T_{\mathrm{D}1}$ is independent of $I_{\mathrm{L}}$ and constant at 300$\pm$20 K which is much higher than $T_{\mathrm{SM}}$ ($<$100 K). Since in the region of $I_{\mathrm{L}}<24$  mJ/pulse cm$^{2}$ the calculated result of LITD fails to account for the experimental data, other desorption mechanisms should be operative. Especially with $I_{\mathrm{L}}<10$ mJ/pulse cm$^{2}$, the LITD yield of Xe is negligible because $T_{\mathrm{SM}}$ is too low for the thermal activation of Xe desorption. Therefore, only non-thermal PSD of Xe atoms from Au(001) is operative in this region of $I_{\mathrm{L}}$. As shown in the inset of $Y_{\mathrm{1}}$ as a function of $I_{\mathrm{L}}$ in Fig.~\ref{td} (b), $Y_{\mathrm{1}}$ with 6.4 eV photons linearly increases with increasing $I_{\mathrm{L}}$, indicating that the observed nonthermal PSD is a one-photon process. The nonthermal PSD cross section $\sigma_{\mathrm{PSD}}$ was deduced to be 10$^{-21}$$-$10$^{-22}$ cm$^{2}$ by comparing the nonthermal PSD yield with the LITD yield of the Xe monolayer.

\subsection{Coverage dependence}

\begin{figure}
\begin{center}
\includegraphics[scale=.6, clip]{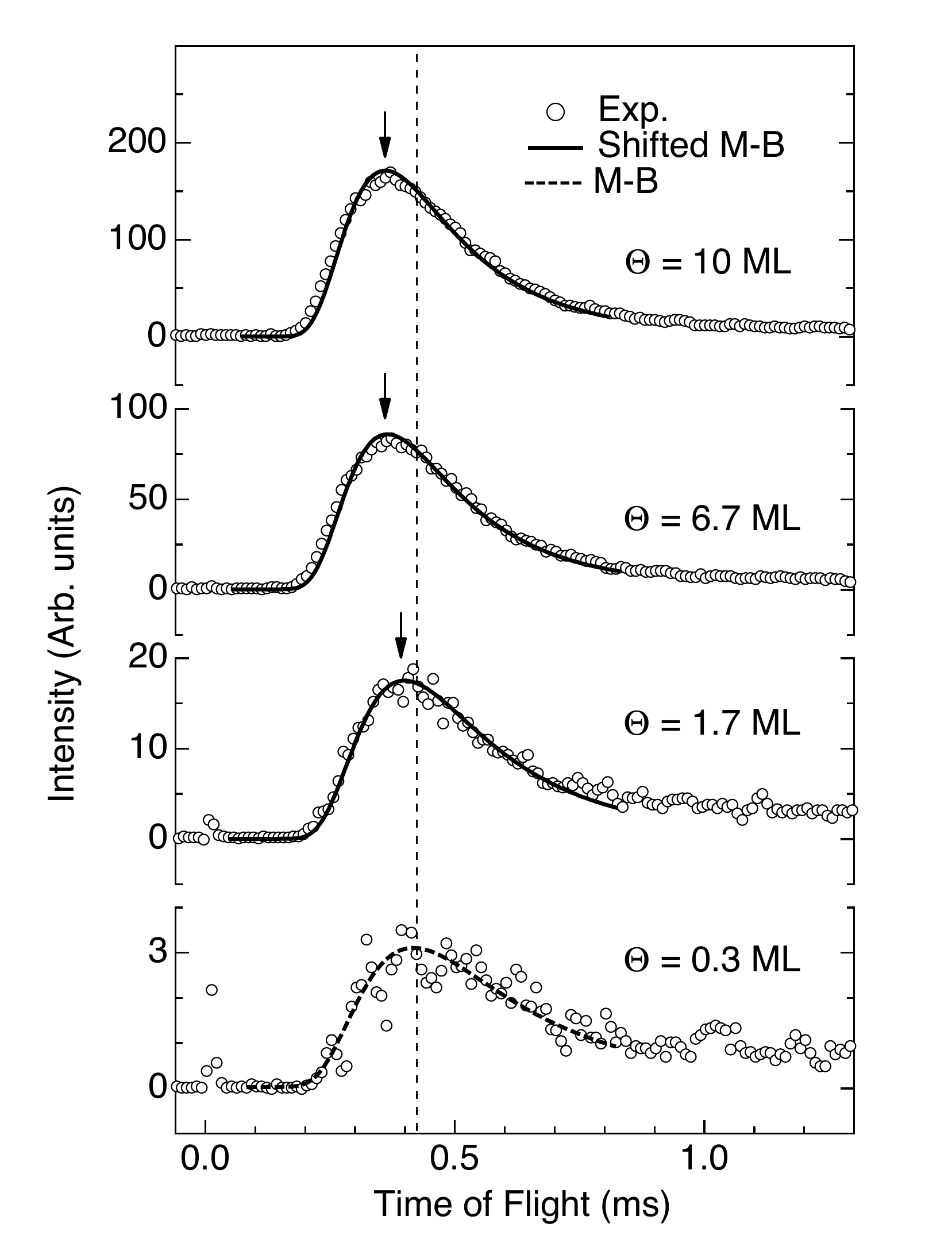} 
\caption{Time-of-flight (TOF) spectra of Xe from Au surfaces following pulsed laser irradiations. The Xe coverage $\Theta$ are 10, 6.7, 1.7 and 0.3 ML from top to bottom, respectively. The dashed curves and the solid curves are the Maxwell-Boltzmann (M-B) velocity distribution and the shifted M-B velocity distribution fitted to the experimental results, respectively. The vertical dashed line and the arrows indicate the peak position of each spectrum.  \label{tof2}}
\end{center}
\end{figure}

\begin{figure}
\begin{center}
\includegraphics[scale=.8, clip]{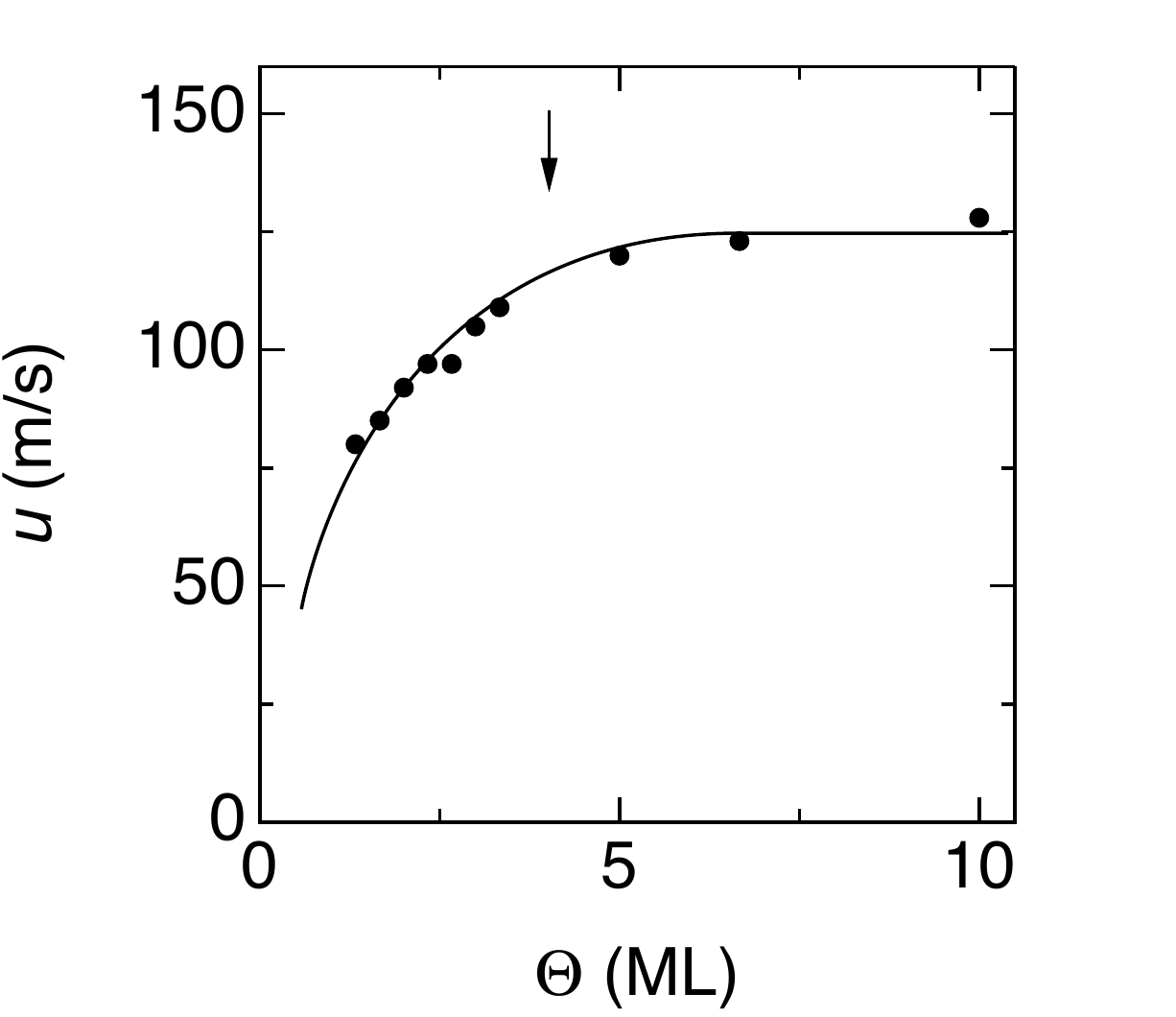}
\caption{The stream velocity $u$ of the desorbed Xe atoms from Au surfaces following the pulsed laser irradiations as a function of the Xe coverage $\Theta$. The data are obtained by analyzing the time-of-flight spectrum with the shifted Maxwell-Boltzmann velocity distribution. The solid line is a guide for eyes. $u$ become constant at above 4 ML which is denoted by the arrow.\label{tud}}
\end{center}
\end{figure}

We obtained a series of TOF spectra of desorbed Xe from Au surfaces following pulsed laser irradiation as shown in Fig. \ref{tof2}. The TOF spectrum of Xe from the Au surface at $\Theta = 0.3$ ML is shown at the bottom of Fig. \ref{tof2}. The vertical dashed line at 0.42 ms shows the peak position of the spectrum. The spectrum was well analyzed with a Maxwell-Boltzmann velocity distribution in a flux from a thermal source described as, \cite{Comsa}
\begin{equation}
J(v)dv=Av^{3}\exp\left(-\frac{mv^{2}}{2kT_{\mathrm{D}}}\right)dv,\label{mb}
\end{equation}
where $m$, $k$ and $v$ are the mass of a Xe atom, the Boltzmann constant and the velocity of Xe atoms, respectively, and $A$ and $T_{\mathrm{D}}$ are fitting parameters. In the analysis, we convert Eq. (\ref{mb}) to a TOF function for the density sensitive detector, which results in the form $f(t)dt=a t^{-4}\exp(-bt^{-2})dt$. By fitting Eq. (\ref{mb}) to the TOF of Xe at $\Theta=0.3$ ML, we obtained $T_{\mathrm{D}}$ of 255 K as illustrated by the dashed curve at the bottom of Fig. \ref{tof2}.

At the top and middle of Fig. \ref{tof2}, the TOF spectra of desorbed Xe at $\Theta=10$ to 1.7 ML are shown. We note that the peak positions of those TOF spectra, which are indicated by the arrows, are shifted towards smaller values with increasing $\Theta$ compared with that of $\Theta$ = 0.3 ML. It was found that the peak positions of the TOF spectra become constant at about 0.34 ms at $\Theta\geq5.0$ ML. 

\clearpage
\section{Nuclear resonant scattering by $^{83}$Kr/TiO$_{2}$(110)}
In this section, the results of experiments conducted at SPring-8 BL09XU are presented. The results include x-ray reflectivity curves which roughly informs us the thickness of the Kr film prepared on the TiO$_{2}$(110) cooled at 20 K. With this result, it is confirmed in situ that Kr is present on the TiO$_{2}$ surface with a thickness of a few to a few hundreds of nano meters by varying the exposure time. Then, with a Kr film of about 200 nm, the resonant absorption of x ray by $^{83}$Kr is explored and found. Here, the energy width of the incident radiation on the sample is clarified. Then, the coverage dependence of the resonance intensity was measured. With this result, it become possible for us to estimate the thickness of the Kr film in situ at the monolayer resolution. I then prepared a monolayer physisorbed layer of $^{83}$Kr and 5 ML film. Time spectra of NRS of each sample was measured. The spectra are then analyzed with the single exponential function and with the periodic function. The possibility that the observed oscillation results from the hyperfine structure is inspected.

\subsection{X-ray reflectivity curves}
\begin{figure}
\begin{center}
\includegraphics[scale=1.8, clip]{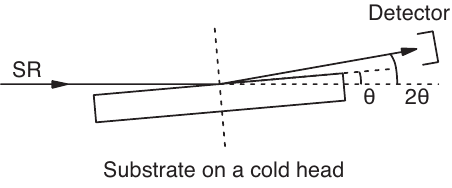}\\
\includegraphics[scale=0.6, clip]{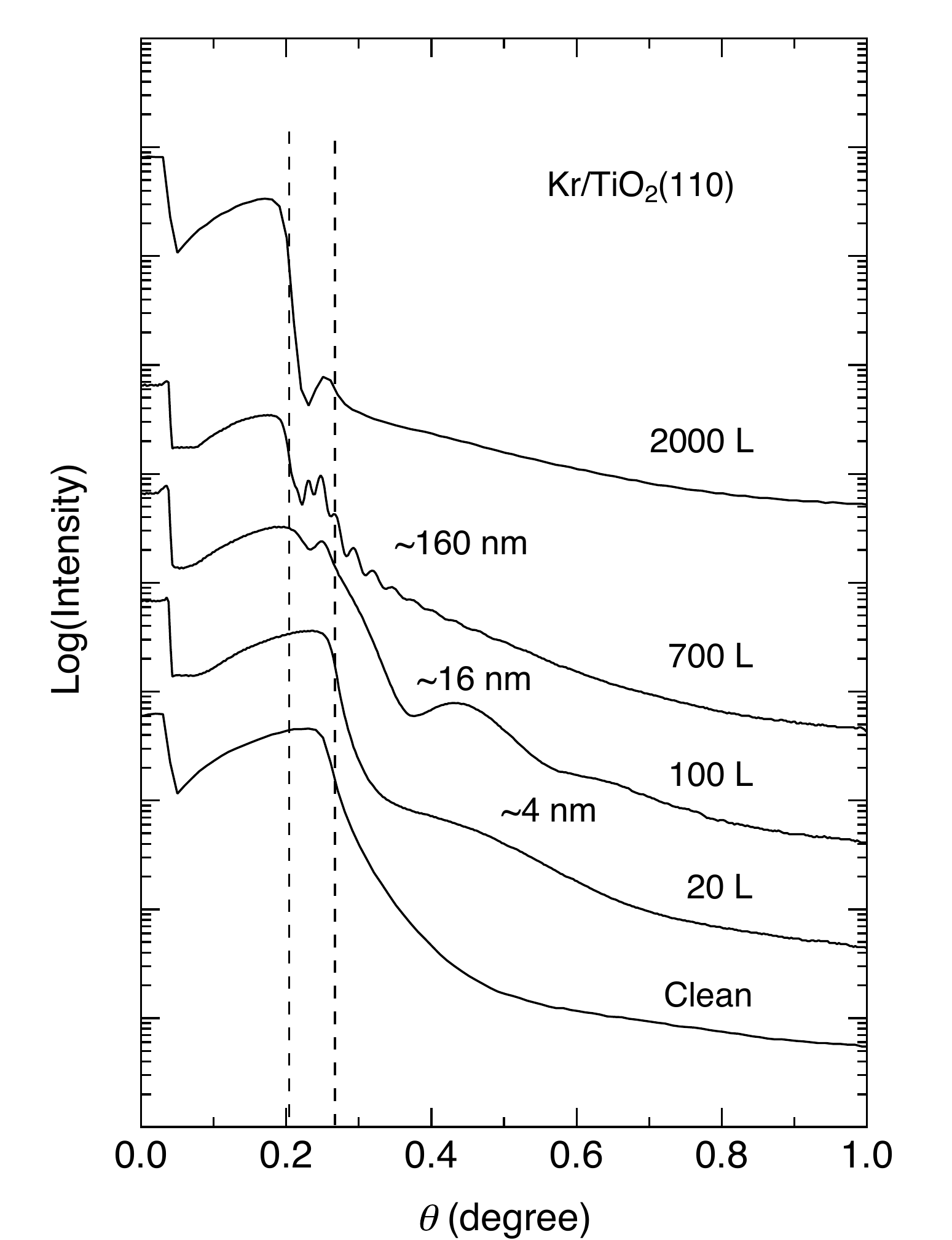}
\caption{X ray reflectivity curves as a function of the thickness of the Kr physisorbed layer. Two vertical dashed line indicate the position of the critical angle of the clean TiO$_{2}$(110) surface and thick Kr film with exposure of 2000 L.\label{xr}}
\end{center}
\end{figure}

In order to confirm that Kr is adsorbed on TiO$_{2}$(110), a series of x ray reflectivity curves are measured with the $\theta$ and $2\theta$ configuration. The results of x ray reflectivity curves as a function of initial Kr exposure was displayed in Fig. \ref{xr}. In this experiment, natural Kr gas was used because the resonance is not involved. As can be seen from the bottom curve of Fig. \ref{xr}, on the clean surface of TiO$_{2}$(110) which was annealed at 250 K, the reflectivity drops at about 0.26$^{\circ}$. This is consistent with a calculated value of 0.25$^{\circ}$ based on the electron density in rutile TiO$_{2}$. With increasing exposure of Kr gas from 20 L to 2000 L, the x ray reflectivity curve changes its shape as seen in the curves of Fig. \ref{xr} from bottom to the top. As represented by the third and forth curve from the bottom, formation of the thin film of Kr is confirmed by the interference pattern. The thickness of Kr film is tentatively estimated using a Bragg reflection model and the oscillation periods that are experimentally obtained. The estimated thickness is shown in Fig. \ref{xr}. This result fairly well coincides with the exposure Kr assuming the condensation probability to be unity. Furthermore, with the exposure of 2000 L, the oscillation disappears indicating that the thickness of Kr film is so large that the reflection of x ray is completed on Kr film with a negligible reflection from the substrate. This assumption is validated by the fact that the critical angle at the top curve is around 0.21$^{\circ}$, which is in very good agreement with the calculated result based on the electron density in solid Kr. With these results, it is confirmed that it is possible to prepare a thick Kr film on the TiO$_{2}$(110) surface. This is mandatory for looking for the resonance position of $^{83}$Kr in the following experiment.

\subsection{Resonant spectrum}
\begin{figure}
\begin{center}
\includegraphics[scale=0.65, trim=0 0 20 0]{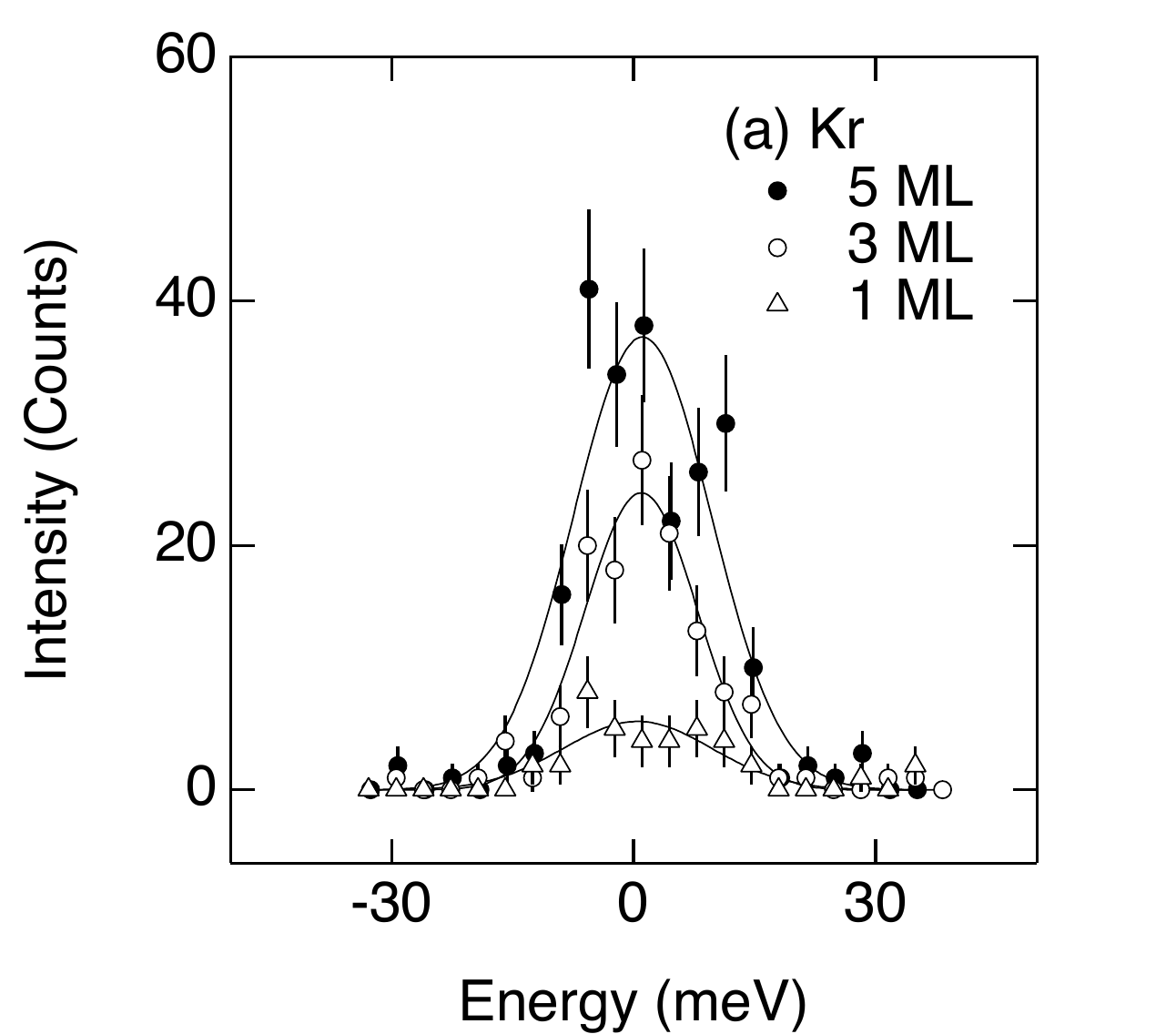}\\
\includegraphics[scale=0.65]{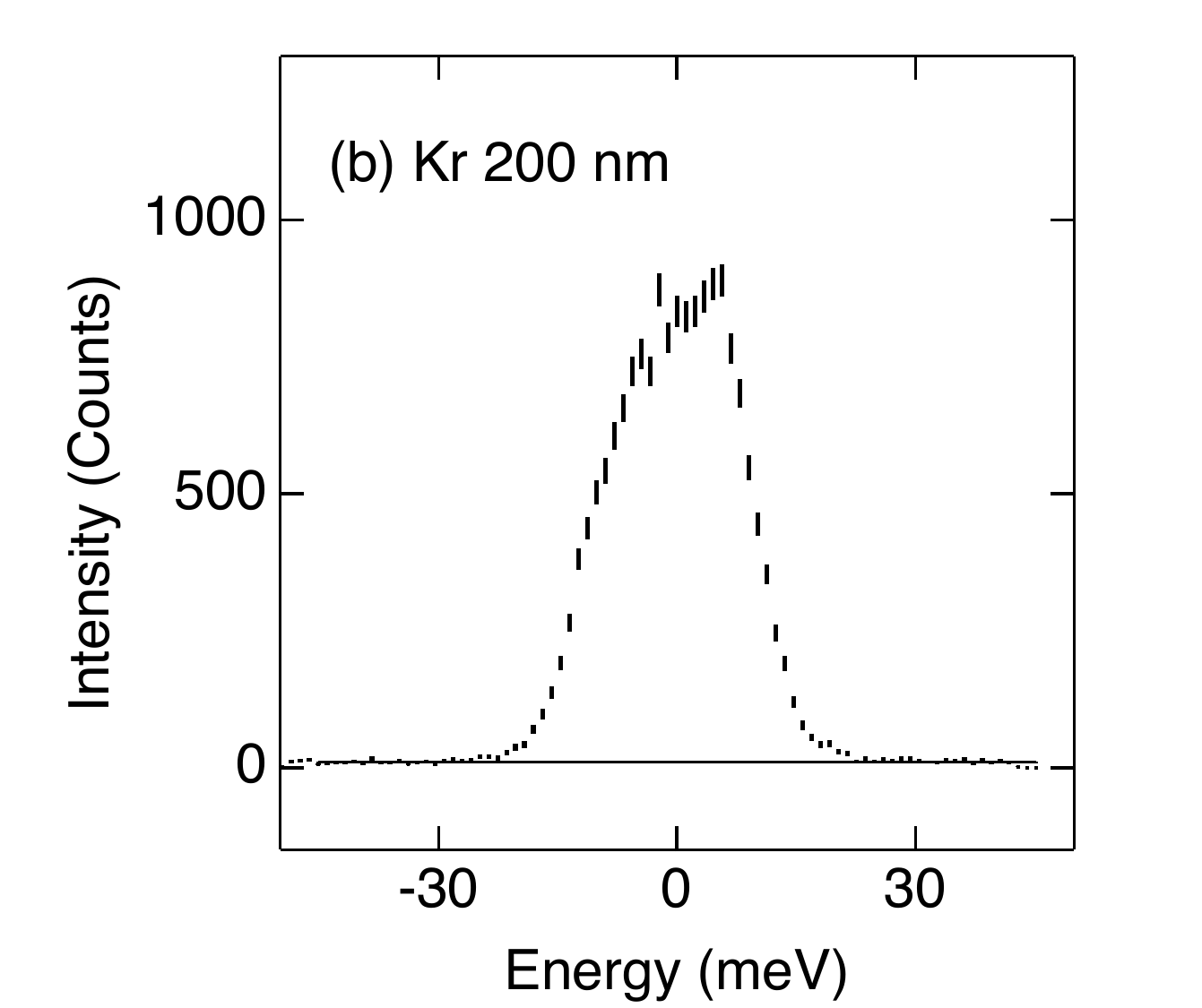}
\caption{Energy spectrum of nuclear resonant scattering of $^{83}$Kr at the nuclear transition of $I=9/2$ to 7/2. The excitation energy is 9.4 keV. The width of the spectrum is about 30 meV in full width half maximum. \label{sp}}
\end{center}
\end{figure}

The resonance of $^{83}$Kr nuclear levels has not been achieved in SPring-8 BL09XU. It had been achieved in KEK, Japan with a monochromator developed in KEK, which was used in the present study with a courtesy of Dr. X. W. Zhang \cite{Zhang}. The energy width of the monochromator was reported to be 25 meV \cite{Zhang}. First a very thick Kr thin film was prepared on the TiO$_{2}$ surface so as to increase the count rate. By varying the incident energy, a resonance of $^{83}$Kr was found as shown in Fig. \ref{sp}. The width of the energy spectrum is determined by a convolution of the natural width of the spectrum and the energy resolution of the monochromator. In the present case the natural width of the spectrum is about 3 neV and the energy resolution of the monochromator is about 25 meV, which is much larger compared to the natural width. Therefore, the observed spectrum naturally shows the energy resolution of the monochromator. As reported in the previous study, the energy resolution is about 25 - 30 meV in full width half maximum. The slight difference of the spectrum shape arise from the different characteristics of the synchrotron radiation. This is the first observation of the resonance of $^{83}$Kr nucleus in the glancing angle regime so far. This is also the first observation of the resonance of $^{83}$Kr nucleus in SPring-8.

\begin{figure}
\begin{center}
\includegraphics[scale=0.6, clip]{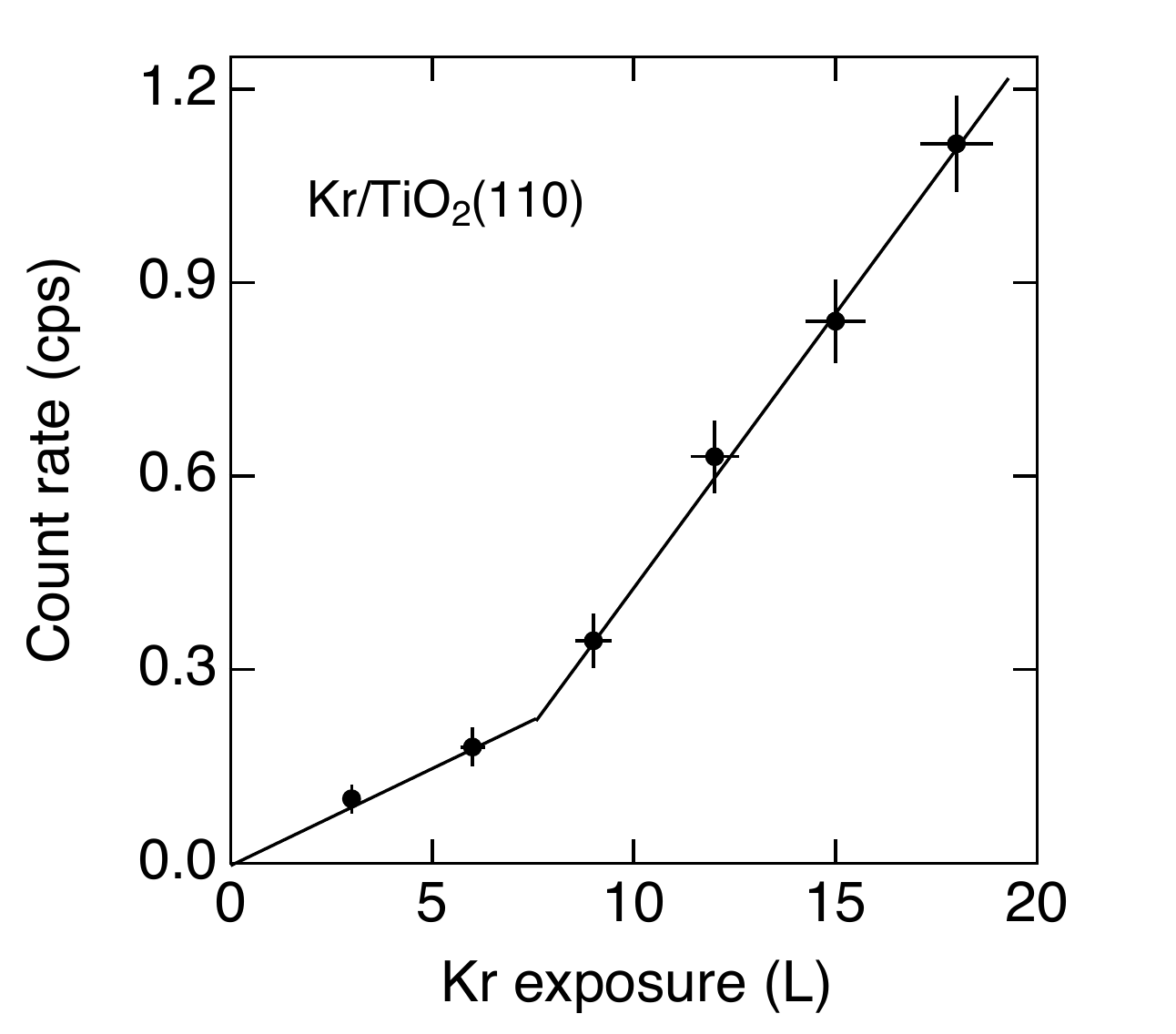}
\caption{NRS delay intensity as a function of the Kr exposure on a TiO$_{2}$ surface. The experimental data are plotted by dots and error bars. The data are fitted with two lines.\label{cdelay}}
\end{center}
\end{figure}

As we have succeeded in observing the resonance of $^{83}$Kr nucleus, we further reduced the thickness of the prepared Kr layers and measured the delay intensity as a function of the Kr exposure. In this experiment, isotope enriched $^{83}$Kr gas is used for the better count rate and the simplicity of the analysis. The result is shown in Fig. \ref{cdelay}. The experimental data are well fitted with two solid lines of which the slopes are different.

The result is understood in the following way. It has been accepted that the sticking probability of physisorbed species on to the bare surfaces is less than unity, whereas the condensation probability of physisorbed species on to the physisorbed layers (i.e. the formation of second and higher layers) is close to unity. Considering this point, the slope at the Kr exposure above 10 L results from the full condensation of $^{83}$Kr atoms. In the case of Kr gas, given that the condensation probability is unity, the exposure of 2.5 L corresponds to the formation of monolayer on TiO$_{2}$(110) which is commensurate to the substrate. Thus, the following relation should hold in the multi-layer region of Kr, that is $(\mathrm{Count\ rate})=a\times(\mathrm{Kr\ exposure})/2.5$. By analyzing the experimental data with the relation, we obtained a value of $a$ to be 0.28 cps/ML. This result is well coincide with the fact that the region where the slope is small is about 0.28 cps from 0, indicating that in this region the initial sticking probability is small compared to unity. The slope in the region of Kr exposure from 0 to 7 L is 3 time smaller than that in the region above 7 L. This indicates that the initial sticking probability is 0.3.

The present result shows that with Kr exposure of 0 to 7 L, the sub-mono to monolayer Kr is formed. With Kr exposure greater than 7 L, multi-layer growth of Kr layer is achieve. In the way, it was become possible to prepare a monolayer Kr and multi-layer Kr on TiO$_{2}$(110) surfaces. The thickness can be monitored in situ with NRS intensity.

\subsection{Time spectra}
\begin{figure}
\begin{center}
\includegraphics[scale=.6, clip]{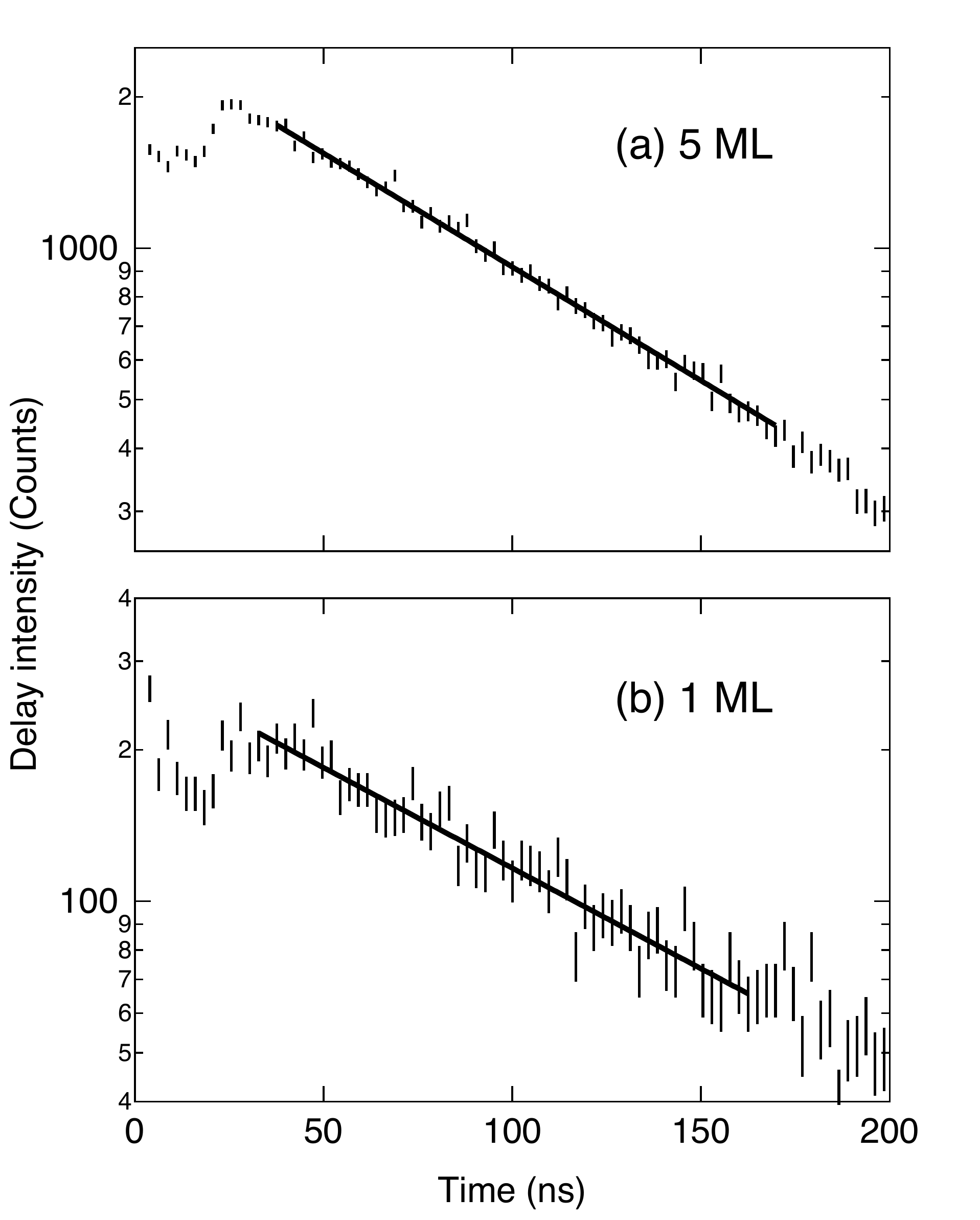}
\caption{Time spectrum of nuclear resonant scattering from Kr physisorbed layer on a TiO$_{2}$ surface. The coverage of Kr layer is (a) 5 ML and (b) 1 ML, respectively.\label{ts}}
\end{center}
\end{figure}

In order to investigate the hyperfine structure of Kr at the surface, we prepared the Kr monolayer and 5 ML on TiO$_{2}$(110) surfaces and measured the time spectra of NRS. The results are shown in Fig. \ref{ts}. In Fig. \ref{ts} (a) and (b), the time spectra of NRS from $^{83}$Kr at 5 ML and 1 ML are shown, respectively. The spectra show single exponential decay. Both spectra are well fitted with a single exponential function as
\begin{equation}
I(t)=A\exp\left(-\frac{t}{\tau}\right),\label{singlexp}
\end{equation}
where $A$ and $\tau$ are a pre-factor and a time constant. By fitting Eq. (\ref{singlexp}) to the experimental date shown in Fig. \ref{ts} (a) and (b), we obtained the values of $\tau$ to be 96 ns for 5 ML and 109 ns for 1 ML Kr layer. The natural lifetime of the first excited state of $^{83}$Kr is 212 ns \cite{Ruby1963}. 

Compared to the natural lifetime of $^{83}$Kr, both experimentally observed values are small. This is assumed to be due to the effect called speed-up, which is caused by the multiple scattering in the lattice. The effect is also referred to as dynamical effect. The dynamical effect is determined by the effective thickness of the sample which is a product of the Lamb-M\"ossbauer factor $f_{\mathrm{LM}}$ and the column density of the sample nuclei \cite{Rohlsberger}. In case of the transmission geometry, the dynamical beat is analytically described with a Bessel function including $f_{\mathrm{LM}}$. Unlike the case of transmission geometry, the dynamical effect is not simple in case of the reflection geometry. Qualitatively, the larger the effective thickness of the sample is, the larger the speed up is. The present result is consistent with this trend. The thiner sample of 1 ML Kr shows a less speed up compared to the speed up of the thicker sample of 5 ML Kr.

Although the analysis of the time spectrum of NRS by 1 ML $^{83}$Kr on TiO$_{2}$(110) using Eq. (\ref{singlexp}) is not bad, a systematic deviation of the experimental data from the analytical line can be noticed in Fig. \ref{ts} (b) as compared to the agreement between the analytical line and the experimental results in Fig. \ref{ts} (a). This minor signal seen in Fig. \ref{ts} (b) might be due to the quantum beat due to the hyperfine structure of $^{83}$Kr nuclei in the vicinity of the surface. The possibility of the observation of quantum beat structure is discussed in the next section.

\chapter{Discussion}
In this chapter, the experimental results presented in the previous chapter are discussed. In the first section, the desorption dynamics of Xe from the Au(001) surface is discussed, which involves the fluence dependence and the coverage dependence of the time-of-flight spectra. In the next section, the results of NRS of $^{83}$Kr at the surface is discussed. The discussion concentrates on the effects of EFG on the $^{83}$Kr and whether it appears as a quantum beat in the time spectrum.

\section{Non-thermal desorption of Xe}
In this section, the desorption dynamics of Xe from the Au(001) surface is discussed. In the first subsection, the explanation for the desorption path which showed a wavelength dependence is discussed. It is proposed that the transient Xe$^{-}$ formation is involved in the laser desorption of Xe at 6.4 eV photons. In the following subsection, the desorption dynamics is classically calculated and compared with the experimental results. It was found that the calculated results and the experimental results show a very good agreement with each other, indicating the proposed model is valid. 

\begin{figure}
\begin{center}
\includegraphics[scale=1.2, clip]{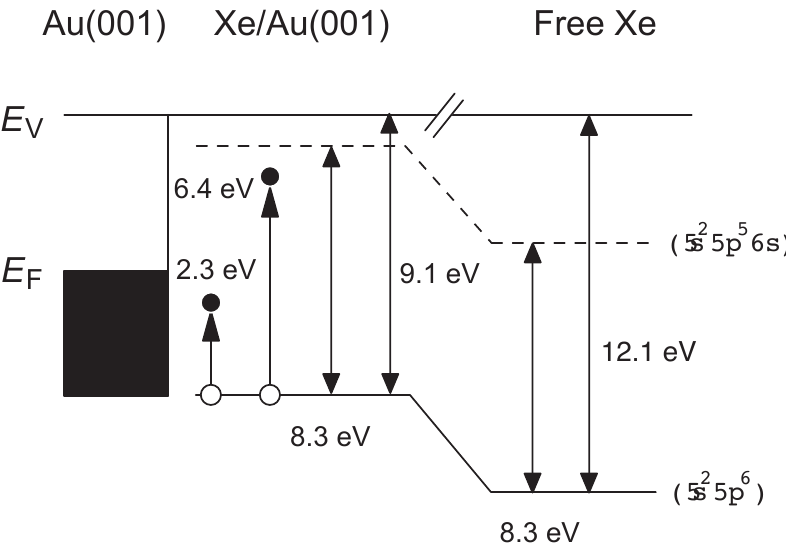} 
\caption{A schematic drawing of electronic levels of Xe on Au(001) surfaces and photo-excitation of Xe. The ionization energy and excitation energy of free Xe is denoted as 12.1 and 8.3 eV, respectively. The ionization energy and excitation energy of Xe on Au(001) is reduced due to adsorption on to metal surface.\label{eledi1}}
\end{center}
\end{figure}

\begin{figure}
\begin{center}
\includegraphics[scale=1.2, clip]{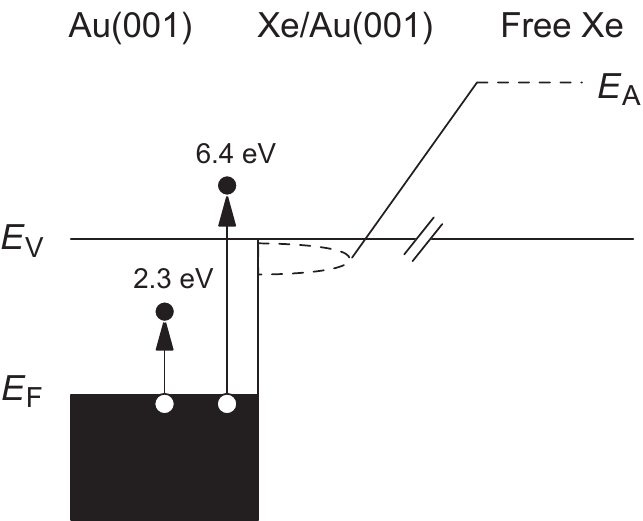} 
\caption{A schematic drawing of electronic levels of Xe on Au(001) surfaces and photo-excitation of substrate electrons. The electron affinity of gaseous Xe atoms is negative and unbound. The electron affinity level may be stabilized due to the metal proximity effect.\label{eledi2}}
\end{center}
\end{figure}

\subsection{Excitation paths}
We first discuss the initial excitation of the non-thermal PSD of Xe from Au(001) upon irradiation of 6.4 eV photons. The following discussion is schematically shown in Fig. \ref{eledi1}. As the initial excitation, we argue that the negative ion state of Xe is formed via the photoexcitation of the substrate electron \cite{Richter}. Other excitation pathways can be excluded for the following reasons. The first-excitation energy of Xe from the ground state (5$s^{2}$5$p^{6}$) to the metastable state (5$s^{2}$5$p^{5}$6$s$) is 8.3 eV \cite{Feulner1987}. When Xe is condensed into a two-dimensional layer on a surface, the excitation energy might be modified, as denoted by the surface exciton. The value is, however, reported to be little modified in the monolayer adsorption regime \cite{Schonhense}, suggesting that such excitation is unlikely to occur at 6.4 eV. The first ionization energy is 12.1 eV \cite{Feulner1984, Moog}, which is also unreachable with 6.4 eV photons even though it is reduced due to the image charge effect by $\sim$2.9 eV \cite{Schonhense}. Xe desorption from Ag nanoparticles (AgNP) or Si(001), on the other hand, is also reported to occur via surface-plasmon excitation of AgNP at 2.3$-$4.0 eV photons \cite{Watanabe2007} and localized surface phonon excitation of Si(001) at 1.1$-$6.4 eV photons \cite{Watanabe2000}. Desorption via direct excitation from the bound state to a continuum state took place at a photon energy of lower than 1 eV \cite{Pearlstine, Rao}. All these desorption mechanisms can be ruled out because the nonthermal PSD of Xe was observed only at 6.4 eV and not at 2.3 eV photoirradiation.

The work function of the Au(001) surface is 5.0 eV, which is reduced by $\sim$0.5 eV with Xe adsorption. Therefore, the electronic states nearby the vacuum level are accessible with the hot electrons from the substrate band created by 6.4 eV photoexcitation, and not by 2.3 eV as schematically shown in Fig.~\ref{eledi2}. Although the electron affinity of Xe atoms in the gas phase is known to be negative \cite{Buckman}, the following studies suggest stabilization of the affinity level due to Xe condensation. Bulk Xe has a conduction band minimum (CBM) at 0.5 eV below the vacuum level \cite{Schwentner1975, Schwentner1973}. Haberland \textit{et al.} found that the ground-state Xe$_{N}$ clusters are able to bind an electron stably with $N>6$ \cite{Haberland}, of which the electron affinity is calculated to be a few meV \cite{Stampfli, Martyna}. These studies suggest that interaction with neighboring atoms lowers the electron affinity level of Xe due to the mixing between the unoccupied orbitals. Furthermore, the image charge effect on metal surfaces shifts the electron affinity level downward by $\sim$1.0 eV.

\begin{figure}
\begin{center}
\includegraphics[scale=.6, clip]{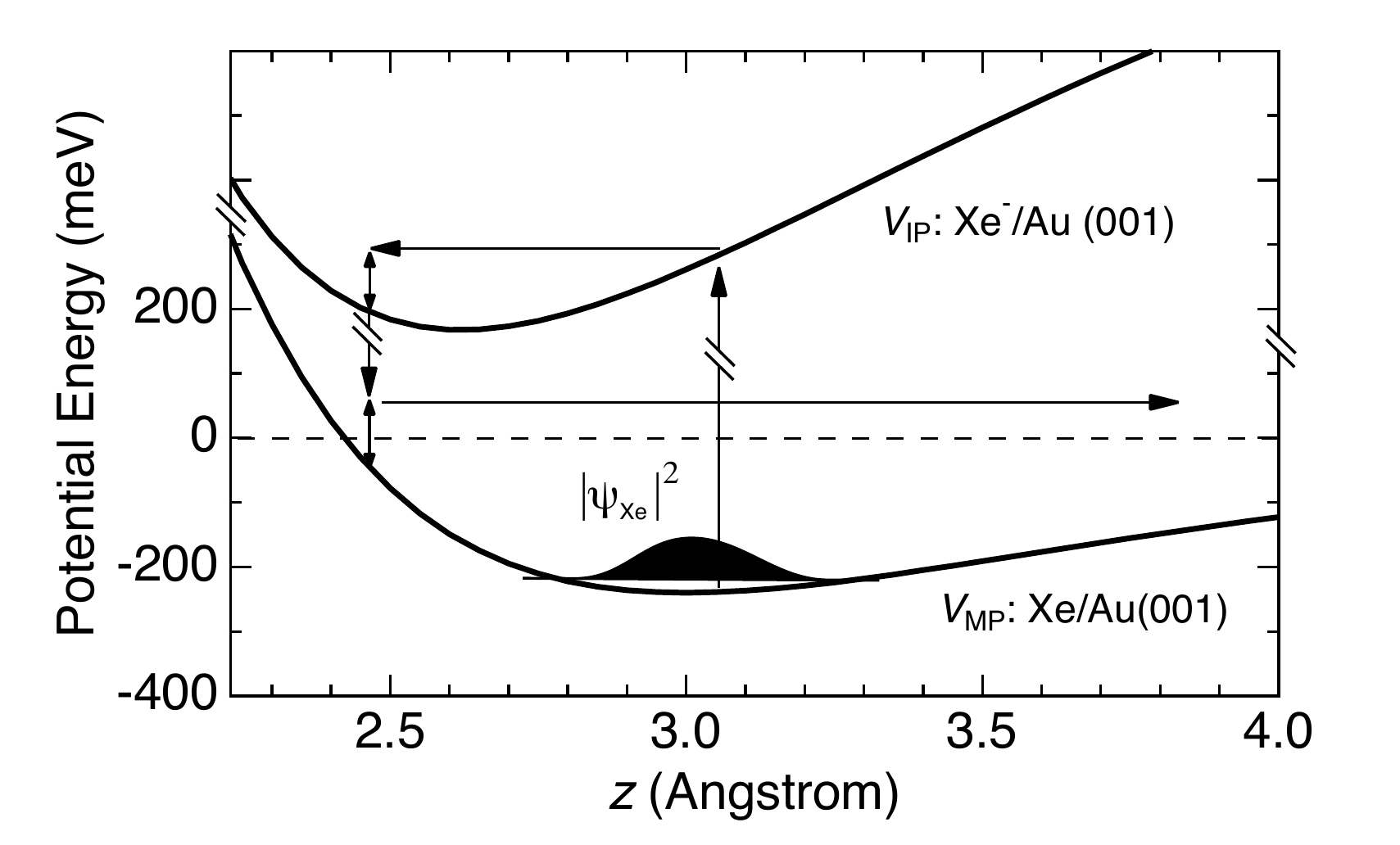} 
\caption{Adiabatic potential of Xe and Xe$^{-}$ on Au(001) and a schematics of Antoniewicz model of neutral desorption used for the model calculations.\label{ant1}}
\end{center}
\end{figure}

\subsection{Transient Xe$^{-}$ formation}

Assuming that the excitation intermediate is the negative ion state, a plausible desorption mechanism is the Antoniewicz model \cite{Antoniewicz}. In the model, for the appreciable desorption to occur, the lifetime of the Xe$^{-}$ state is required to be long enough. We tentatively estimated the lifetime ($\tau$) of Xe$^{-}$ on Au(001) that reproduces the experimentally observed values of $T_{\mathrm{D1}}$ and $\sigma_{\mathrm{PSD}}$. $\sigma_{\mathrm{PSD}}$ is a product of the photoionization cross section $\sigma_{\mathrm{PI}}$ and desorption probability $P_{\mathrm{D}}$. We assume, as a first approximation, that the $\sigma_{\mathrm{PI}}$ is as large as $\sim$$10^{-16}$ cm$^{2}$ \cite{Moog}. TOF of desorbing Xe and $P_{\mathrm{D}}$ are calculated on the basis of the Antoniewicz model and classical kinetics, as is depicted in Fig.~\ref{ant1}. Initially, Xe atoms are trapped at the bottom of the physisorption well described by a Morse potential of the form 
\begin{equation}
V_{\mathrm{MP}}(z)=D[1-\exp\{-\alpha(z-z_{0})\}]^{2}, 
\end{equation}
where $D$, $\alpha$ and $z_{0}$ represent the depth (240 meV) \cite{Mcelhiney}, the width (14 nm$^{-1}$) and the position (3.0 \AA) \cite{Silva}, respectively. The distribution of the initial position of Xe is accounted for as described in Ref.~\cite{Moog}. Upon Xe$^{-}$ formation, the adiabatic potential of the Xe atom evolves into the form 
\begin{align}
V_{\mathrm{Ion}}(z)=V_{\mathrm{MP}}&(z)+V_{\mathrm{IP}}(z)+\Delta E,\\
\intertext{where }
V_{\mathrm{IP}}(z)&=-\frac{e^{2}}{16\pi\epsilon_{0}z}
\end{align}
is the image charge potential and $\Delta E$ is an excitation energy. Due to the image charge attraction, the Xe atom is first attracted toward the surface, and is neutralized at a certain distance from the surface. The nuclear motion on $V_{\mathrm{Ion}}(z)$ is treated classically. If the Xe atom gains enough energy, it escapes the physisorption well leading to desorption. We assume that the neutralization rate of Xe$^{-}$ is described by $R(t)=\exp(-t/\tau)/\tau$ independent of $z$.

\begin{figure}
\begin{center}
\includegraphics[scale=.6, clip]{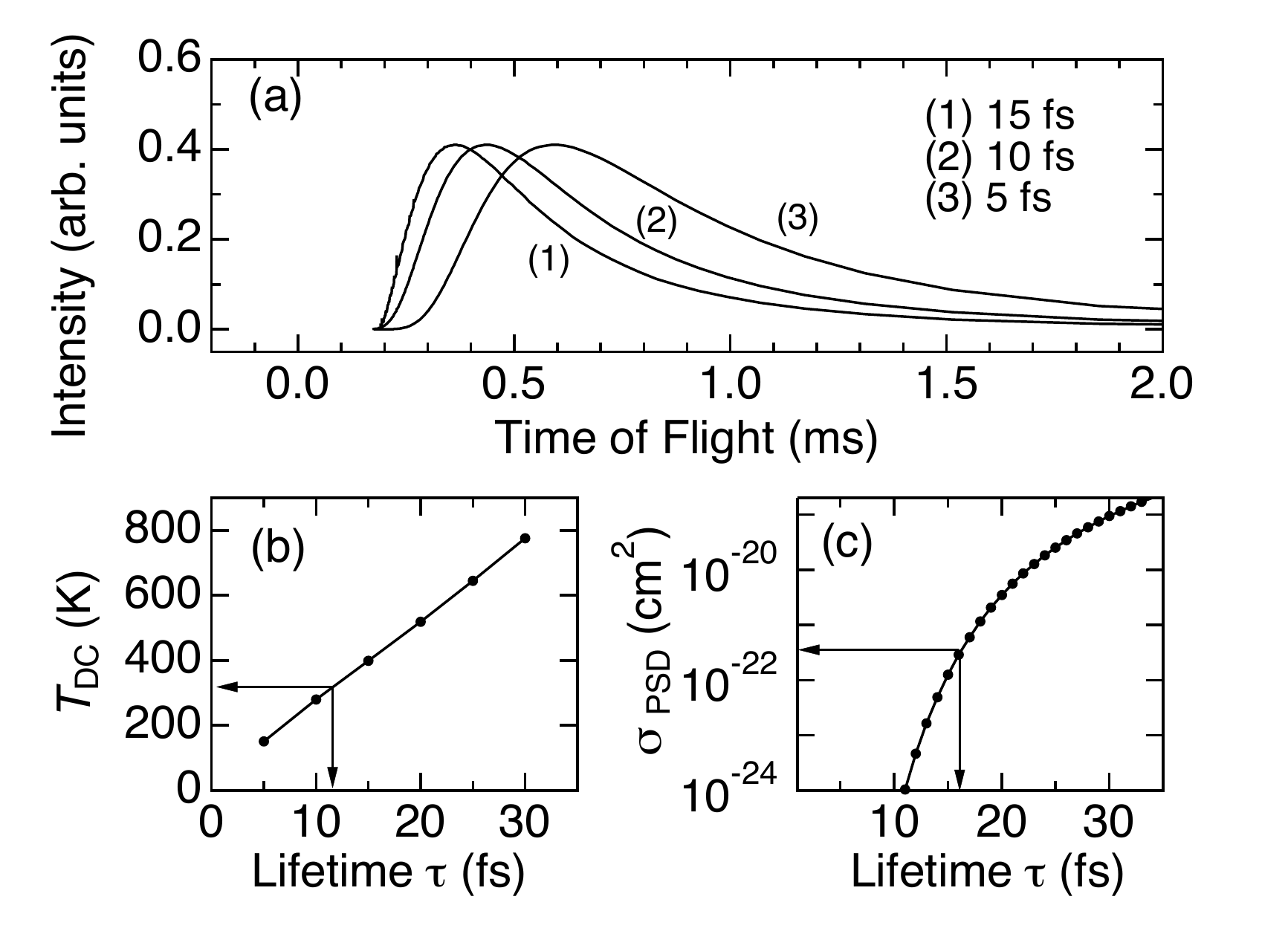} 
\caption{(a) Calculated results of time-of-flight (TOF) of desorbing Xe for several Xe$^{-}$ lifetimes. (b) Translational temperature ($T_{\mathrm{DC}}$) of the calculated TOF obtained by analyzing with a Maxwell-Boltzmann distribution as a function of Xe$^{-}$ lifetime. (c) Calculated result of the non-thermal PSD cross section as a function of Xe$^{-}$ lifetime. \label{ant2}}
\end{center}
\end{figure}

Figure~\ref{ant2} (a) shows the TOF results of Xe calculated for lifetimes of 5$-$15 fs. Each TOF is well expressed by a single MB velocity distribution with a translational temperature $T_{\mathrm{DC}}$. Figure~\ref{ant2} (b) shows the obtained $T_{\mathrm{DC}}$ of the calculated TOF as a function of $\tau$, where $\tau=$ $\sim$13 fs reproduces the experimentally observed $T_{\mathrm{D1}}$ of $300\pm20$ K. Figure~\ref{ant2} (c) shows the calculated result of $\sigma_{\mathrm{PSD}}$ as a function of $\tau$, where $\tau=17\pm5$ fs reproduces the experimentally observed $\sigma_{\mathrm{PSD}}$ of 10$^{-21}$$-$10$^{-22}$ cm$^{2}$. It is worth emphasizing that the two experimental data of $T_{\mathrm{D1}}$ and $\sigma_{\mathrm{PSD}}$ are well reproduced by a common $\tau$ value of $\sim$15 fs based on the Antoniewicz model. The fact is indicative of the validity of the present model. Walkup \textit{et al.} have shown that the classical adiabatic potential concerning the image charge potential is essentially correct, although it is slightly different from the one obtained by a quantum-mechanical treatment. They have furthermore shown that the classical treatment of nuclear motion is valid as long as distribution of the initial Xe position is accounted for and that it qualitatively reproduces the kinetic-energy distribution \cite{Walkup}. It is noted that the affinity level of Xe should lie below the vacuum level for the $\tau$ to be as long as 15 fs.

The obtained value of $\tau$ $\sim$15 fs corresponds to the linewidth of 70$-$120 meV for the Xe$^{-}$ state. Padowitz \textit{et al.} found that in using the two-photon photoemission spectroscopy, the image charge state on clean Ag(111) is shifted by Xe adsorption due to the coupling with the Xe orbitals \cite{Padowitz, Merry}. The linewidth obtained in the present study is similar to the  value of 25$-$50 meV observed for the image charge states ($n=1, 2, 3$) on Xe/Ag(111) at 0.6$-$0.1 eV below the vacuum level. Hence, we suggest the image charge state of Xe/Au(001) is resonanced with the affinity level of Xe, which causes the PSD. We note that a smaller estimation of $\sigma_{\mathrm{PI}}$ $\sim$10$^{-18}$ cm$^{2}$ in the model calculation results in a linewidth of $\sim$35 meV.

As already mentioned above, two possible mechanisms of Xe$^{-}$ stabilization are hybridization of unoccupied orbitals and the image charge effect. Since unoccupied orbitals have an extended feature compared with occupied orbitals, unoccupied states could be appreciably hybridized with substrate states even in a weakly bound physisorption well. In addition to these two factors, we discuss another possible reason for the Xe$^{-}$ formation on a metal surface. In the gas phase, contrary to the ground state Xe (5$s^{2}$5$p^{6}$), metastable Xe$^{*}$ (5$s^{2}$5$p^{5}$6$s$) binds an electron to form a transient Xe$^{-}$ (5$s^{2}$5$p^{5}$6$s^{2}$) with a large cross section ($\sim$10$^{-16}$ cm$^{2}$) \cite{Blagoev}. In the gas phase, the 5$s^{2}$5$p^{5}$6$s$ state is located at 8.3 eV above the ground state. A recent density functional study \cite{Silva} has shown that the Xe adsorption on a metal surface results in a partial depletion of the occupied Xe 5$p_{\mathrm{z}}$ state and a partial occupation of the previously unoccupied Xe 6$s$ and 5$d$ states. This indicates mixing of the 5$s^{2}$5$p^{5}$6$s$ state upon adsorption on a metal surface, which may contribute to the stabilization of the Xe$^{-}$ state. 

Lastly, we comment on the result of an earlier study on the nonthermal PSD of Xe from Ru(001) surfaces \cite{Feulner1987}. In the study, no significant desorption was observed from Xe mono and multilayers following 7$-$30 eV photoirradiations, whereas desorption from Ar mono- and multilayers and Kr multilayers were observed. As discussed in the present paper, Xe$^{-}$ is expected to be formed following the photoirradiations of $h\nu>$ 6 eV, and subsequently Xe desorption is expected to occur. Although the desorption cross section is not mentioned in Ref.~\cite{Feulner1987}, we suspect $\sigma_{\mathrm{PSD}}$ of $\sim$10$^{-22}$ cm$^{2}$ was too small for the signal to be detected in their experimental condition. Arakawa \textit{et al.} reported that the absolute yield of the PSD from solid Ar following 12$-$50 eV photons is as large as $\sim$0.1 atoms/photon \cite{Arakawa2}, indicating that the cross section of the Xe PSD via the Xe$^{-}$ formation observed in the present study is several orders of magnitude smaller than those of Ar via exciton excitation.

\clearpage

\section{Post-desorption collision effects}

In the following subsections, the desorption dynamics involving the post-desorption collision effects is discussed. It is concluded that the experimental observation is due to the post-desorption collision effects. Furthermore, the experimental observation is compared with the prediction based on the Knudsen layer formation theory. The consistency between the experimental observation and the theoretical prediction strongly indicates the formation of Knudsen layer in the laser induced thermal desorption.

\subsection{Thermal desorption of Xe}

We first discuss that the observed desorption proceeds only via thermal activation. In the previous study,\cite{Ikeda} we investigated the same system by varying the photon fluence at $\Theta=1$ ML. At the small fluence region where the laser power is below 20 mJ/cm$^{2}$, we observed a desorption of Xe only at a photon energy of 6.4 eV and no desorption at 2.3 eV. We assigned this photodesorption as a non-thermal desorption via a transient Xe$^{-}$ formation. At a large fluence region where laser power exceeds 60 mJ/cm$^{2}$ we observed, on the other hand, that thermal desorption is the dominant desorption process regardless of the photon energy. These observations are confirmed by comparing the experimentally observed $T_{\mathrm{D}}$ and Xe desorption yield with the results of LITD simulations. Therefore, we regard that the initial desorption is only thermally activated in the present experiment.

\subsection{Analysis of TOF spectra}

In order to analyze the TOF spectra at $\Theta>0.3$ ML, we introduce a shifted Maxwell-Boltzmann velocity distribution which is characterized by $u$. Generally, the velocity distribution in a desorption flux shows an angular distribution and is expressed by the elliptical distribution which involves an angular dependent translational temperature as described in Ref. \cite{SiboldJAP1993}. In the present experiment, the detector was fixed in the surface normal direction. Therefore, the translational temperature only in this direction is detected in the present experiment. In this case, the shifted Maxwell-Boltzmann velocity distribution in a flux from a thermal source is expressed as,\cite{KellySS1988, Sibold1991}
\begin{equation}
J(v)dv=Av^{3}\exp\left\{-\frac{m(v-u)^{2}}{2kT_{\mathrm{D}}}\right\}dv.\label{smb}
\end{equation}
Under the condition of $u=0$, Eq. (\ref{smb}) is identical to Eq. (\ref{mb}). In the analysis, Eq. (\ref{smb}) was also converted to a TOF function for the density sensitive detector. 

The TOF spectra at $\Theta > 0.3$ ML were well analyzed using Eq. (\ref{smb}). However, we found it difficult to obtain a unique set of $T_{\mathrm{D}}$ and $u$. The spectra were well reproduced with an arbitrary value of $T_{\mathrm{D}}$ from 165 to 310 K by adjusting $u$ from 125 to 0 m/s. We note that, with a fixed value of $T_{\mathrm{D}}$ between 165 and 310 K, the obtained value of $u$ monotonically increases with increasing $\Theta$ from 0.3 to 4 ML and becomes constant at $\Theta > 4$ ML. For example, Fig. \ref{tud} shows u as a function of $\Theta$ obtained by using Eq. (\ref{smb}) with $T_{\mathrm{D}}$ fixed at 165 K, of which the fitting curves are shown as solid curves in Fig. \ref{tof}.

\subsection{Moderate desorption at $\Theta\simeq0$ ML}

At $\Theta\simeq0$, the observed feature of the TOF can be rationalized with a model employing thermal desorption followed by collision-free flow. At the instance of thermal desorption, the desorbed Xe gas is in thermal equilibrium with the surface at the temperature $T_{\mathrm{S}}$. Here, it is reasonable to consider the half-range Maxwell-Boltzmann velocity distribution for desorbed Xe expressed as \cite{KellySS1988}
\begin{align}
f(\mathbf{v})d\mathbf{v}=A\exp\left(-\frac{m\mathbf{v}^{2}}{2kT_{\mathrm{S}}}\right)d\mathbf{v},&\notag\\
v_{z}>0& \label{mbv}
\end{align}
where $A$ and $\mathbf{v}$ are a normalization factor and the velocity vector of Xe atoms in the Cartesian coordinate, respectively. We note that in Eq. (\ref{mbv}) the velocity component in the $z$ direction has a distribution only at $v_{z}>0$,\cite{KellySS1988} assuming the $z$ axis to be the surface normal direction. Given the desorption period is much shorter than the flight time and the irradiation diameter is much smaller than the flight distance, Eq. (\ref{mbv}) can be, as a velocity distribution in a flux from a thermal source, transformed to the same form as Eq. (\ref{mb}). Thus, we notice that $T_{\mathrm{D}}=T_{\mathrm{S}}$ at $\Theta\simeq0$.\cite{Wedler, Cowin1978}

For a qualitative analysis, we estimated the surface temperature during the present LITD experiment using the first order desorption rate equation, assuming that the desorption activation energy is 240 meV\cite{Mcelhiney} and that the desorption period is 4 ns.\cite{Wedler, Brand} The calculated results show that the required surface temperature is 260 K, which is in good agreement with the obtained value of $T_{\mathrm{S}}=255$ K at $\Theta=0.3$ ML. The agreement indicates that the model of the thermal desorption followed by the collision-free flow well describes the observed TOF at $\Theta\simeq0$, in good accordance with the results by Cowin \cite{Cowin1978} and Wedler \cite{Wedler}.

\subsection{Intensive desorption at $\Theta>4$ ML}

At $\Theta>4$ ML, we discuss that the desorption flow is described as an intense flow, where the post-desorption collision modifies the velocity distribution.\cite{NoorbatchaPRB1987, NoorbatchaJCP1987} Several theoretical studies have speculated that the so-called Knudsen layer is formed in the vicinity of the surface as a result of collisions.\cite{KellySS1988, KellyNIMB1988, KellyPRA1992, Sibold1991} As schematically drawn in Fig. \ref{KL}, in the Knudsen layer model, the initial velocity distribution is in thermal equilibrium with the surface at $T_{\mathrm{S}}$ except that there is no distribution at $v_{z}<0$ as in Eq. (\ref{mbv}). As a result of a significant number of post-desorption collisions, the half-range velocity distribution at $z=0$ becomes thermally equilibrated at some distance from the surface to a full-range Maxwell-Boltzmann velocity distribution with a stream velocity $u$.\cite{Ytrehus, Cercignani, KellySS1988} The model also requires that, at the end of the Knudsen layer, the temperature of the flow becomes identical ($T_{\mathrm{K}}$) in all degrees of freedom. Hence, at the end of the Knudsen layer, the velocity distribution may be described as \cite{Ytrehus, Cercignani, KellySS1988}
\begin{equation}
f(\mathbf{v}) d\mathbf{v}=A\exp\left[-\frac{m\{v_{x}^{2}+v_{y}^{2}+(v_{z}-u_{\mathrm{K}})^{2}\}}{2kT_{\mathrm{K}}}\right] d\mathbf{v},\label{smbv}
\end{equation}
where $v_{i}$, $u_{\mathrm{K}}$ and $T_{\mathrm{K}}$ are the velocity component in the Cartesian coordinate, stream velocity and the temperature at the end of the Knudsen layer, respectively. Under the condition that the flight distance is sufficiently longer than the thickness of the Knudsen layer besides the conditions described above, Eq. (\ref{smbv}) is converted, as a velocity distribution from a thermal source, to the identical form to Eq. (\ref{smb}). Thus, we see that $T_{\mathrm{D}}=T_{\mathrm{K}}$ and $u=u_{\mathrm{K}}$ at $\Theta>4$ ML. Equation (\ref{smb}) was shown to well describe the experimental results of the TOFs at $\Theta>4$ ML. This fact, along with the convergence feature of $u$, indicates that the experimental result at $\Theta>4$ ML may be rationalized with the Knudsen layer formation model.

\subsection{Mach number of the flow}

We further examine the results at $\Theta>4$ ML by quantitatively comparing the obtained results of $u_{\mathrm{K}}$ with previous theoretical reports on Knudsen layer formation in the steady strong evaporation \cite{Ytrehus, Cercignani} and the pulsed desorption \cite{Sibold1991} from plane surfaces. We compare $u_{\mathrm{K}}$ in terms of the Mach number $M$ at the end of the Knudsen layer. $M$ is defined as
\begin{equation}
M=\frac{u_{\mathrm{K}}}{c}=u_{\mathrm{K}}\sqrt{\frac{m}{\gamma kT_{\mathrm{K}}}},\label{mach}
\end{equation}
where $c$ and $\gamma$ are the local velocity of sound and the heat capacity ratio of the gas, respectively, the latter of which is 5/3 for monoatomic gas as in the present case.

\begin{table}
\caption{\label{tablem}Mach number $M$ and Translational temperature $T_{\mathrm{D}}$ at the end of Knudsen layer obtained in the present study and previously reported values by simulation and theory.}
\begin{center}
\begin{tabular}{lccc} \toprule
Method&$M$&Condition\\ \midrule
Experiment (Present study)& 0.96 & Pulsed flow\\
Simulation \cite{Sibold1991} & 1.0 & Pulsed flow\\
Simulation \cite{SiboldPFFD1993} & 1.0 & Steady flow\\
Theory \cite{Ytrehus, Cercignani, KellySS1988} & 0.99 & Steady flow\\ \bottomrule
\end{tabular}
\end{center}
\end{table}

Ytrehus \cite{Ytrehus} and Cercignani \cite{Cercignani} formulated $M$ and $T_{\mathrm{K}}/T_{\mathrm{S}}$ by finding a solution to the Boltzmann equation under the Knudsen layer formation model assuming the conservation of the particle number, momentum and the energy flux between the surface and the end of the Knudsen layer. Sibold and Urbassek estimated $M$ and $T_{\mathrm{K}}/T_{\mathrm{S}}$ using the Monte Carlo simulation of the Boltzmann equation in one-dimension for both pulsed flow \cite{Sibold1991} and steady flow \cite{SiboldPFFD1993} conditions. Those reports have stated that at the end of the Knudsen layer, $M$ should always be close to unity and that $T_{\mathrm{K}}/T_{\mathrm{S}}$ is at around 0.65.

As discussed in IV. B, it was difficult to unambiguously determine 
the value of $T_{\mathrm{D}}$. Here, we assume the relation of $T_{\mathrm{K}}/T_{\mathrm{S}}=0.65$ following the theoretical studies and evaluate the Mach number. With the value of $T_{\mathrm{S}} = 255 $ K, we obtain the value of $T_{\mathrm{K}}$ to be 165 K. With $T_{\mathrm{K}} = 165 $ K and Eq. (\ref{mach}), we obtain the value of $M$ to be 0.96. As can be seen in Table I, the previously reported values of $M$ show a quantitative agreement with the obtained value of $M$ in the present study. This indicates that the observed trend of $u$ at $\Theta > 4$ ML is consistent with the Knudsen layer formation theory.\cite{KellySS1988}

With adiabatic expansion, $M$ should well exceed unity.\cite{Kelly1990, KellyPRA1992, KellyNIMB1992} We note that we observed the saturation of $M$ at around unity at $\Theta>$ 4 ML in the present experimental conditions. This indicates that the adiabatic expansion is practically absent, although the slight increase of $u$ at $\Theta>$ 4 ML may be due to the adiabatic expansion.

\subsection{Knudsen number of the flow}
In order to generalize the present result, we tentatively consider mean gas density $\bar{n}$, mean free path $\lambda$ and Knudsen number Kn in the vicinity of the surface at the moment of desorption as a function of $\Theta$. Kn is defined by \cite{Sibold1991}
\begin{equation}
\mathrm{Kn}=\frac{\lambda}{l_{z}}=\frac{1}{\sqrt{2}\bar{n}\sigma l_{z}},\label{KN}
\end{equation}
where $l_{z}$ and $\sigma$ are the representative length of the system and the Van der Waals collision cross section of Xe ($1.0\times10^{-19}$ m$^{2}$), respectively.

Here, we simply regard that $l_{z}$ is the mean thickness of the gas cloud above the surface. It, then, reads that $l_{z}=\bar{v}_{z}\tau$, where $\bar{v}_{z}$ is the mean thermal velocity of desorbed Xe in the $z$ direction and $\tau$ is the mean desorption period. We further take $\bar{v}_{z}=\sqrt{2kT_{\mathrm{S}}/\pi m}$ and estimate that $\bar{n}=\Theta/l_{z}$. By substituting $\bar{n}$ in Eq. (\ref{KN}), we obtain a simple relation $\mathrm{Kn}=a/(\sqrt{2}\tilde{\Theta})$ as introduced in Ref. \cite{Sibold1991}. $\tilde{\Theta}$ and $a$ are the relative coverage $\tilde{\Theta}=\Theta/\Theta_{\mathrm{S}}$ with the monolayer saturation coverage $\Theta_{\mathrm{S}}$ and the ratio of the area occupied by an atom at $\Theta_{\mathrm{S}}$ to $\sigma$ described as $a=1/(\sigma\Theta_{\mathrm{S}})$, respectively.

Now, we see that Kn depends simply on $\tilde{\Theta}$ and $a$. Therefore, we can discuss the required condition for the formation of Knudsen layer in LITD for general systems in terms of Kn. We fixed $\tau=4$ ns. On the basis of the result employing the LITD simulation, $\tau$ showed little dependence on $\Theta$ in the region concerned in the present study.

In Table \ref{tablek}, we summarized the obtained $\bar{n}$, $\lambda$ and Kn as a function of $\Theta$. Kn can also be understood as an inverse of mean collision times per each atoms. In the present study, we observed the formation of the Knudsen layer at $\Theta>4$ ML, which corresponds to Kn $<$ 0.39 and to more than $\sim2.6$ collisions per desorbing atom. The manifestation of collision effects is observed at $\Theta>0.3$ ML, which corresponds to Kn $>5.2$. The results coincide with the previous theoretical work by Sibold and Urbassek \cite{Sibold1991} that the Knudsen layer is formed at $\Theta=2.5$ ML and that the collision effect already appears while the Knudsen layer is not formed at $\Theta=0.25$ ML, respectively. The obtained values of Kn also agree well with the previous experimental observation of the collision effect in LITD of D$_{2}$/W at $\Theta=1$ ML reported by Cowin \textit{et al.} \cite{Cowin1978}. They also coincide with the theoretically estimated collision number of 2.9 per atom at $\Theta=1$ ML reported by Noorbatcha \textit{et al.}.\cite{NoorbatchaJCP1987, NoorbatchaPRB1987}

\begin{table}
\caption{\label{tablek}Density $\bar{n}$, mean free path $\lambda$ and Knudsen number Kn in the vicinity of the surface at the moment of laser desorption as a function of $\Theta$.}
\begin{center}
\begin{tabular}{cccc}\toprule
$\Theta$ (ML)&$\bar{n}$ (10$^{24}$ atoms/m$^{3}$)&$\lambda$ (nm)&Kn\\ \midrule
0.3&3.3&2114&5.2\\
4.0&45&159&0.39\\
10&111&63&0.16\\ \bottomrule
\end{tabular}
\end{center}
\end{table}

Lastly, we note the observed feature at $\Theta\simeq1$ ML, where the Knudsen layer is not formed. Thus, the velocity distributions are not necessarily well fitted with Eq. (\ref{smb}). Nevertheless, they were very well analyzed with Eq. (\ref{smb}) in all $\Theta$ at $\Theta\simeq1$ ML. The result is in good agreement with the simulated results by Sibold and Urbassek,\cite{Sibold1991} although any analytical formulation for the flow at $\Theta\simeq1$ ML has not been presented so far. Hence, further theoretical and experimental works are required for elucidating the mechanism for the modification of velocity distribution at $\Theta\simeq1$ ML due to the post-desorption collision as pointed out in Ref. \cite{Sibold1991}.

\clearpage

\section{Quadrupole splitting of $^{83}$Kr on TiO$_{2}$(110)}

In this section, the observation of hyperfine structure of $^{83}$Kr at the surface of the rutile TiO$_{2}$(110) shown in Fig. \ref{ts} (b) is discussed. The spectrum is well fitted to a single exponential decay with a time constant $\tau=109$ ns. Nontheless, we also notice that a systematic deviation of the experimental data from the fitting line in Fig. \ref{ts} (b). Hence, we tentatively discuss the possibility of observation of hyperfine structure by comparing the experimental results and the simulated results. We discuss all the possibility that can be thought of at this moment. The discussion concentrates on the following topics (1) the simulation of the quantum beat structure, (2) the comparison between the experimental and simulation results, (3) Several stories that can potentially explain the present result in terms of the symmetry of the electric field gradient at the surface and the variation of the adsorption sites on the sample surface.

The experimental observation of Fig. \ref{ts} (a) and (b) is a great step forward as an experimental technique. Previously, Baron and coworkers observed the time spectra of NRS by gaseous Kr, which do not involve the M\"ossbauer effect \cite{Baron}. Johnson and coworkers succeeded in observing the time spectra of bulk and monolayer Kr condensed on the exfoliated graphite which possess a large surface area \cite{Johnson}.  Both experiments are carried out in the transmission regime with a large volume of the sample nuclei. The present study is an experimental step forward in the sense that they are carried out in a reflection regime and that the sample is well-defined multi- and monolayer films on an atomically flat single crystal surface. Application of NRS on a well-defined sample is plausible, for a simple hyperfine structure may be observed.

\subsection{Quantum beat}

The quantum beat structure may well be observed based on the following reasons. The EFG at the surface may well be large enough, for at the surface the translational symmetry of the lattice is broken. Although the surface orientation of the sample used in the present study is (110) which is not axially symmetry with respect to the surface normal vector, the position of Kr atom adsorbed on the surface is assumed to be at $\sim3$ \AA\ away from the surface atoms as estimated in recent studies based on first principle calculations \cite{Gomes2005, Gomes2010}. Actually it has been reported that the potential corrugation in the surface parallel direction is as small as $1/10$ to $1/100$ on metal and semiconductor surfaces \cite{Meixner1993, Meixner1994, Ellis, Nabighian, Thomas}. Therefore, as compared to $V_{ZZ}$, both $V_{XX}$ and $V_{YY}$ should be small and close to each other. We discuss that it is a good approximation to take $\eta=0$. In the end, in the present experimental setup, the direction of magnetic vector of the synchrotron radiation is parallel to the surface normal direction. Hence, only $\Delta m=0$ transitions are expected to be allowed in the present experimental setup.

\begin{figure}
\begin{center}
\includegraphics[scale=.6, clip]{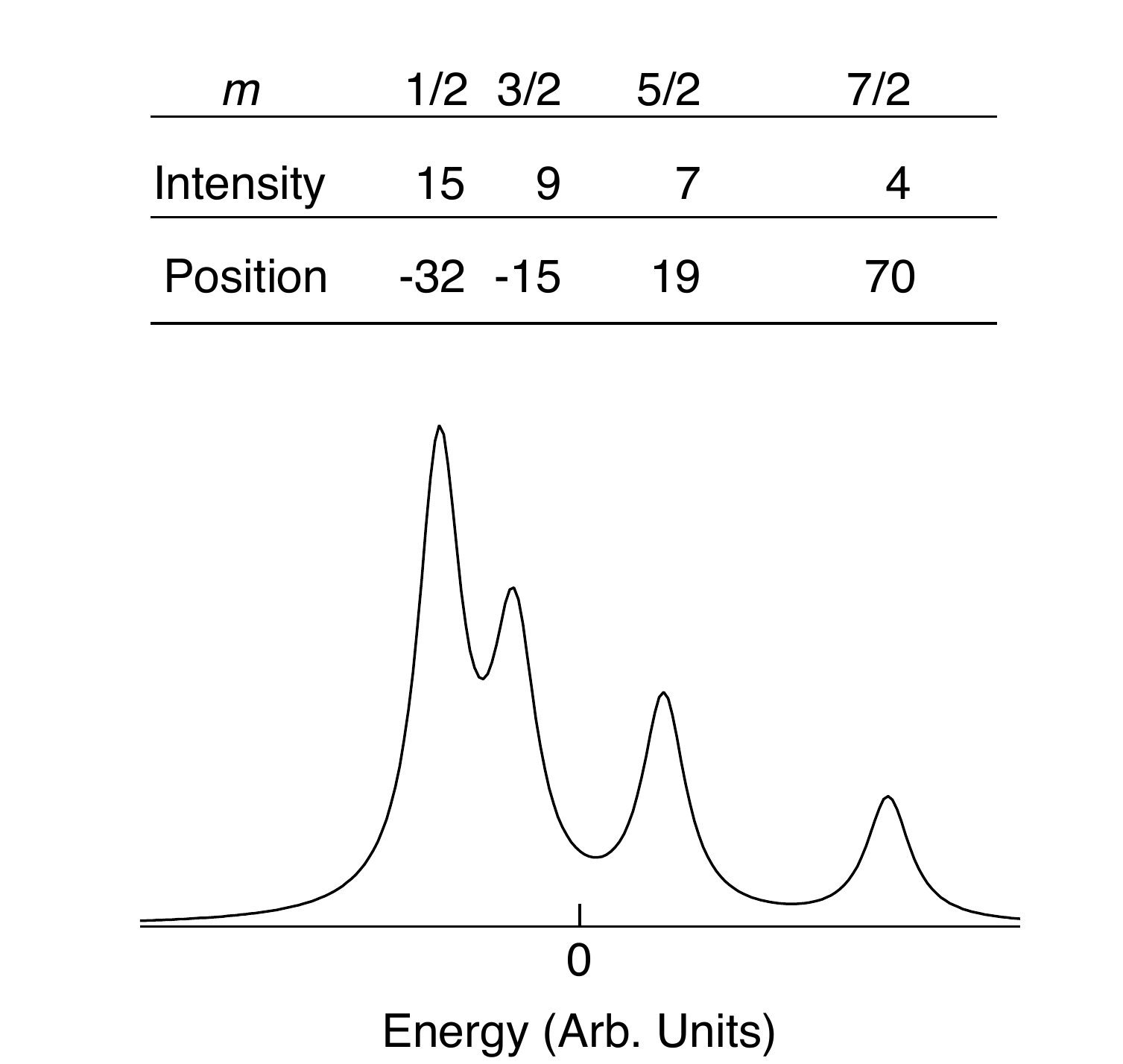}
\caption{The simulated M\"ossbauer absorption spectrum of $^{83}$Kr with excitation from $I=9/2$ to $I=7/2$ only with $\Delta m=0$ transitions. The four peaks are present. The relative position of the respective peaks are obtained from Eq. (\ref{QS2}). The relative intensity of the respective peaks are square of the appropriate Clebsch-Gordan coefficient \cite{Ruby1966}. The ratio of the quadrupole splitting and the width of each line is displayed arbitrary. \label{QS3}}
\end{center}
\end{figure}

Using Eq. (\ref{QS2}), we simulate the M\"ossbauer absorption spectrum as shown in Fig. \ref{QS3}. Only four lines of eleven lines are present because only $\Delta m =0$ transitions are allowed as indicated by Fig. \ref{POL}. The selection rule applies only when the $V_{ZZ}$ is a principal axis of the electric field gradient on the surface and the axial symmetry is satisfied (i.e. $V_{XX}=V_{YY}$). If not, the eleven transitions should appear. For the simplicity of the discussion, we first think of the four transitions. The schematic picture of the transitions are shown in Fig. \ref{eleven} and Fig. \ref{POL}. The relative positions are based on the calculated results using Eq. (\ref{QS2}) and the assumption that the ratio of quadrupole moment of the ground and the excited state $R=Q(7/2)/Q(9/2)$ to be 2 as reported by previous studies \cite{Kolk1, Holloway}. The relative intensity of each line is the square of the proper Clebsch-Gordan coefficient which is taken from \cite{Ruby1963}. The ratio of the width of the resonant line and the energy splitting is determined arbitrary.

\begin{figure}
\begin{center}
\includegraphics[scale=.65, clip]{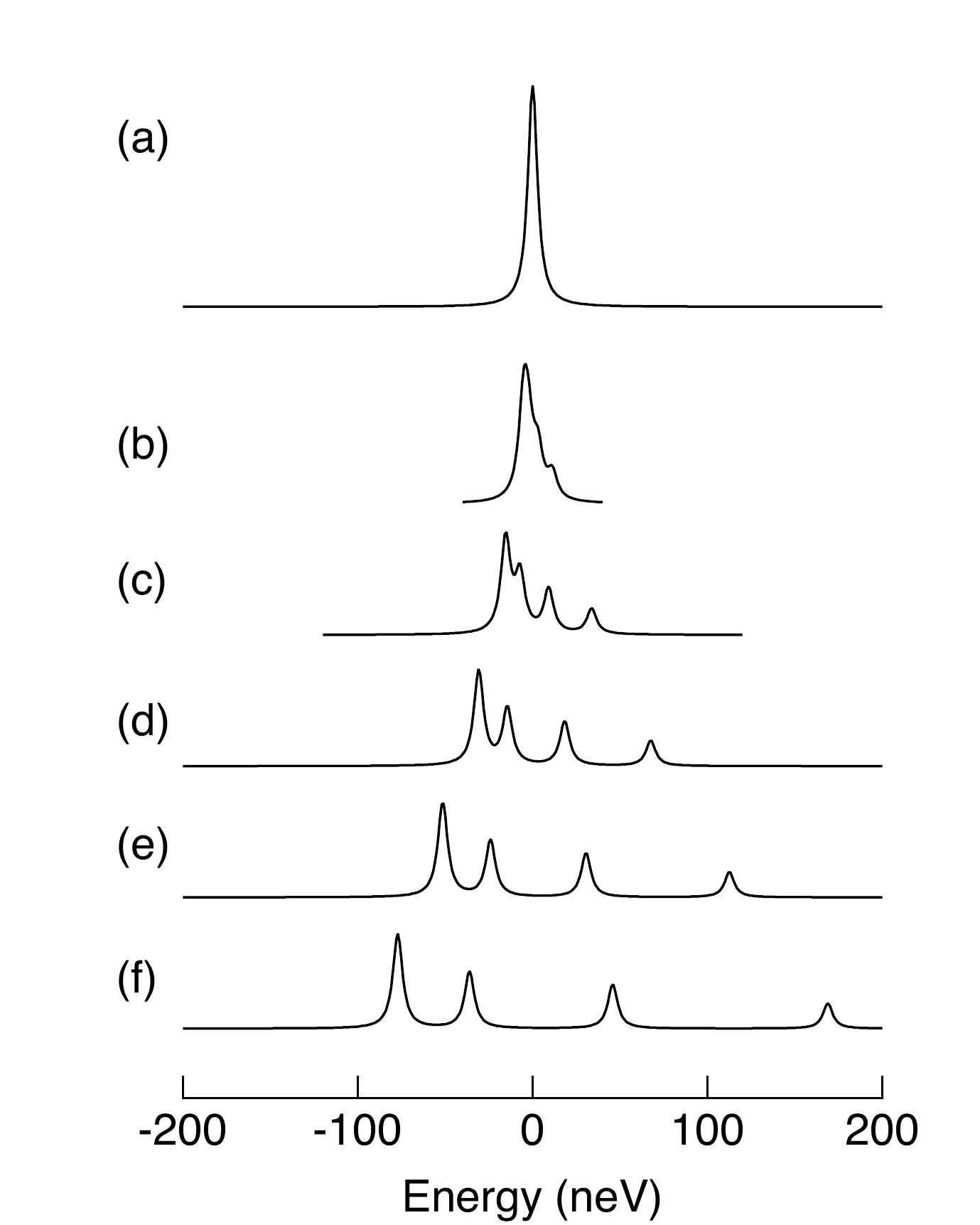}
\caption{The simulated M\"ossbauer absorption spectrum of $^{83}$Kr with only $\Delta m=0$ transitions as a function of the applied electric field gradient $V_{ZZ}$. $V_{ZZ}=$ (a) 0 V/m$^{2}$, (b) $1.0\times10^{21}$ V/m$^{2}$, (c) $3.0\times10^{21}$ V/m$^{2}$,  (d) $6.0\times10^{21}$ V/m$^{2}$, (e) $1.0\times10^{22}$ V/m$^{2}$, (f) $1.5\times10^{22}$ V/m$^{2}$. The line width of each lines are broadened to 6.6 neV compared to the natural line width of 3.3 neV, in order to simulate the dynamical effect that is observed experimentally. \label{QS4}}
\end{center}
\end{figure}

\begin{figure}
\begin{center}
\includegraphics[scale=.65, clip]{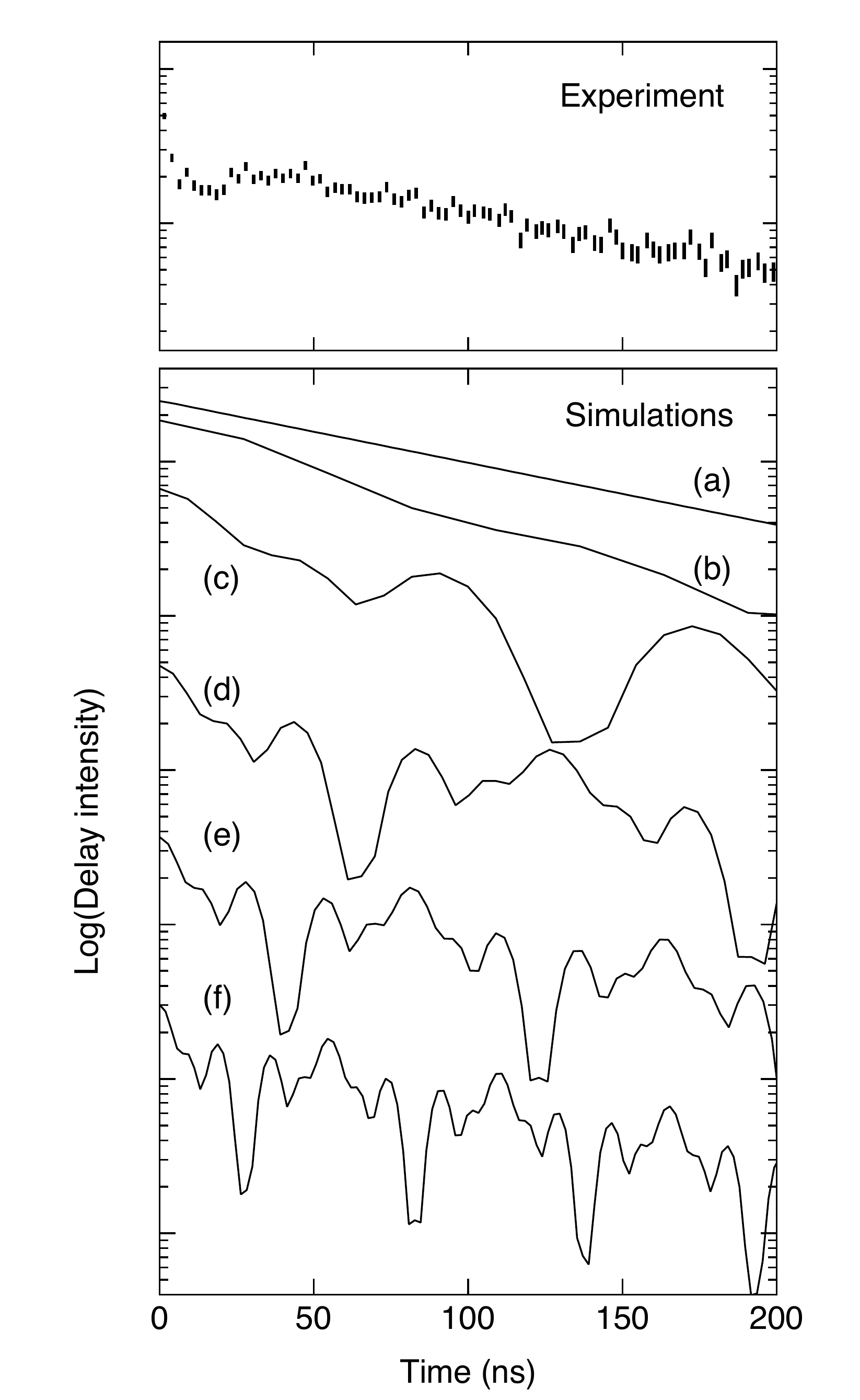}
\caption{Experimentally obtained (top) and the simulated time spectra of NRS by $^{83}$Kr with only $\Delta m=0$ transitions as a function of the applied electric field gradient $V_{ZZ}$. $V_{ZZ}=$ (a) 0 V/m$^{2}$, (b) $1.0\times10^{21}$ V/m$^{2}$, (c) $3.0\times10^{21}$ V/m$^{2}$,  (d) $6.0\times10^{21}$ V/m$^{2}$, (e) $1.0\times10^{22}$ V/m$^{2}$, (f) $1.5\times10^{22}$ V/m$^{2}$. \label{QS5}}
\end{center}
\end{figure}

In order to compare the experimental result with the theoretical predictions, we further simulated M\"ossbauer absorption spectra and the time spectra of NRS as a function of the electric field gradient in the surface normal direction $V_{ZZ}$ using Eq. (\ref{QS2}). The natural line width of the $^{83}$Kr is reported to be 3.3 neV \cite{Ruby1963} which corresponds to the natural life time of 212 ns. We experimentally obtained the value of $\tau=109$ ns for the monolayer Kr. In the calculation, we assume that the line width is broadened to be 6.6 neV due to the effect of the multiple scattering. The energy shift $\Delta E$ due to the quadrupole splitting can be estimated using Eq. (\ref{QS2}) as 
\begin{align}
\Delta E&=E(I^{*}, m^{*})-E(I, m)\notag\\
&=\frac{e^{2}qQ(I^{*})}{4I^{*}(2I^{*}-1)}\left\{3m^{*2}-I^{*}(I^{*}+1)\right\}-\frac{e^{2}qQ(I)}{4I(2I-1)}\left\{3m^{2}-I(I+1)\right\},
\label{QS6}
\end{align}
where $I$, $I^{*}$, $m$, $m^{*}$ are nuclear spin and magnetic quantum numbers, which were taken to be $I=9/2$, $I^{*}=7/2$ and where $eq$ and $Q(I)$ are electric field gradient and quadrupole moment. We take $eq$ to be identical to $V_{ZZ}$ in the following discussion. The dimension of $eq$ is V/m$^{2}$. $Q(I)$ is a quadrupole moment which has a dimension of area. We used a reported value of $Q(\frac{9}{2})$ to be 0.27 barn (where 1 barn = 10$^{-24}$ cm$^{2}$ = 10$^{-28}$ m$^{2}$) \cite{Faust}. As the value of the ratio $R=Q(\frac{7}{2})/Q(\frac{9}{2})$, we used $R=2$ as reported in Ref. \cite{Kolk1, Holloway}. We varied the value of $eq=V_{ZZ}$ from 0 V/m$^{2}$ to $1.5\times10^{22}$ V/m$^{2}$. In order to simulate the time spectra of NRS from Kr$^{83}$ under EFG, we further obtained the discrete Fourier transform of the simulated energy spectra. The results of the simulations of energy spectra and time spectra of NRS along with a comparison with the experimental result are shown in Fig. \ref{QS4} and Fig. \ref{QS5}, respectively.

Figure \ref{QS4} shows the calculated results of the M\"ossbauer absorption spectra. As we can see from Fig. \ref{QS4}, the larger the EFG, the lager the splitting between each line. As indicated in Eq. (\ref{QS2}) and Eq. (\ref{QS6}), the quadrupole splitting is a function of $m^{2}$ rather than $m$. Thus, the evolution of the splitting are not symmetric with regard to the center of energy. Each line in Fig. \ref{QS4} is assigned as $(\pm1/2\to\pm1/2)$, $(\pm3/2\to\pm3/2)$, $(\pm5/2\to\pm5/2)$ and $(\pm7/2\to\pm7/2)$ from the left to right, respectively, where $(m\to m^{*})$.

We turn next to the calculated results of the time spectra of NRS from $^{83}$Kr which are shown in Fig. \ref{QS5}. On top of Fig. \ref{QS5}, the experimental result of the time spectrum of NRS by monolayer $^{83}$Kr on TiO$_{2}$(110) are shown. From Fig. \ref{QS5} (a) to (f), the calculated result of the time spectra under the conditions that $eq=0$ to $1.5\times10^{22}$ V/m$^{2}$. The experimental data at $t=0$ to 25 ns are distorted due to the recovery of the detector. We look at the data in the region of $t>25$ ns.

At a glance of Fig. \ref{QS5}, the experimental result is in most good agreement with the calculated result of Fig. \ref{QS5} (a) and (b) which are the results under the condition of $eq=0.0$ to $1.0\times10^{21}$ V/m$^{2}$. The calculated results of Fig. \ref{QS5} (c) to (f) show features of sharp dips and strong oscillations originating from the interference between the four split lines in Fig. \ref{QS4}. Although the experimental data in Fig. \ref{ts} (b) shows a systematic deviation from the single exponential line, the deviation is not explained by the four lines due to the quadruple splitting of the nuclei of $^{83}$Kr shown in Fig. \ref{QS4}. These comparison indicate that the experimental result does not suggest a picture that all Kr atoms sit in an identical adsorption site and feeling the identical EFG and at the same time that the principal axis is in the direction of surface normal with an axial symmetry.


To conclude this subsection, we observed that the experimentally obtained time spectrum in Fig. \ref{ts} (b) is not in good agreement with the calculated results employing four lines which are split due to the quadrupole splitting of the nuclei of $^{83}$Kr under the large electric field gradient at the surface shown in Fig. \ref{QS4}.    Nevertheless, we note that the current comparison does not necessarily indicate that the EFG is too small to be observed. It is true that it is a one of the greatest possibility where the EFG was too small to be observed. There are other remaining possibilities that are not rejected at this moment.

The main topics is whether there is a large EFG which is applied on physisorbed atoms at the solid surfaces or not. As can be estimated from the Fig. \ref{QS4}, the effect of EFG can be observed if $V_{ZZ}\ge \sim2.0\times10^{21}$ V/m$^{2}$. Beside the speculation that there is a large EFG at the surface applied on the physisorbed atoms, in the current discussion, we put two more assumptions. One is the homogeneity of the EFG applied on the nuclei of $^{83}$Kr at the surface. The other concerns the orientation and the symmetry of the EFG at the surface which are assumed to possess a principal axis in the surface normal direction (i.e. $V_{ZZ}\ge V_{YY}\ge V_{XX}$) and assumed to be axially symmetric (i.e. $V_{XX}=V_{YY}$). The latter assumption is a basis of the reduction of the transition lines from 11 to 4 with a polarized synchrotron radiation. In the current discussion, the realization of these two assumptions at the same time is disagreed with.  

Therefore, it is possible that either one of the two assumptions are true or that both two assumptions are not true even though the EFG at the surface was large enough to be observed. We further point out these three possibilities remained at this moment as potential explanations for the current result. The first possibility is based on the removal of the first assumption, where the inhomogeneous EFG is applied on the nuclei of $^{83}$Kr which leads to various yields of quadrupole splittings. In this condition, the time spectra are expressed as a sum of the time spectra of which the applied EFG is various which thus will be a complicated time spectra. Compared to the relatively simple time spectra simulated and shown in Fig. \ref{QS4} (a) to (f), the time spectra which is a sum of these various frequencies are inclined to be flat because any major dips in the spectra will not survive. These kind of situations are possible if the $^{83}$Kr atoms adsorb on the surface with an incommensurate adsorption structure \cite{Bruch}. The second possibility is based on the removal of the second assumption, where, the principal axis of the EFG is not parallel to the surface normal direction or the EFG in the surface parallel directions are not axially symmetric although the applied EFG is homogeneous. The principal axis is not parallel to the surface normal direction in cases where the EFG is stronger in surface normal direction. Even if the principal axis is parallel to the surface normal direction, the EFG needs not be axially symmetric on the (110) surface where the lattice constants in $x$ and $y$ directions are not identical. In this condition, the number of transitions is not reduced to 4 but remains to be 11. Although the time spectra are not simulated yet, the time spectra with a 11 transitions are expected to be more complicated than those in Fig. \ref{QS5}. The resultant time spectra may well be as flat as the observed spectrum.

\subsection{Electric field gradient}

In order to estimate the absolute value of the EFG at the surface, a tentative calculation of the electric filed at the TiO$_{2}$(110) surface was carried out. The calculated results suggest that there is a large EFG that may well be observed in the experiment. The calculation was carried out by taking sum of the Coulomb potential and by taking the derivative and the second derivative of it based on the assumption that each atom behave like a point charge of $+4$ and $-2$ which are located at the lattice sites of Ti$^{4+}$ and O$^{2-}$. With analytical calculations, the potential easily diverges to the infinity. To avoid this divergence of the summation, the Ewald-Kornfeld method was used \cite{Kittel} and a series of numerical calculations was conducted. The calculation was carried out by Takeyasu \cite{Takeyasu}.

\begin{figure}
\begin{center}
\includegraphics[scale=.6, clip]{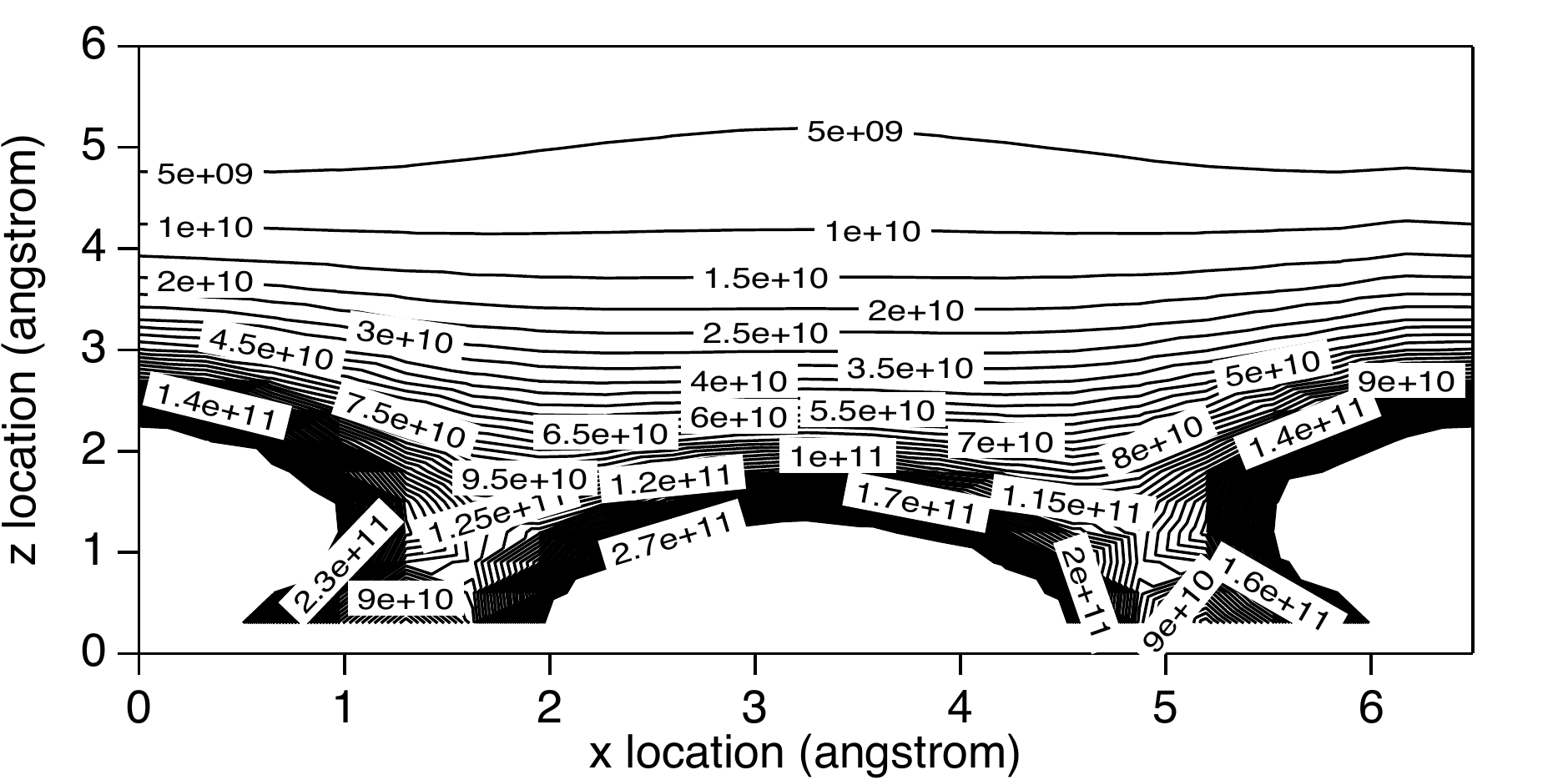}
\caption{A cross sectional view of the calculated electric field in the surface normal direction on TiO$_{2}$(110) surfaces. The unit of the figures in the contour plot is V/m. \label{EFG}}
\end{center}
\end{figure}

The representative result of the calculation is shown in Fig. \ref{EFG}. Figure \ref{EFG} shows a cross sectional contour plot of the electric field at the vicinity of the TiO$_{2}$(110) surface. The cross section is parallel to the surface normal direction. The cross section is orienting in a (100) direction if seen from the position $x=0$ to the position $x=6$ \AA. There is a titan atom at the position of ($x$, $z$)=(3.3 \AA, 0) in Fig. \ref{EFG}. There are oxygen atoms at the both sides of Fig. \ref{EFG}.

Previous report on the adsorption structure of Ar and Xe on TiO$_{2}$(110) both indicates that rare gas atoms adsorb on the outer most titan atoms. The distance of Ar and Xe varies from 3 \AA \ to 4 \AA\ \cite{Gomes2005, Gomes2010}. Absolute value of the electric field gradient at this position can be obtained by taking a derivative of the electric field in terms of $x$ and $z$, which roughly yields to be about an order of 10$^{20}$ V/m$^{2}$. The criteria for the EFG applied on the nuclei was about an order larger as 10$^{21}$ V/m$^{2}$. This deviation of an order or two may be compensated if one consider a factor called the Sternheimer's anti-shielding factor, which describes that the EFG applied on an atom is multiplied by the anti-shielding factor which is then applied on the nuclei due to the distortion of valence electrons \cite{Sternheimer1966}. The anti-shielding factor yields from 10 to more than 100 depending on the element \cite{Fukai}.

Another feature observed is that as the distance from the surface atoms is increased, the derivative of the electric field parallel to the surface plane become small as compared to that in surface normal direction. This observation supports the relation of $V_{ZZ}\ge V_{YY}\ge V_{XX}$.

\chapter{Concluding remarks}
In this chapter, conclusions for the study of laser desorption of Xe and for the nuclear resonant scattering study of Kr are described. The excitation path and the desorption dynamics are determined for the non-thermal desorption of Xe. The observation of Knudsen layer formation in laser induced thermal desorption of Xe is confirmed. The potential explanations for the observed time spectra of NRS are unambiguously presented. Lastly, future prospects for each experiments are mentioned.

\section{Laser desorption of Xe}
In the followings, the conclusion of the laser desorption studies are summarized. Firstly, the implication of the stabilization of the affinity level of the physisorbed Xe is mentioned. Secondly, the formation of the Knudsen layer due to the post-desorption collision effects are described.

\subsection{Stabilization of the affinity level}
We have investigated the PSD of Xe on Au(001) at photon energies of 2.3 and 6.4 eV \cite{Ikeda}.  With decreasing pump laser fluence, the desorption was found to undergo transition from thermal to non-thermal regimes.  The non-thermal PSD of Xe occurred only at 6.4 eV as a one-photon process, and the desorption proceeds via the Antoniewicz model with transient negative ion formation.  On the basis of the model calculation, the lifetime of Xe$ ^ { - } $ is estimated to be $\sim$15 fs. These results strongly suggest that the affinity level of Xe is substantially stabilized by the metal proximity effect.

\subsection{Knudsen layer formation}
We observed the TOFs of Xe from a Au(001) surface by LITD at a wide range of $\Theta$. At $\Theta\simeq0.3$ ML, the TOF is well analyzed by the Maxwell-Boltzmann velocity distribution at 255 K indicating thermal desorption followed by the collision-free flow. At $\Theta>0.3$ ML, the TOF is well analyzed by the shifted Maxwell-Boltzmann velocity distribution, which is regarded as the manifestations of the post-desorption collision effects. At $\Theta>4$ ML, the value of $u$ under the assumption that $T_{\mathrm{K}}/T_{\mathrm{S}}=0.65$ became constant at around 125 m/s, which corresponds to $M=0.96$, being in good agreement with the prediction by the Knudsen layer formation theory. In the LITD experiment, we found that the collision effect already appears at Kn $\simeq5.2$ and that the Knudsen layer is formed at Kn $<0.39$, which corresponds to a mean collision number of greater than 2.6 per atom for general systems.

\clearpage
\section{Nuclear resonant scattering by $^{83}$Kr}

NRS by physisorbed Kr on a solid surface in both multi-layer and monolayer regime are achieved for the first time thanks to the experimental realization of the glancing angle regime with an UHV chamber equipped with a cold head. The time spectra of NRS by 5 ML Kr shows a decay time of 96 ns whereas the decay time at 1 ML was 109 ns, indicating a dynamical effect is in play. The time spectrum at 1 ML Kr is compared with the simulated time spectra based on the quadrupole splitting and the selection rule where only 4 of 11 transitions are allowed. The experimentally obtained spectrum is not in good agreement with the simulated result with large electric field gradient (EFG). The comparison indicates the following possibilities.
\begin{enumerate}
\item EFG at the surface is too small ($<\sim10^{20}$ V/m$^{2}$) for the quadrupole splitting to be observed in the time spectra of NRS.
\item Although the electric field gradient is large enough for the quadrupole splitting to be observed, either one or neither of the following assumptions (that are used in the simulations) are fulfilled in the experiment.
\begin{enumerate}
\item The adsorption site of Kr at 1 ML is only one where the identical EFG is applied on all the $^{83}$Kr nuclei. 
\item The main axis of the EFG is parallel to the surface normal direction. The EFG in the surface parallel direction is axially symmetric.
\end{enumerate}
\end{enumerate}
The EFG at the surface of TiO$_{2}$(110) was estimated based on a numerical calculation. The result shows the absolute value of EFG was large enough ($>\sim10^{20}$ V/m$^{2}$) for the time spectra of NRS to be observed, although the axial symmetry on (110) surface is not positively confirmed.

\clearpage

\section{Future prospects}

As for the non-thermal photon stimulated desorption of Xe, I proposed a stabilization of Xe$^{-}$ state at the vicinity of the metal surface. This is a condensation effect. Further investigation on this effect may be possible if one investigate a coverage dependence of the photon stimulated desorption yield of Xe or if one observe the electron affinity level of Xe at surface more directly using methods such as two-photon photoemission spectroscopy or scanning tunneling spectroscopy on the similar system.

The potential specialty of Au surface may also be interesting as compared to Pt, Ag or else. Watanabe reported that there was no signifiant photon stimulated desorption of Xe from Pt(111) surfaces \cite{Kazuo}. Considering the inconsistency of this report and the present result \cite{Ikeda}, the fundamental difference between Au surface and Pt surface may be found by further investigations. The speciality may arise from the electronic structure of each surface originating from the $s$ electrons and $d$ electrons, respectively.

Angular resolved experiments possess a larger impact on the study of the collision effects in laser induced thermal desorption, in order to more vividly take a hold of the whole picture of the Knudsen layer formation. Another fascinating advanced experiment of the laser desorption should concern the desorption of binary mixture of different atoms or molecules. It has been theoretically predicted that the fast desorption of a mixture of heavy and light atoms will results in a jet effect, where heavy atoms gets larger and center oriented momentum in the surface normal direction \cite{Urbassek}. 

Cometary comma and its tails are assumed to be connected with a Knudsen layer \cite{Davidsson}. Thus, further exploration of the collision effects in laser induced thermal desorption has a potentially impact on the planetary science. For the simplicity of the phenomena, the present experiment was carried out using a mono atomic gas. The real comet is known to consists mainly of amorphous solid water \cite{Crovisier}. Therefore, the dynamics of the generation of the dust tail of comets may be clarified by an investigation on the rapid desorption of H$_{2}$O molecules from the surfaces, which may be realized in the experiment room by using the laser induced thermal desorption of water ice.

The present study showed that the NRS of Kr in the monolayer regime is able to be recorded. Next, the goal should be the use of the technique for the measurements of properties of materials. Further, detailed preparation of the sample surface including the adsorption structure and the axially symmetric substrate (e.g. (111) or (001) surfaces) are even more plausible for the experiment. The EFG of the solid surface will be observable with such a preparation.

The beauty of the adsorption of Kr on the surface is that they are not destructive at all. Therefore, the use of Kr as a probe of the surface properties is a promising way. The interaction between solid surfaces and the Kr atoms is only Van der Waals interactions. Hence, by varying the sample substrates, the EFG at respective surfaces such as metal, insulator and semiconductor surfaces can be compared by use of the NRS by Kr at each surfaces.

\backmatter

\begin{thebibliography}{100}
\expandafter\ifx\csname urlstyle\endcsname\relax
  \providecommand{\doi}[1]{doi:\discretionary{}{}{}#1}\else
  \providecommand{\doi}{doi:\discretionary{}{}{}\begingroup
  \urlstyle{rm}\Url}\fi

\bibitem{Sakurai}
J.~J. Sakurai, \emph{Modern Quantum Mechanics Revised Edition} (Addison-Wesley
  Publishing Company, 1994).

\bibitem{Zangwill}
A.~Zangwill, \emph{Physics at Surfaces} (Cambridge University Press, 1988).

\bibitem{Schonhense}
G.~Sch\"onhense, A.~Eyers, and U.~Heinzmann, Phys. Rev. Lett. \textbf{56}, 512
  (1986).

\bibitem{Eigler}
D.~M. Eigler and E.~K. Schweizer, Nature \textbf{344}, 524 (1990).

\bibitem{Weiss}
P.~S. Weiss and D.~M. Eigler, Phys. Rev. Lett. \textbf{69}, 2240 (1992).

\bibitem{Silva}
J.~L. F.~D. Silva, C.~Stampfl, and M.~Scheffler, Phys. Rev. Lett. \textbf{90},
  066104 (2003).

\bibitem{Kelkkanen}
A.~K. Kelkkanen, B.~I. Lundqvist, and J.~K. Norskov, Phys. Rev. B \textbf{83},
  113401 (2011).

\bibitem{Chen}
D.-L. Chen, W.~A. Al-Saidi, and J.~K. Johnson, Phys. Rev. B \textbf{84}, 241405
  (2011).

\bibitem{Bruch}
L.~W. Bruch, R.~D. Diehl, and J.~A. Venables, Rev. Mod. Phys. \textbf{79}, 1381
  (2007).

\bibitem{Mandel}
T.~Mandel, G.~Kaindl, M.~Domke, W.~Fischer, and W.~D. Schneider, Phys. Rev.
  Lett. \textbf{55}, 1638 (1985).

\bibitem{YCChen}
Y.~C. Chen, J.~E. Cunningham, and C.~P. Flynn, Phys. Rev. B \textbf{30}, 7317
  (1984).

\bibitem{Muller}
J.~E. Muller, Phys. Rev. Lett. \textbf{65}, 3021 (1990).

\bibitem{Bagus}
P.~S. Bagus, V.~Staemmler, and C.~Woll, Phys. Rev. Lett. \textbf{89}, 096104
  (2002).

\bibitem{Silva2008}
J.~L. F.~Da Silva and C.~Stampfl, Phys. Rev. B \textbf{77}, 045401 (2008).

\bibitem{Silva2}
J.~L. F.~Da Silva, C.~Stampfl, and M.~Scheffler, Phys. Rev. B \textbf{72},
  075424 (2005).

\bibitem{Ho1}
H.-L. Dai and W.~Ho, \emph{Laser Spectroscopy and Photo-chemistry on Metal
  Surfaces Part I and Part II} (World Scientific, Singapore, 1995).

\bibitem{Chuang}
T.~J. Chuang, Surf. Sci. Rep. \textbf{3}, 1 (1983).

\bibitem{Fukutani2003}
K.~Fukutani, K.~Yoshida, M.~Wilde, W.~A. Dino, M.~Matsumoto, and T.~Okano,
  Phys. Rev. Lett. \textbf{90}, 1972 (2003).

\bibitem{Wedler}
G.~Wedler and H.~Ruhmann, Surf. Sci. \textbf{121}, 464 (1982).

\bibitem{Hussla}
I.~Hussla, H.~Coufal, F.~Trager, and T.~J. Chuang, Can. J. Phys. \textbf{64},
  1070 (1986).

\bibitem{Yim}
C.~M. Yim, K.~L. Man, X.~Xiao, and M.~S. Altman, Phys. Rev. B \textbf{78},
  155439 (2008).

\bibitem{Schwalb}
C.~H. Schwalb, M.~Lawrenz, M.~D\"urr, and U.~H\"ofer, Phys. Rev. B \textbf{75},
  085439 (2007).

\bibitem{Feulner1987}
P.~Feulner, T.~Muller, A.~Puschmann, and D.~Menzel, Phys. Rev. Lett.
  \textbf{59}, 791 (1987).

\bibitem{Pearlstine}
K.~A. Pearlstine and G.~M. Mcclelland, Surf. Sci. \textbf{134}, 389 (1983).

\bibitem{Richter}
L.~J. Richter and R.~R. Cavanagh, Prog. Surf. Sci. \textbf{39}, 155 (1992).

\bibitem{Antoniewicz}
P.~R. Antoniewicz, Phys. Rev. B \textbf{21}, 3811 (1980).

\bibitem{Buckman}
S.~J. Buckman and C.~W. Clark, Rev. Mod. Phys. \textbf{66}, 539 (1994).

\bibitem{Nicolaides}
C.~A. Nicolaides and G.~Aspromallis, Phys. Rev. A \textbf{44}, 2217 (1991).

\bibitem{Bae}
Y.~K. Bae, J.~R. Peterson, A.~S. Schlachter, and J.~W. Stearns, Phys. Rev.
  Lett. \textbf{54}, 789 (1985).

\bibitem{Hird}
B.~Hird and S.~P. Ali, Can. J. Phys. \textbf{57}, 867 (1979).

\bibitem{Haberland}
H.~Haberland, T.~Kolar, and T.~Reiners, Phys. Rev. Lett. \textbf{63}, 1219
  (1989).

\bibitem{Schwentner1975}
N.~Schwentner, F.~J. Himpsel, V.~Saile, M.~Skibowski, W.~Steinmann, and E.~E.
  Koch, Phys. Rev. Lett. \textbf{34}, 528 (1975).

\bibitem{Nordlander}
P.~Nordlander, Phys. Rev. B \textbf{46}, 2584 (1992).

\bibitem{MGR1}
D.~Menzel and R.~Gomer, J. Chem. Phys. \textbf{41}, 3311 (1964).

\bibitem{MGR2}
P.~A. Redhead, Can. J. Phys. \textbf{42}, 886 (1964).

\bibitem{Avouris}
P.~Avouris and R.~E. Wlkup, Annu. Rev. Phys. Chem. \textbf{40}, 173 (1989).

\bibitem{Zheng}
C.~Z. Zheng, C.~K. Yeung, M.~M.~T. Loy, and X.~Xiao, Phys. Rev. Lett.
  \textbf{97}, 166101 (2006).

\bibitem{Wang2009}
X.~Wang, Y.~Y. Fei, and X.~D. Zhu, Chem. Phys. Lett. \textbf{481}, 58 (2009).

\bibitem{Klass}
K.~Klass, G.~Mette, J.~G\"udde, M.~D\"urr, and U.~H\"ofer, Phys. Rev. B
  \textbf{83} (2011).

\bibitem{Toker}
G.~Toker and M.~Asscher, Phys. Rev. Lett. \textbf{107}, 167402 (2011).

\bibitem{Kools1992}
J.~C.~S. Kools, T.~S. Baller, S.~T. Dezwart, and J.~Dieleman, J. Appl. Phys.
  \textbf{71}, 4547 (1992).

\bibitem{Kools1993}
J.~C.~S. Kools, E.~van~de Riet, and J.~Dieleman, Appl. Surf. Sci. \textbf{69},
  133 (1993).

\bibitem{Konomi2009}
I.~Konomi, T.~Motohiro, and T.~Asaoka, J. Appl. Phys. \textbf{106} (2009).

\bibitem{Konomi2010}
I.~Konomi, T.~Motohiro, T.~Kobayashi, and T.~Asaoka, Appl. Surf. Sci.
  \textbf{256}, 4959 (2010).

\bibitem{Singh}
R.~K. Singh and J.~Narayan, Phys. Rev. B \textbf{41}, 8843 (1990).

\bibitem{Puretzky}
A.~A. Puretzky, D.~B. Geohegan, G.~B. Hurst, M.~V. Buchanan, and B.~S.
  Luk'yanchuk, Phys. Rev. Lett. \textbf{83}, 444 (1999).

\bibitem{Cowin1978}
J.~P. Cowin, D.~J. Auerbach, C.~Becker, and L.~Wharton, Surf. Sci. \textbf{78},
  545 (1978).

\bibitem{NoorbatchaJCP1987}
I.~Noorbatcha, R.~R. Lucchese, and Y.~Zeiri, J. Chem. Phys. \textbf{86}, 5816
  (1987).

\bibitem{NoorbatchaPRB1987}
I.~Noorbatcha, R.~R. Lucchese, and Y.~Zeiri, Phys. Rev. B \textbf{36}, 4978
  (1987).

\bibitem{NoorbatchaSS1988}
I.~Noorbatcha, R.~R. Lucchese, and Y.~Zeiri, Surf. Sci. \textbf{200}, 113
  (1988).

\bibitem{Ytrehus}
T.~Ytrehus, in \emph{Rarefied Gas Dynamics}, edited by J.~L. Potter (AIAA, New
  York, 1977), p. 1197.

\bibitem{Cercignani}
C.~Cercignani, in \emph{Rarefied Gas Dynamics}, edited by S.~S. Fisher (AIAA,
  New York, 1981), p. 305.

\bibitem{Davidsson}
B.~Davidsson, Space Sci. Rev. \textbf{138}, 207 (2008).

\bibitem{KellySS1988}
R.~Kelly and R.~W. Dreyfus, Surf. Sci. \textbf{198}, 263 (1988).

\bibitem{KellyNIMB1988}
R.~Kelly and R.~W. Dreyfus, Nucl. Instrum. Methods B \textbf{32}, 341 (1988).

\bibitem{Stein}
O.~Stein, Z.~Lin, L.~V. Zhigilei, and M.~Asscher, J. Phys. Chem. A
  \textbf{115}, 6250 (2011).

\bibitem{Taborek}
P.~Taborek, Phys. Rev. Lett. \textbf{48}, 1737 (1982).

\bibitem{Burgess}
D.~Burgess, R.~Viswanathan, I.~Hussla, P.~C. Stair, and E.~Weitz, J. Chem.
  Phys. \textbf{79}, 5200 (1983).

\bibitem{Cowin1985}
J.~P. Cowin, Phys. Rev. Lett. \textbf{54}, 368 (1985).

\bibitem{HusslaBBPC1986}
I.~Hussla, H.~Coufal, F.~Trager, and T.~J. Chuang, Ber. Bunsen. Phys. Chem.
  \textbf{90}, 240 (1986).

\bibitem{HusslaCJP1986}
I.~Hussla, H.~Coufal, F.~Trager, and T.~J. Chuang, Can. J. Phys. \textbf{64},
  1070 (1986).

\bibitem{Sibold1991}
D.~Sibold and H.~M. Urbassek, Phys. Rev. A \textbf{43}, 6722 (1991).

\bibitem{Mossbauer}
R.~L. M\"ossbauer, Z. Physik \textbf{151}, 124 (1958).

\bibitem{Craig}
P.~P. Craig, J.~G. Dash, A.~D. McGuire, D.~Nagle, and R.~R. Reiswig, Phys. Rev.
  Lett. \textbf{3}, 221 (1959).

\bibitem{Lee}
Jr. L.~L.~Lee, L.~Meyer-Schutzmeister, J.~P. Schiffer, and D.~Vincent, Phys.
  Rev. Lett. \textbf{3}, 223 (1959).

\bibitem{Sano}
Hirotoshi Sano, \emph{M\"ossbauer Spectroscopy The Chemical Applications}
  (Kodansha, Tokyo, 1972).

\bibitem{Smirnov1996}
G.~V. Smirnov, Hyperfine Interact. \textbf{97-8}, 551 (1996).

\bibitem{Smirnov1999}
G.~V. Smirnov, Hyperfine Interact. \textbf{123-124}, 31 (1999).

\bibitem{Trammell1}
G.~T. Trammell and J.~P. Hannon, Phys. Rev. B \textbf{18}, 165 (1978).

\bibitem{Trammell2}
G.~T. Trammell and J.~P. Hannon, Phys. Rev. B \textbf{19}, 3835 (1979).

\bibitem{Gerdau}
E.~Gerdau, R.~R\"uffer, R.~Hollatz, and J.~P. Hannon, Phys. Rev. Lett.
  \textbf{57}, 1141 (1986).

\bibitem{Ruby1963}
S.~L. Ruby, Y.~Hazoni, and M.~Pasternak, Phys. Rev. \textbf{129}, 826 (1963).

\bibitem{Grunsteudel}
H.~Gr\"unsteudel, V.~Rusanov, H.~Winkler, W.~Klaucke, and A.~X. Trautwein,
  Hyperfine Interact. \textbf{122}, 345 (1999).

\bibitem{Sladecek}
M.~Sladecek, B.~Sepiol, J.~Korecki, T.~Slezak, R.~R\"uffer, D.~Kmiec, and
  G.~Vogl, Surf. Sci. \textbf{566}, 372 (2004).

\bibitem{Partykajankowska}
E.~Partykajankowska, B.~Sepiol, M.~Sladecek, D.~Kmiec, J.~Korecki, T.~Slezak,
  M.~Zajac, S.~Stankov, R.~Ruffer, and G.~Vogl, Surf. Sci. \textbf{602}, 1453
  (2008).

\bibitem{Vanhove1}
M.~A.~Van Hove, R.~J. Koestner, P.~C. Stair, J.~P. Biberian, L.~L. Kesmodel,
  I.~Bartos, and G.~A. Somorjai, Surf. Sci. \textbf{103}, 189 (1981).

\bibitem{Vanhove2}
M.~A.~Van Hove, R.~J. Koestner, P.~C. Stair, J.~P. Biberian, L.~L. Kesmodel,
  I.~Bartos, and G.~A. Somorjai, Surf. Sci. \textbf{103}, 218 (1981).

\bibitem{Figuera}
J.~de~la Figuera, M.~A. Gonzalez, R.~Garcia-Martinez, J.~M. Rojo, O.~S. Hernan,
  A.~L.~V. de~Parga, and R.~Miranda, Phys. Rev. B \textbf{58}, 1169 (1998).

\bibitem{Meyer}
R.~Meyer, C.~Lemire, S.~K. Shaikhutdinov, and H.~Freund, Gold Bulletin
  \textbf{37}, 72 (2004).

\bibitem{Kolk1}
B.~Kolk, Phys. Rev. B \textbf{12}, 1620 (1975).

\bibitem{Ruby1966}
S.~L. Ruby and H.~Selig, Phys. Rev. \textbf{147}, 348 (1966).

\bibitem{Das}
T.~P. Das and E.~L. Hahn, \emph{Nuclear Quadrupole Resonance Spectroscopy}
  (Academic Press, New York, London, 1958).

\bibitem{Kolk2}
B.~Kolk, Phys. Rev. B \textbf{12}, 4695 (1975).

\bibitem{KawauchiVac}
T.~Kawauchi, M.~Matsumoto, K.~Fukutani, T.~Okano, Y.~Suetsugu, X.~W. Zhang, and
  Y.~Yoda, Vacuum \textbf{83}, 873 (2009).

\bibitem{Hicks}
J.~M. Hicks, L.~E. Urbach, E.~W. Plummer, and H.-L. Dai, Phys. Rev. Lett.
  \textbf{61}, 2588 (1988).

\bibitem{Mcelhiney}
G.~Mcelhiney and J.~Pritchard, Surf. Sci. \textbf{60}, 397 (1976).

\bibitem{Koehler}
B.~G. Koehler and S.~M. George, Surf. Sci. \textbf{248}, 158 (1991).

\bibitem{Comsa}
G.~Comsa and R.~David, Surf. Sci. Rep. \textbf{5}, 145 (1985).

\bibitem{Zhang}
X.~Zhang, T.~Kawauchi, H.~Fujimoto, and T.~Okano, Jpn. J. Appl. Phys.
  \textbf{45}, L142 (2006).

\bibitem{Rohlsberger}
R.~R\"ohlsberger, \emph{Nuclear Condensed Matter Physics with Synchrotron
  Radiation (Basic Principles Methodology and Applications)} (Springer, 2004).

\bibitem{Feulner1984}
P.~Feulner, D.~Menzel, H.~J. Kreuzer, and Z.~W. Gortel, Phys. Rev. Lett.
  \textbf{53}, 671 (1984).

\bibitem{Moog}
E.~R. Moog, J.~Unguris, and M.~B. Webb, Surf. Sci. \textbf{134}, 849 (1983).

\bibitem{Watanabe2007}
K.~Watanabe, K.~H. Kim, D.~Menzel, and H.-J. Freund, Phys. Rev. Lett.
  \textbf{99}, 225501 (2007).

\bibitem{Watanabe2000}
K.~Watanabe, H.~Kato, and Y.~Matsumoto, Surf. Sci. \textbf{446}, L134 (2000).

\bibitem{Rao}
R.~M. Rao, R.~J. Beuhler, and M.~G. White, J. Chem. Phys. \textbf{109}, 8016
  (1998).

\bibitem{Schwentner1973}
N.~Schwentner, M.~Skibowski, and W.~Steinmann, Phys. Rev. B \textbf{8}, 2965
  (1973).

\bibitem{Stampfli}
P.~Stampfli and K.~H. Bennemann, Phys. Rev. A \textbf{38}, 4431 (1988).

\bibitem{Martyna}
G.~J. Martyna and B.~J. Berne, J. Chem. Phys. \textbf{88}, 4516 (1988).

\bibitem{Walkup}
R.~E. Walkup, P.~Avouris, N.~D. Lang, and R.~Kawai, Phys. Rev. Lett.
  \textbf{63}, 1972 (1989).

\bibitem{Padowitz}
D.~F. Padowitz, W.~R. Merry, R.~E. Jordan, and C.~B. Harris, Phys. Rev. Lett.
  \textbf{69}, 3583 (1992).

\bibitem{Merry}
W.~R. Merry, R.~E. Jordan, D.~F. Padowitz, and C.~B. Harris, Surf. Sci.
  \textbf{295}, 393 (1993).

\bibitem{Blagoev}
A.~Blagoev, I.~Ivanov, T.~Mishonov, and T.~Popov, J. Phys. B \textbf{17}, L647
  (1984).

\bibitem{Arakawa2}
I.~Arakawa, T.~Adachi, T.~Hirayama, and M.~Sakurai, Low Temp. Phys.
  \textbf{29}, 259 (2003).

\bibitem{Ikeda}
A.~Ikeda, M.~Matsumoto, S.~Ogura, K.~Fukutani, and T.~Okano, Phys. Rev. B
  \textbf{84}, 155412 (2011).

\bibitem{SiboldJAP1993}
D.~Sibold and H.~M. Urbassek, J. Appl. Phys. \textbf{73}, 8544 (1993).

\bibitem{Brand}
J.~L. Brand and S.~M. George, Surf. Sci. \textbf{167}, 341 (1986).

\bibitem{KellyPRA1992}
R.~Kelly, Phys. Rev. A \textbf{46}, 860 (1992).

\bibitem{SiboldPFFD1993}
D.~Sibold and H.~M. Urbassek, Phys. Fluids A \textbf{5}, 243 (1993).

\bibitem{Kelly1990}
R.~Kelly, J. Chem. Phys. \textbf{92}, 5047 (1990).

\bibitem{KellyNIMB1992}
R.~Kelly, A.~Miotello, B.~Braren, A.~Gupta, and K.~Casey, Nucl. Instrum.
  Methods B \textbf{65}, 187 (1992).

\bibitem{Baron}
A.~Q.~R. Baron, A.~I. Chumakov, S.~L. Ruby, J.~Arthur, G.~S. Brown, G.~V.
  Smirnov, and U.~Vanburck, Phys. Rev. B \textbf{51}, 16384 (1995).

\bibitem{Johnson}
D.~E. Johnson, D.~P. Siddons, J.~Z. Larese, and J.~B. Hastings, Phys. Rev. B
  \textbf{51}, 7909 (1995).

\bibitem{Gomes2005}
J.~R.~B. Gomes and J.~P.~Prates Ramalho, Phys. Rev. B \textbf{71}, 235421
  (2005).

\bibitem{Gomes2010}
J.~R.~B. Gomes, J.~P.~P. Ramalho, and F.~Illas, Surf. Sci. \textbf{604}, 428
  (2010).

\bibitem{Meixner1993}
D.~L. Meixner and S.~M. George, J. Chem. Phys. \textbf{98}, 9115 (1993).

\bibitem{Meixner1994}
O.~Sneh and S.~M. George, J. Chem. Phys. \textbf{101}, 3287 (1994).

\bibitem{Ellis}
J.~Ellis, A.~P. Graham, and J.~P. Toennies, Phys. Rev. Lett. \textbf{82}, 5072
  (1999).

\bibitem{Nabighian}
E.~Nabighian and X.~D. Zhu, Chem. Phys. Lett. \textbf{316}, 177 (2000).

\bibitem{Thomas}
P.~Thomas, J.~Gray, X.~D. Zhu, and C.~Y. Fong, Chem. Phys. Lett. \textbf{381},
  376 (2003).

\bibitem{Holloway}
J.~H. Holloway, G.~J. Schrobilgen, S.~Bukshpan, W.~Hilbrants, and H.~Dewaard,
  J. Chem. Phys. \textbf{66}, 2627 (1977).

\bibitem{Faust}
W.~L. Faust and L.~Y.~Chow Chiu, Phys. Rev. \textbf{129}, 1214 (1963).

\bibitem{Kittel}
Charles Kittel, \emph{Introduction to Solid State Physics (8th edition)} (John
  Wiley \& Sons, Inc, 2005).

\bibitem{Takeyasu}
K.~Takeyasu, Unpublished .

\bibitem{Sternheimer1966}
R.~M. Sternheimer, Phys. Rev. \textbf{146}, 140 (1966).

\bibitem{Fukai}
Y.~Fukai, Phys. Lett. A \textbf{34}, 425 (1971).

\bibitem{Kazuo}
Kazuo~Watanabe (private communication) .

\bibitem{Urbassek}
H.~M. Urbassek and D.~Sibold, Phys. Rev. Lett. \textbf{70}, 1886 (1993).

\bibitem{Crovisier}
J.~Crovisier and T.~Encrenaz, \emph{Comet Science} (Cambridge University Press,
  2000).

\end{thebibliography}

\end{document}